\documentclass[12pt]{elsarticle}
\usepackage{lineno,hyperref}
\modulolinenumbers[5]

\biboptions{authoryear}

\usepackage{setspace}
\usepackage{amscd,amsfonts}
\usepackage{amsmath,amsthm,bbm,bm,mathrsfs}
\usepackage{epsfig,graphics,subfigure,longtable,float}
\usepackage[english]{babel}
\usepackage[usenames]{color}
\usepackage[top=1.2in, bottom=1.6in, left=1.1in, right=1.1in]{geometry}
\usepackage{multirow}
\usepackage{ctable}
\usepackage{makecell}
\usepackage{tabularx}
\usepackage{rotating}
\usepackage{tabularray}
\usepackage{comment}
\usepackage[toc,page]{appendix}
\usepackage{enumitem}
\setlist[enumerate]{label*=\arabic*.}

\usepackage{caption}
\newtheorem{defn}{Definition}
\newtheorem{prop}{Proposition}
\newtheorem{rem}{Remark}
\usepackage{rotating}

\begin{document}

\begin{frontmatter}

\title{First--order integer--valued autoregressive processes with Generalized Katz innovations}

\author[2]{Ovielt Baltodano Lopez}
\ead{ovielt.baltodano@unive.it}
\author[1]{Federico Bassetti\corref{cor1}}
\ead{federico.bassetti@polimi.it}
\cortext[cor1]{Corresponding author}
\author[2]{Giulia Carallo}
\ead{giulia.carallo@unive.it}
\author[2]{Roberto Casarin}
\ead{r.casarin@unive.it}

\address[1]{Polytechnic University of Milan, Italy}
\address[2]{Ca' Foscari University of Venice, Fondamenta San Giobbe, 873, 30121 Venice, Italy.}

\begin{abstract}
\textcolor{black}{A new integer--valued autoregressive process (INAR) with Generalised Lagrangian Katz (GLK) innovations is defined. This process family provides a flexible modelling framework for count data, allowing for under and over--dispersion, asymmetry, and excess of kurtosis and includes standard INAR models such as Generalized Poisson and Negative Binomial as special cases. We show that the GLK--INAR process is discrete semi--self--decomposable, infinite divisible, stable by aggregation and provides stationarity conditions. Some extensions are discussed, such as the Markov--Switching and the zero--inflated GLK--INARs. A Bayesian inference framework and an efficient posterior approximation procedure are introduced. The proposed models are applied to 130 time series from Google Trend, which proxy the worldwide public concern about climate change. New evidence is found of heterogeneity across time, countries and keywords in the persistence, uncertainty, and long--run public awareness level.}
\end{abstract}

\begin{keyword}
Bayesian inference \sep Big data \sep Counts time series \sep Climate Risk \sep Generalized Lagrangian Katz distribution \sep MCMC
\end{keyword}

\end{frontmatter}


\doublespacing

\section{Introduction}\label{sec:Intro}
In the recent years there has been a large interest in discrete--time integer--valued models, also due to increased availability of count data in very diverse fields including finance \citep{LieMolPoh2006,AknoucheAlmohaimeedDimitrakopoulos2021}, economics \citep{FreCab2004,Berri2020}, social sciences \citep{PedKar2011}, sports \citep{ShaMoy2016}, image processing \citep{Afrifa2022} and oceanography \citep{CunVasBou2018}. Among the modelling approaches, integer--valued autoregressive processes (INAR), introduced independently by \cite{AlOAlz1987} and \cite{McK1985}, become \textcolor{black}{very} popular. The stochastic construction of the INAR relies on the binomial thinning operator and the properties of the model on the discrete self--decomposability of the stationary distribution of the process \citep{Ste79}. See \cite{ScoWeiGou2015} for a review.

The original INAR model has been studied further in \cite{AlOAlz1987} and extended in different directions. \citep{McK1986} introduced an INAR model with negative--Binomial and geometric marginal distributions, \cite{DuLi1991} extended the INAR(1) model of \cite{AlOAlz1987} to the higher order INAR$(p)$. \cite{AlOAly1992} introduced a negative--binomial INAR with a new iterated thinning operator. Other extensions of the INAR process have been made to include a seasonal structure in the model \citep[e.g., see][]{Bouetal16}. INAR models with values in the set of signed integers have been propose firstly by \cite{KimPar2008} and generalised by \cite{AlzOma2014} and \cite{AndKar2014}. \cite{Fre2010} proposed a true integer--valued autoregressive model (TINAR(1)). More flexible INAR models have been introduced by assuming more flexible distributions for the innovations terms. \cite{AlzAlO1993} propose integer--valued ARMA process with Generalized Poisson marginals and \cite{KimLee2017} introduced INAR with Katz innovations.

This paper introduces a general class of INARs with Generalized Lagrangian Katz innovations. The Lagrangian Katz family is a flexible distribution and naturally arises as first crossing probabilities, which is a common problem in actuarial mathematics, e.g. claim number distribution in cascading processes or ruin probability in discrete--time risk models \citep[e.g., see][ ch. 12]{ConFam2006}. It has been extended further by \cite{jan98} and \cite{jan99}, which introduced the four--parameter generalized P\'{o}lya--Eggenberger (GPED) distributions of the first and second kind. \cite{jan98}  showed that both families contain the Lagrangian Katz distribution as a special case. 
\textcolor{black}{We consider the four-parameters GPED of the first kind, also known as Generalized Lagrangian Katz (GLK). The resulting process family provides a flexible modelling framework for count data, allowing for under and over--dispersion, asymmetry, and excess of kurtosis and includes standard INAR models such as Generalized Poisson and Negative Binomial as special cases.} \textcolor{black}{Further extensions are provided, such as the Markov--Switching and the zero--inflated GLK--INARs,  to account for different sources of model instability and excess of zeros.  
}


Various approaches to inference have traditionally been presented for count data models, such as the conditional likelihood approach, generalized method of moments and Yule--Walker approach. See \cite{WeiKim2013} for a review. Despite the popularity gained in recent years by Bayesian methods, the applications to count data models are still limited \cite[e.g., see][]{CabMar05, NeSu07, Droetal16, ShaZha18, Garetal20}. Thus, we provide a Bayesian inference procedure for our model and illustrate the procedure's efficiency on a synthetic dataset. 
The Bayesian approach to inference entirely considers parameter uncertainty in the prior knowledge about a random process. It allows for imposing parameter restrictions by specifying the prior distribution \citep{Chen2016}. The posterior distribution of the parameters quantifies uncertainty in the estimation \citep{Chen17}, which can be included in the prediction.  The inference from the Bayesian perspective may result in richer inferences in the case of small samples \citep{Garayetal20} and extra--sample information and in robust inference in the presence of outliers \citep{fried2015retrospective}. Finally, model selection for both nested and non--nested models can be easily carried out.

\textcolor{black}{We illustrate the model's flexibility with an application to an original Google Trend dataset of 130 time--series measuring the public concern about climate change in different countries. The contrasting features of the series, such as excess of zeros, outliers, and regimes, are common in count data and provide a challenging and diversified ground for illustrating the robustness and flexibility of the GLK--INAR model}. Assessing public awareness and knowledge of a specific topic and understanding the dynamics of social consciousness allows for designing more effective public policies. For this reason, researchers measured and studied the level of awareness about the effects of climate change in different sectors of society such as households \citep{Frondel2017}, winegrowers \citep{bat2009}, farmers \citep{Fahad2018}, mountain peoples \citep{Ullah2018}. Most of these studies rely on surveys conducted in a specific geographical area and sector of society, with a few exceptions. For example, \cite{Zieg2017} proposed a cross--country analysis of climate change beliefs and attitudes. \cite{lineman2015talking} provided a broader and global perspective by exploiting the potentiality of big data provided by Google Trend. This extended climate change perception literature along two lines. First, we consider a multi--country dataset, including country--specific measures to capture worldwide heterogeneity in public awareness. Moreover, we offer a model--based approach and an inference procedure to analyze these measures.   

The paper is organized as follows. Section 2 introduces the GLK family and INAR process with some extensions such as the Markov--Switching GLK--INAR. Section 3 proposes a Bayesian inference procedure and provides some simulation results. Section 4 provides some illustrations on a multi--country Google Trend dataset related to climate change. Section 5 concludes.

\section{INAR(1) with generalized Katz innovations}\label{sec:INARtheory}
\subsection{Generalized Lagrangian Katz family}
The probability mass function (pmf), $P(X=x)=p_x$, of the Generalized Lagrangian Katz (GLK) is
\begin{equation}\label{pmf}
p_x = 
\frac{1}{x!}\beta^{x}\frac{a}{c} \frac{1}{(\frac{a}{c}+x\frac{b}{c}+x) }(1-\beta)^{\frac{a}{c}+x\frac{b}{c}} \left( \frac{a}{c}+x\frac{b}{c} +1\right)_{{x}\uparrow}  
\end{equation}
$x=0,1,2,\ldots$, where  $(x)_{{k} \uparrow} = x(x+1) \ldots (x+k-1)$ is the rising factorial with the convention that $(x)_0=1$, 
and $a >0$, $c >0$, $b \geq -c$ and $0 < \beta < 1$ are the parameters  \citep{ConFam2006}. 
We denote the distribution with $\mathcal{GLK}(a,b,c,\beta)$. 
We notice that for $-c<b<0$ some additional constraints 
 on the parameters are needed to have all the 
 $p_x\geq 0$. See the discussion at the beginning of 
 Subsection \ref{SubSec:prior} and Appendix  \ref{App:ThProof} in the Supplementary.
GLK distributions have probability generating function (pgf) 
\begin{equation*}
H(u)=\sum_{x=0}^{\infty}p_x u^{x}
\end{equation*}
which satisfies:
\begin{equation}\label{pgf1}
H(u)=(1-\beta+\beta z)^{a/c},\,\, z=u(1-\beta+\beta z)^{b/c+1},
\end{equation}
or alternatively
\begin{equation}\label{pgf2}
H(u)=\left((1-\beta)/(1-\beta z)\right)^{a/c},\,\, z=u\left((1-\beta z)/(1-\beta)\right)^{b/c},
\end{equation}
see \cite{jan98}.

\begin{rem}\label{rem:GLKpmf}
Building on the Lagrangian expansion, \cite{jan98} introduced the Generalized Polya Eggenberger distribution. \citep{ConFam2006} argued that since the distribution is unrelated to the Polya, it should be named Generalized Lagrangian Katz distribution. As shown in Appendix \ref{App:ThProof} in the Supplementary Material, it is possible to derive the Generalized Polya Eggenberger / Generalized Lagrangian Katz as a particular  "generalized Lagrangian distribution". 
\end{rem}

The GLK distribution family is very general and includes some well--known distributions and new distributions that have yet to be used in count data modelling.

\begin{itemize}
\item The Lagrangian Katz distribution $\mathcal{LK}(a,b,\beta)$ is obtained 
 by replacing $c$ with $\beta$ 
 (which is called Generalized Katz in \citep{ConFam2006}). 
 \item The Katz distribution $\mathcal{K}(a,\beta)$ is obtained  for $b=0$ and by replacing $c$ with $\beta$,
  \citep{katz1965}.
  \item  The Polya--Eggenberger distribution $\mathcal{PE}(a,c,\beta)$ is obtained for $b=0$, 
   \citep{jan98}. Note that the Katz distribution in \cite{ConFam2006}, Tab. 2.1, is not the Katz distribution of \cite{katz1965}, it corresponds instead to the Generalized Polya Eggenberger of the first type (GPED$_{1}$--I) of \cite{jan98} and can be obtain as the limit of the zero--truncated GLK for $a\rightarrow -c$.
 \item  The Generalized Negative Binomial distribution $\mathcal{GNB}(r,\gamma,p)$ is obtained for $c=1$, $a=r$, $b=\gamma-1$ and $\beta=p$ .
\item The Negative Binomial distribution $\mathcal{NB}(r,p)$ is obtained for $b=0$ $\beta=1-p$ and $r=a/c$. 
 \item The Binomial distribution $\mathcal{B}in(n,p)$ is obtained  for $c=1$, $b=-1$, $a=n\in\mathbb{N}$ and $\beta=p$
  \item The Generalized Poisson (GP) distribution $\mathcal{GP}(\theta,\lambda)$ for $c \rightarrow 0$ s.t. $b/a = \lambda$ and $a\beta/c = \theta>0$ with $0<\lambda<\theta^{-1}$. The GP limit of the GLK distribution is stated in \citep{ConFam2006} without proof. In  Appendix \ref{Appendix:LKlimit} in the Supplementary Material, we provide a proof. 
 \item   The Poisson distribution $\mathcal{P}(\theta)$ for $c \rightarrow 0$, $b \rightarrow 0$ s.t. $a\beta/c = \theta$.
\end{itemize}

\begin{figure}[t]
\begin{center}
\renewcommand{\arraystretch}{2}
\setlength{\tabcolsep}{5pt}
\begin{tabular}{cc}
\includegraphics[scale=0.47]{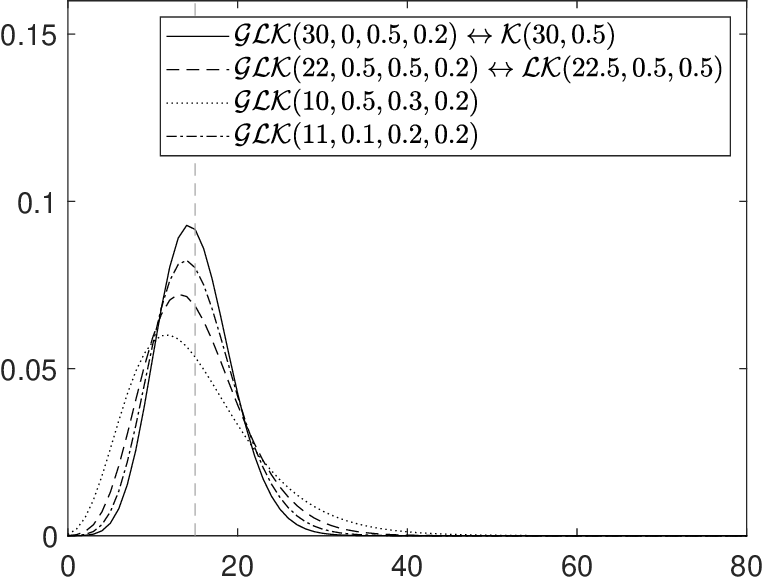}&
\includegraphics[scale=0.47]{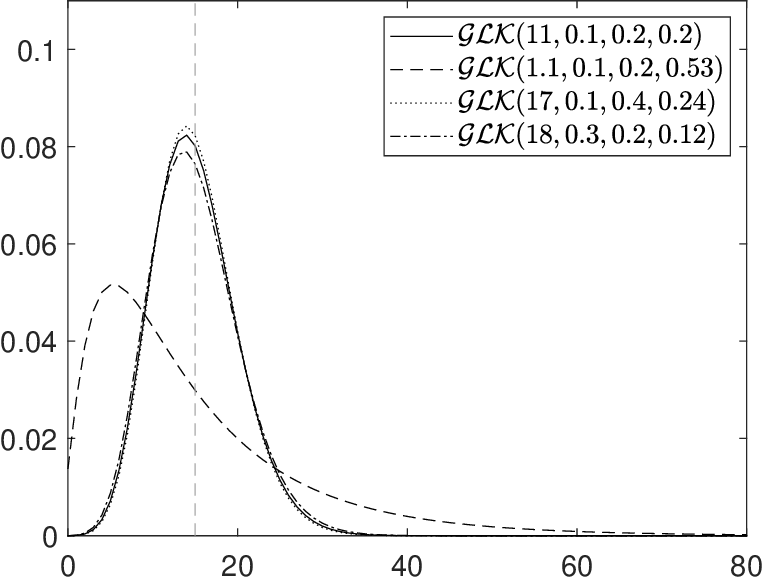}\\
\includegraphics[scale=0.47]{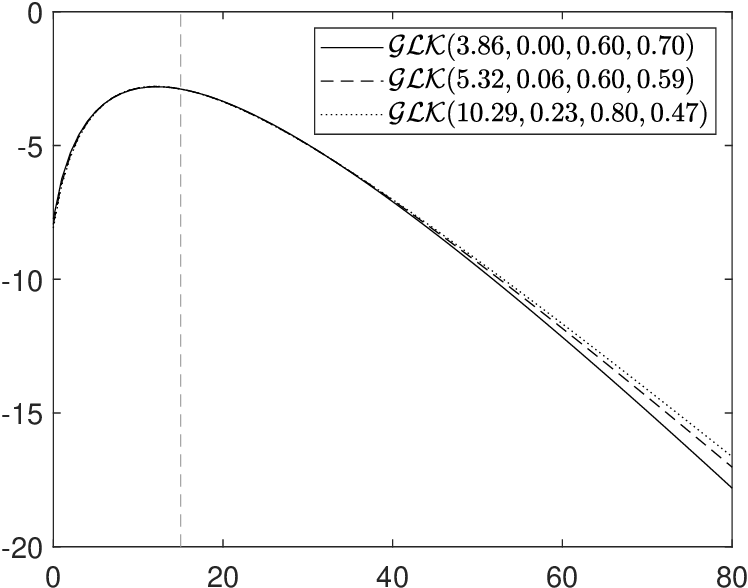}&
\includegraphics[scale=0.47]{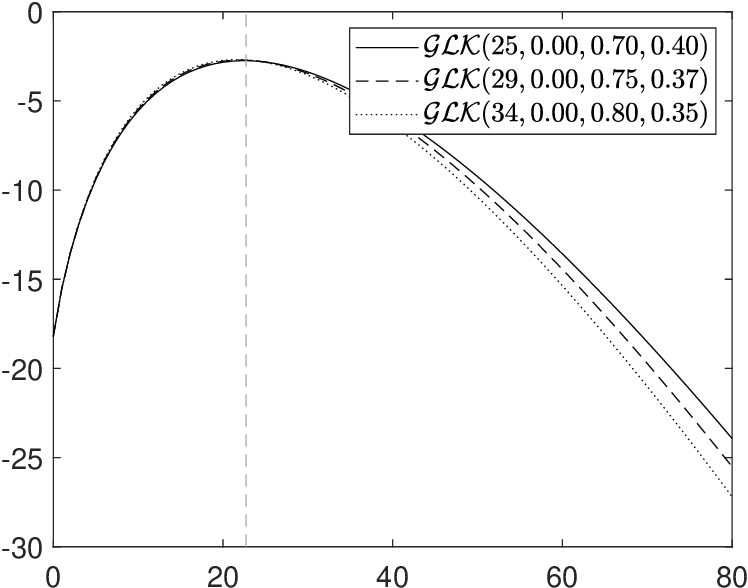}\\
\end{tabular}
\end{center}
\caption{Probability mass function of the Generalized Lagrangian Katz for different parameter settings. Top left: comparison between $\mathcal{LK}(a,c)$, $\mathcal{LK}(a,b,\beta)$ and $\mathcal{GLK}(a,b,c,\beta)$. Top right: sensitivity of $\mathcal{GLK}(a,b,c,\beta)$ with respect to the parameters. Bottom: effect of the parameters on the tails (log scale) for a $\mathcal{GLK}(a,b,c,\beta)$ with over--dispersion ($VMR=50/15$, left) and under--dispersion ($VMR=13/15$, right). In each plot the distribution mean (vertical dashed line).}\label{fig:pmf}
\end{figure}

The probability mass function of the GLK for different parameter settings is given in Fig. \ref{fig:pmf}. In the top--left plot, we compare 
$\mathcal{K}(a,c)$, $\mathcal{LK}(a,b,\beta)$
and $\mathcal{GLK}(a,b,c,\beta)$ with the same mean. The top--right plot illustrates the sensitivity of the $\mathcal{GLK}(a,b,c,\beta)$ pmf with respect to the different parameters. All distributions have the same mean (vertical dashed line). The bottom plots illustrate the effects of the parameters on the tails (log--scale) for a $\mathcal{GLK}(a,b,c,\beta)$ with over--dispersion $VMR=50/15$ (left) and under--dispersion $VMR=13/15$ (right). 

We provide in Appendix \ref{App:Prop} in the Supplementary Material some useful moments of the GLK distributions, which can be used to derive the following measures of dispersion. The standard deviation to the mean ratio returns the coefficient of variation. From the results in Appendix \ref{App:Prop} in the Supplementary Material it follows that the coefficient of variation is $CV =\left((1-\beta)/(a\theta\kappa)\right)^{1/2}$
where $\kappa=1 -\beta - b \beta/c$ and $\theta=\beta/c$, assuming $\kappa>0$.
The Fisher index is given by the  variance--to--mean ratio $VMR =(1-\beta)/(\kappa^{2})$ which does not depend on the parameter $a$. For a given $\beta$, following the values of $\kappa$ ($b$ and $c$), the distribution allows for various degrees of dispersion: not dispersed ($VMR = 0$), under--dispersed ($VMR < 1$), equally dispersed ($VMR = 1$) and over--dispersed ($VMR > 1$).

We conclude this section with another important property.
\begin{prop}\label{Th:infdivGLK}
A random variable $X\sim\mathcal{GLK}(a,b,c,\beta)$ is infinite divisible, 
in particular $X \stackrel{\mathcal{L}}{=} \sum_{j=1}^n X_{jn}$ where $ X_{jn} \stackrel{iid}{\sim} \mathcal{GLK}(a/n,b,c,\beta)$.
\end{prop}

\begin{proof}
From the pgf of a GLK given in Eq. \ref{pgf2}
\begin{equation}
\mathbb{E}(X)=\mathbb{E}\left(u^X\right)=\left(\frac{1-\beta}{1-\beta z}\right)^{\frac{a}{c}}=\prod_{j=1}^{n}\left(\frac{1-\beta}{1-\beta z}\right)^{\frac{a}{nc}}
\end{equation}
which is the pgf of the sum of $n$ independent GLKs with distribution $\mathcal{GLK}(a/n,b,c,\beta)$ where $a/n>0$ according to the definition of GLK.
\end{proof}

\subsection{A INAR(1) process}
The Generalized Katz INAR(1) process (GLK--INAR(1)) is defined using the binomial thinning operator, $\circ $. The binomial thinning for a non--negative discrete random variable $X$ is defined as
\begin{equation*}
\alpha \circ X = \sum_{i=1}^{X} B_{i}(\alpha)
\end{equation*}
where $B_{i}(\alpha)$ are iid Bernoulli r.v.s with success probability $P(B_{i}(\alpha)=1)=\alpha$.
\begin{defn}[GLK--INAR process]
For $\alpha \in (0,1)$, 
the GLK--INAR(1) process is defined by
\begin{equation*}\label{def:INAR}
X_{t} = \alpha \circ X_{t-1} +\varepsilon_{t}, \qquad t \in \mathbb{Z}
\end{equation*}
where $\varepsilon_{t}$ are iid random variables with Generalized Lagrangian Katz distribution $\mathcal{GLK}(a,b,c,\beta)$, independent of $X_{s}$, $s \leq t-1$. 
\end{defn}

\begin{figure}[h!]
\begin{center}
\renewcommand{\arraystretch}{1}
\setlength{\tabcolsep}{5pt}
\begin{tabular}{cc}
\multicolumn{2}{c}{\small (a) Over--dispersed innovations}\vspace{8pt}\\
\includegraphics[scale=0.47]{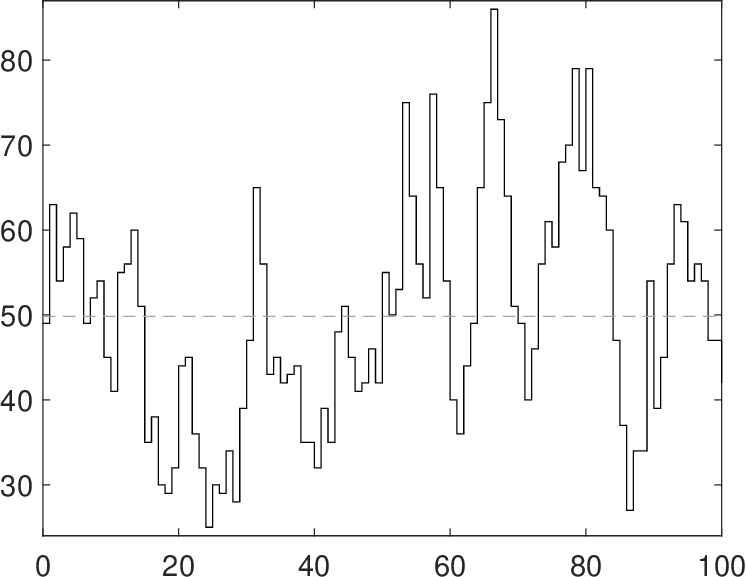}&
\includegraphics[scale=0.47]{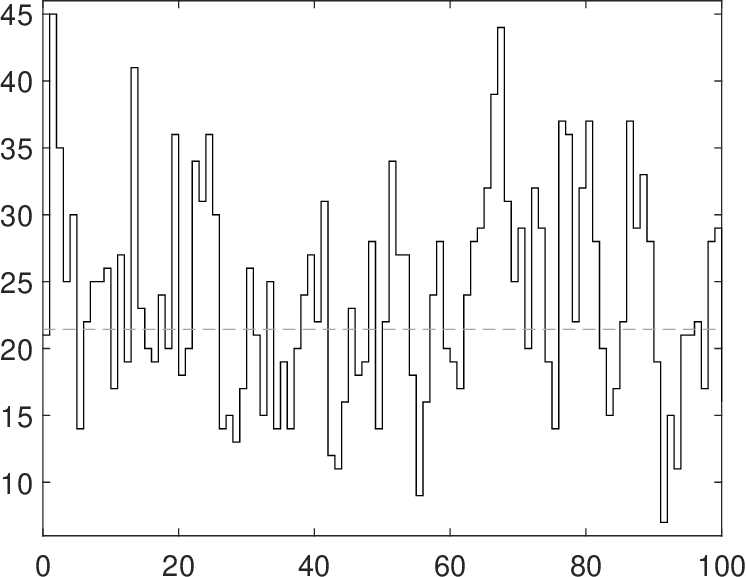}\\
\multicolumn{2}{c}{\small (a) Under-dispersed innovations}\vspace{8pt}\\
\includegraphics[scale=0.47]{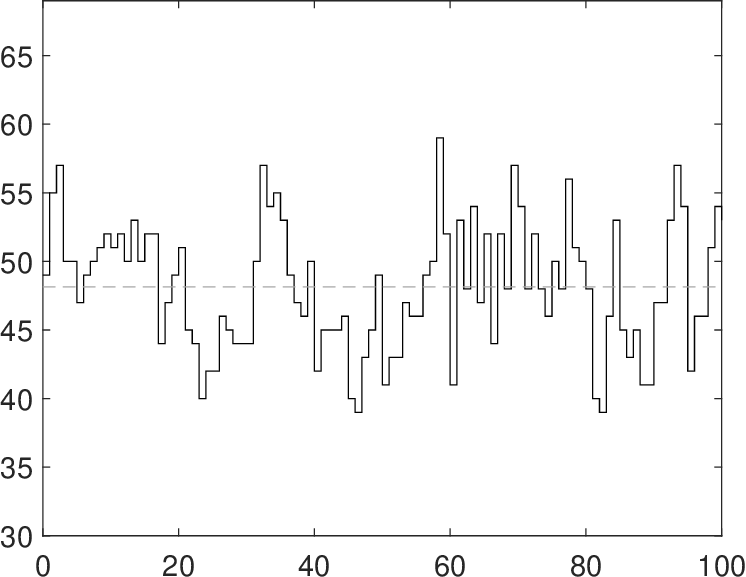}&
\includegraphics[scale=0.47]{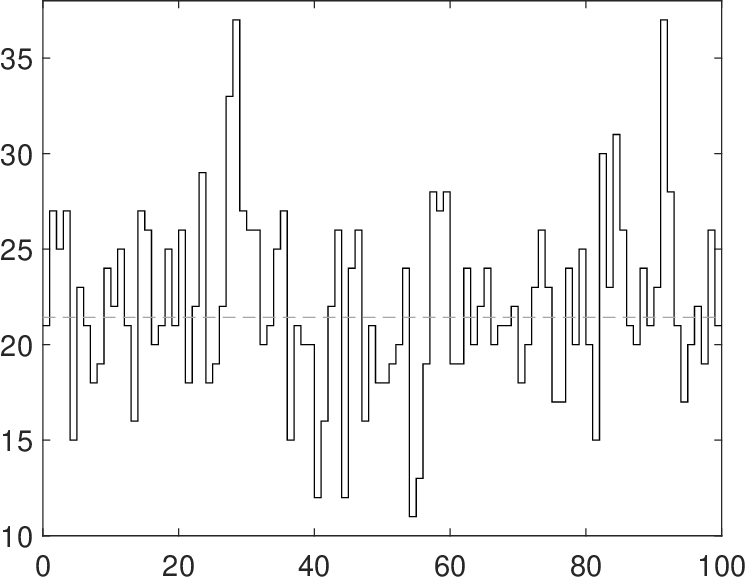}\\
\end{tabular}
\end{center}
\caption{Trajectories of the GLK--INAR(1) in the high ($\alpha=0.7$, left column) and low persistence ($\alpha=0.3$, right column) regimes. The trajectories in the over- and under-dispersion settings are in the rows. In all plots, the empirical mean of the observations (dashed line).}\label{fig:inartraj}
\end{figure}

Figure \ref{fig:inartraj} provides some trajectories of $T=100$ points each, simulated from a GLK--INAR(1) with innovation distributions given by the solid lines in the bottom plots of Fig. \ref{fig:pmf}, that are $\mathcal{GLK}(3.86$, $0$, $0.60$, $0.70)$ (overdispersion) and  $\mathcal{GLK}(25.00$, $0.00$, $0.70$, $0.42)$ (underdispersion). The trajectories correspond to the two parameter settings we find the empirical application to climate change discussed in Section \ref{sec:climate}, that are: (i) high persistence setting ($\alpha=0.7$, left); (ii) low persistence setting ($\alpha=0.3$, right). In all plots, 
the empirical mean of the observations is reported (dashed line) as a reference to illustrate the different levels of persistence in the trajectories.

Thanks to the general parametric family assumed, by setting $b=0$, $c=\beta=\theta_1$ and $a=\theta_2$, our GLK--INAR(1) nests the INARKF(1) of \cite{KimLee2017} as special case. The GLK--INAR(1) naturally nests the Poisson INAR(1) of \cite{AlOAlz1987}, the Negative Binomial INAR(1) of \cite{AlOAly1992} (NBINAR(1)), and the Generalized Poisson INAR(1) of \cite{AlzAlO1993}.

As for any INAR process, the GLK--INAR(1) has the following representation
\begin{equation*}\label{sumrepINAR}
X_{t+k}=\alpha^{k} \circ X_{t}+ \sum_{j=0}^{k-1}\alpha^j\circ\varepsilon_{t+1-j}
\end{equation*}
and its conditional pgf can be written as 
\begin{equation*}\label{conditionalPGF}
H_{X_{t+k}|X_{t}}(u)=(1-\alpha^k+\alpha^k u)^{X_{t}}\prod_{j=0}^{k-1}H(1-\alpha^j+\alpha^j u)
\end{equation*}
where $H(u)$ is  defined in Eq. \ref{pgf1} or in Eq. \ref{pgf2}. Starting from the general results on INAR processes given in \cite{AlOAlz1988}, 
 one easily obtains explicit expressions for 
 the conditional mean and variance of the  GLK--INAR(1):
\begin{eqnarray}\label{momentINAR}
&&\mathbb{E}(X_{t+k}|X_t)=\alpha^{k} X_{t}+ \frac{1-\alpha^{k-1}}{1-\alpha}\frac{a\theta}{\kappa}\\
&&\mathbb{V}(X_{t+k}|X_t)= \frac{1-\alpha^{2k}}{1-\alpha^2} \left(\frac{a(1-\beta)\theta}{\kappa^3}-\frac{a\theta}{\kappa}\right)\nonumber\\
&&\hspace{70pt}+(\alpha^{k}-\alpha^{2k}) X_{t}+\frac{1-\alpha^{k}}{1-\alpha}\frac{a\theta}{\kappa}
\end{eqnarray}
where $\kappa=1 -\beta - b \beta/c$ and $\theta=\beta/c$.

\begin{rem}\label{rem:specialcase}
Setting $b=0$, $c=\beta=\theta_1$ and $a=\theta_2$ the results in \cite{KimLee2017} Th. 2.2 are obtained.
\end{rem}

\begin{rem}\label{rem:lim}
Since $\alpha <1$, $\underset{k\rightarrow\infty}{\lim}\mathbb{E}(X_{t+k}|X_t)=a\theta/(\kappa(1-\alpha)$ and $\underset{k\rightarrow\infty}{\lim}\mathbb{V}(X_{t+k}|X_t)=a\theta((1-\beta)+\alpha\kappa^2)/((1-\alpha^2)\kappa^3)$.
\end{rem}

The process $\{X_t\}_{t\in\mathbb{Z}}$  is a Markov Chain on $\mathbb{N}$  and  the  
 transition probability  $P_{i,j} = \mathbb{P}(X_{t}=j | X_{t-1} = i)$ satisfies 
\begin{eqnarray*}\label{transition}
&&P_{i,j} = \sum_{k=0}^{\min(i,j)} \mathbb{P}(\alpha \circ X_{t-1} = k|X_{t-1} = i)\mathbb{P}(\varepsilon = j-k)\nonumber\\
&&= \sum_{k=0}^{\min(i,j)}  \binom{i}{k} \alpha^{k}(1-\alpha)^{i-k}p_{j-k}
\end{eqnarray*}
where $p_x$ is the pmf given in Eq. \ref{pmf}. 

In the next Proposition, we summarize some of the asymptotic properties of a GLK--INAR(1). These properties follow from general results in 
\cite{AlOAlz1988} and \cite{Schweer2014}.

\begin{prop}\label{Th:stationarity}
Assume that $\{X_t\}_{t\in\mathbb{Z}}$ is a  GLK--INAR(1). 
\begin{itemize}
\item[(i)]
The process $\{X_t\}_{t\in\mathbb{Z}}$ is an irreducible, aperiodic and positive recurrent Markov chain.
Hence there is a unique stationary distribution for the process $\{X_t\}_{t\in\mathbb{Z}}$. 
\item[(ii)] The marginal distribution of the stationary process  $\{X_t\}_{t\in\mathbb{Z}}$ is infinitely divisible.
\end{itemize}
\end{prop}

\begin{proof}
By Proposition \ref{Th:infdivGLK}, the distribution of the innovations is infinitely divisible, and hence, it is a compound Poisson distribution, see, e.g. Lemma 2.1 in \cite{Ste79}. Hence, both (i) and (ii) follow from Theorem 3.2.1 in  \cite{Schweer2014}. In point of fact, (i) is true for any INAR process, see
\cite{AlOAlz1987}.  An alternative derivation of (ii) is as follows. 
Since at stationarity the process satisfies $X=\alpha\circ X +\varepsilon$, where $\varepsilon\sim\mathcal{GLK}(a,b,c,\beta)$, and the innovation terms are infinite divisible by Proposition \ref{Th:infdivGLK}, the stationary distribution satisfies the definition of discrete semi--self--decomposability given in \cite{bouzar2008semi}.
Theorem 2 in \cite{bouzar2008semi} yields that it is also infinitely divisible.
\end{proof}

Since the GLK distribution satisfies the convolution property \cite[see][ Th. 8]{jan98}, then the GLK--INAR(1) is stable by aggregation as stated in the following
\begin{prop}\label{Th:convolution}
Let $\{X_{jt}\}_{t\in\mathbb{Z}}$ with $j=1,2,\ldots,J$ be a sequence of independent GLK--INAR(1) which satisfy:
\begin{equation*}
X_{jt}=\alpha \circ X_{jt-1}+\varepsilon_{jt},\quad \varepsilon_{jt}\sim \mathcal{GLK}(a_j,b,c,\beta)
\end{equation*}
The process $Y_t=X_{1t}+\ldots+X_{Jt}$ is GLK--INAR(1) which satisfies:
\begin{equation*}
Y_{t}=\alpha \circ Y_{t-1}+\varepsilon_{t},\quad \varepsilon_{t}\sim \mathcal{GLK}(a_1+\ldots+a_J,b,c,\beta)
\end{equation*}
\end{prop}

Below, we state explicit closed-form expressions for unconditional moments of the process.

\begin{prop}\label{Th:stationaryMomentsCor}
Let $\mu_\varepsilon$, $\mu_\varepsilon^{(2)}$ and $\sigma^{2}_{\varepsilon}$ the mean, second order non-central moment and variance given in Prop. \ref{Th:Moment} for a $\mathcal{GLK}(a,b,c,\beta)$. For a GLK--INAR(1) process, the following unconditional moments can be derived:
\begin{enumerate}[label=(\roman*)]\setcounter{enumi}{0}
\item $\mu_X=\mathbb{E}(X_t)=\mu_{\varepsilon}/(1-\alpha)$
\item $\mu_X^{(2)}=\mathbb{E}(X_t^2)=(\alpha\mu_{\varepsilon}+2\alpha \mu_{\varepsilon}^{2}/(1-\alpha)+\mu_{\varepsilon}^{(2)})/(1-\alpha^2)$
\item $\mathbb{E}(X_tX_{t-k})=\alpha \mathbb{E}(X_{t-1}X_{t-k})+\mu_{\varepsilon}\mu_X$
\item Higher-order non-central moments can be derived using the formula:
\begin{equation*}
\mu_{X}^{(m)}=\sum_{i=0}^{m}\sum_{k=0}^{i-1}\sum_{l=0}^{i-k}{i \choose k}(1-\alpha^i)^{-1}S(m,i)s(i-k,l)\alpha^{k}\mu_{X}^{({k})}\mu_{\varepsilon}^{(l)}
\end{equation*}
where $s(m,k)$ and $S(m,k)$ denote the Stirling's numbers of the I and II kind, respectively.
\end{enumerate}
\end{prop}

\begin{proof}
First- and second-order moments are known from \cite{AlOAlz1987}
for general INAR. Specifying the GLK innovations gives (i)--(iii). 
High-order moments can be computed similarly. See e.g. 
\cite{weiss2013}.  {For the sake of completeness, details are given in 
Appendix \ref{AppendixThmMoments} in the Supplementary Material.}
\end{proof}

From the previous proposition,
under the  assumption 
$\kappa=1 -\beta - b \beta/c>0$
one obtains the unconditional variance of the process $\sigma^{2}_X=\mathbb{V}(X_t)=(\sigma_{\varepsilon}^{2}+\alpha\mu_{\varepsilon})/(1-\alpha^2)$ and the dispersion index of the process
\begin{equation*}
VMR_{X}=\frac{\sigma^2_X}{\mu_X}=\frac{VMR_{\varepsilon}+\alpha}{1+\alpha}=\frac{1}{1+\alpha}\left(\alpha+\frac{1-\beta}{(1-\beta-b\beta/c)^2}\right)
\end{equation*}
where $VMR_{\varepsilon}=\sigma_{\varepsilon}^{2}/\mu_{\varepsilon}$ is the innovation index of dispersion. It follows that there is under-- or over--dispersion in the marginal distribution, $VMR_{X}<1$ and $VMR_{X}>1$, if and only if there is under- or over--dispersion in the innovation, $VMR_{\varepsilon}<1$ or $VMR_{\varepsilon}<1$ respectively.

The autocorrelation function is
\begin{equation*}
\gamma_k=\mathbb{C}ov(X_{t},X_{t-k})=\mathbb{E}(X_{t}X_{t-k})-\mu_X^2=\alpha^{k}\sigma^{2}_X
\end{equation*}
as in the INAR(1) process \cite[e.g., see][]{AlOAlz1987}.
{\color{black}
\subsection{A Markov--switching GLK--INAR(1) process}

The GLK--INAR(1) process can be extended to account for various sources of model instability such as structural breaks, regimes and outliers by introducing a time--varying parameter setting \citep[see for example][]{malyshkina2009markov}. A parsimonious approach is to assume a finite set of regimes $k=1,\ldots, K$  corresponding to different parameter configurations, i.e.
\begin{equation}
X_{t} = \alpha(S_t) \circ X_{t-1} +\varepsilon_{t}, \qquad t \in \mathbb{Z},
\end{equation}
with $\varepsilon_t|S_t\sim\mathcal{GLK}(a(S_t),b(S_t),c(S_t),\beta(S_t))$, where the thinning coefficient and the $GLK$ parameters of the error term  $\psi(S_t)=(\alpha(S_t),$ $a(S_t),$ $b(S_t),c(S_t),$ $\beta(S_t))'$ are time--varying  $\psi(S_t)=\sum_{k=1}^K\mathbb{I}(S_{t}=k)\psi_k$, where $\psi_k=(a(k),b(k),$ $c(k),\beta(k))'$. The $S_t \in \{1,\ldots,K\}$ for $t \in \mathbb{Z}$ denotes a hidden Markov--chain process with transition probabilities $\mathbb{P}(S_t=j|S_{t-1}=i)=\pi_{ij}$ for $i,j\in \{1,\ldots,K\}$. From now on, this extension is denoted with MS--GLK--INAR(1).

A special case, which is relevant for a common issue in count data series, is the large proportion of zeros \citep[e.g.,][]{maiti2015time}. The excess of zero, which leads to over--dispersion, can be handled by assuming a zero--inflated GLK--INAR(1). This can be defined by assuming that in one of the regimes, e.g.  $S_t=1$, there is complete thinning $\alpha_1\to 0$, and the error distribution is a Dirac centred at zero. Some alternatives where the GLK collapses to a Dirac include the (Negative) Binomial distribution with parameters $c_1=1$, $b_1=-1$ ($b_1=0$), $a_1=1$ and $\beta_1\to0$. 

The transition probabilities provide information on the persistence of these events. If the probability is independent on past information, i.e. $\mathbb{P}(S_t=j|S_{t-1}=i)=\pi_{j}$ for all $i,j\in \{1,2\}$, then the zero--inflated regime is transitory. If the zero--inflated regime is persistent, then the duration of the regime is captured by a large $\pi_{11}$. Other regimes ($S_t\neq 1$) with low mean and/or large variance can also generate zeroes. This is convenient in some applications, such as in epidemiology where zeroes from $S_t\neq 1$ can be interpreted as under--reported cases of a particular disease \citep[e.g.,][]{douwes2022zero}.


\subsection{Possible extensions of the GLK--INAR}

The GLK--INAR can be extended to include more general auto--correlation structures and to the multivariate setting. The process can be modified to allow for multiple lags building on the specification strategy used in \citet{NeSu07}. In particular, a GLK integer--valued ARMA of order $p$ and $q$, i.e. GLK--INARMA$(p,q)$ can be specified using independent thinning operators.

\begin{defn}[GLK--INARMA$(p,q)$ process]
	Let $\alpha_\ell \in (0,1)$ for $\ell=1,\ldots,p$  and $\zeta_r\in (0,1)$ for $r=1,\ldots,q$, 
	the GLK--INARMA(p,q) process is defined as
	\begin{equation*}\label{def:INARMA}
		X_{t} = \sum_{\ell=1}^p\alpha_\ell \circ X_{t-\ell}+\sum_{r=1}^q\zeta_r \circ\varepsilon_{t-r} +\varepsilon_{t}, \qquad t \in \mathbb{Z}
	\end{equation*}
	where $\varepsilon_{t}$ are iid random variables with Generalized Lagrangian Katz distribution $\mathcal{GLK}(a,b,c,\beta)$ and $\sum_{\ell=1}^p\alpha_\ell<1$ and $\sum_{r=1}^q\zeta_r<1$. 
\end{defn}

Compared to GLK--INAR(1), in the GLK--INARMA$(p,q)$, a further restriction on the autoregressive parameters is required for stationarity, although a weaker condition can be used. For alternatives specification strategies such as combined INAR (CINAR) see for example \citet{mckenzie2003ch,weiss2008combined}.

For the case of random integer vectors, a multivariate GLK--INAR(1) (GLK--MINAR(1)) can be used by introducing a thinning matrix operator.

\begin{defn}[GLK--MINAR(1) process]
	Let $\mathbf{X}_{t}=(X_{1t},\ldots,X_{Jt})'$ be a random integer vector and $\mathbf{A}=(\alpha_{ij})_{i,j=1}^{J}$, where $\alpha_{ij}\in (0,1)$, 
	the GLK--MINAR(1) process can be defined by
	\begin{equation*}\label{def:MINAR}
		\mathbf{X}_{t} = \mathbf{A} \circ \mathbf{X}_{t-1}+\boldsymbol{\varepsilon}_{t}, \qquad t \in \mathbb{Z}
	\end{equation*}
	where $\boldsymbol{\varepsilon}_{t}=(\varepsilon_{1t},\ldots,\varepsilon_{Jt})'$ is iid  Generalized Lagrangian Katz distributions, and $\mathbf{A} \circ \mathbf{X}_{t-1}$ is the thinning matrix operator, such that for each $i=1,\ldots,J$,
	$$
	X_{it}=\sum_{j=1}^J\alpha_{ij}\circ X_{jt-1}+ \varepsilon_{it},
		$$
	and $\alpha_{ij}\circ X_{jt-1}$ refers to the binomial thinning operator.		 
\end{defn}

The independence between the innovation terms of the different equations allows us to estimate them separately. This assumption can be relaxed by adding common GLK errors to the equations, or under some special cases of the GLK, a joint distribution can be introduced, such as the bivariate Katz's or Poisson distribution  \citep{PedKar2011,diafouka2022bivariate}.
}

\section{Bayesian inference}

\subsection{Prior distribution}\label{SubSec:prior}
With the construction in Eq. \ref{pmf}, the constraint $\sum_{x\geq 0} p_x =1$ is guaranteed by the condition $H(1)=1$. Nevertheless, some constraints on the parameters are needed to have all the $p_x:>0$.
Three different cases are discussed below (details are given in Appendix  \ref{App:ThProof} in the Supplementary Material.).
\begin{itemize}
\item For parameter values $a>0,b\geq 0,c >0$ the pmf are positive. Moreover, for $a/c, b/c\in\mathbb{N}$ the extended binomial coefficient $((a+bx)/c+1)_{x\uparrow}/x!$ coincides with the standard binomial coefficient ${\frac{a+bx}{c}+x \choose x}$ \cite[][p. 8]{ConFam2006}. 
\item  For $-c<b<0$, $a/c,b/c\in\mathbb{N}$ and $(c-a)/(c+b) \leq (a+c)/|b|$,  the pmf are positive for $x < x^*=(a+c)/|b|$, while $p_x=0$ for $x \geq x^*$.  
\item If $-c<b<0$ but the additional constraints of the previous point are not satisfied, the terms appearing in the product 
$({(a+bx)}/{c}+1)_{x\uparrow}$ change \textcolor{black}{sign}, and there is no guarantee that the result is positive. Indeed, for all the $x$ such that $x>\max\{(a+1)/|b|,(c-a)/(c+b),2\}$ one has $(a+bx)/c+1<0$ and $(a+bx)/c+x-1>0$ and hence there is an integer $q=q_x$ such that $(a+bx)/c+m<0$ for $1 \leq m \leq q$ and $(a+bx)/c+m>0$ for $q+1 \leq m \leq k-1$. Hence 
\begin{eqnarray}
&&\frac{1}{\frac{a+bx}{c}+x}
\Big(\frac{(a+bx)}{c}+1\Big)_{x\uparrow}=(-1)^{q} \prod_{m=1}^{q} \Big|\frac{(a+bx)}{c}\nonumber\\
&&+m\Big|  \prod_{m=q+1}^{k-1} \Big|\frac{(a+bx)}{c}+m\Big|\nonumber
\end{eqnarray} 
which is negative whenever $q_x$ is odd. 
For example take $a=10$, $b=-1$ and $c=2$, for $x=20$ one has $q_{20}=5$, which shows that $p_{20}<0$ which clearly is impossible. 
\end{itemize}

\begin{rem}
It should be noted that alternative definitions for $-c<b<0$ can be considered. 
For example one can set to 0 the $p_x<0$, i.e.  when $x>\max\{(a+1)/|b|$. In this case re-scaling  the $p_x$ is necessary to get $\sum_{x=0}^{x^*} p_x=1$. The resulting pmf is not a generalized Lagrangian distribution (due to the truncation and rescaling), and the normalizing constant is not in closed form. See, for example, \cite{mccabe2020distributions} for a discussion on the parameter values for the Katz distributions. 
\end{rem}
 In a Bayesian framework, the parameter constraints can be easily included in the inference process through a suitable choice of the prior distributions. We assume:
\begin{eqnarray*}
&&\alpha\sim\mathcal{B}e(\kappa_{\alpha},\tau_{\alpha}),\quad a\sim \mathcal{G}a(\kappa_a,\tau_a),\quad b\sim \mathcal{G}a(\kappa_b,\tau_b),\\
&&c\sim\mathcal{G}a(\kappa_c,\tau_c),\quad \beta \sim \mathcal{B}e(\kappa_\beta,\tau_\beta)
\end{eqnarray*}
where $\mathcal{B}e(\kappa,\tau)$ is the beta distribution with shape parameters $\kappa$ and $\tau$ and $\mathcal{G}a(\kappa,\tau)$ the gamma distribution with shape and scale parameters $\kappa$ and $\tau$, respectively. In the empirical applications we assume a non-informative hyper-parameter setting for $\alpha$ and $\beta$, that is $\kappa_\alpha=\tau_\alpha=\kappa_\beta=\tau_\beta=1$ and an informative prior for $a,b$ and $c$ with $\kappa_a=\tau_a=1$, $\kappa_b=\kappa_c=2$ and $\tau_b=\tau_c=1/2$.

{\color{black} In the case of Markov--switching specification of the GLK--INAR(1), the same prior is assumed for the regime--specific parameters 
$h(\psi_k)=\mathcal{B}e(\kappa_{\alpha},\tau_{\alpha})$ $\mathcal{G}a(\kappa_a,\tau_a)$ $ \mathcal{G}a(\kappa_b,\tau_b)$ $\mathcal{G}a(\kappa_c,\tau_c) \mathcal{B}e(\kappa_\beta,\tau_\beta)$ for $k=1,\ldots, K$. 
For the transition probabilities of the allocation variable, we assume a symmetric Dirichlet prior for each row of the transition matrix, i.e. $\pi_{i\cdot}\sim \mathcal{D}(1/K,\ldots,1/K)$ with concentration parameter $1/K$, where $\pi_{i\cdot}=(\pi_{i1},\ldots,\pi_{iK})$ for $i=1,\ldots, K$. }

\subsection{Posterior distribution}
Let $x_{1},\ldots,x_{T}$ be a sequence of observations for the GLK--INAR(1) process, then the joint posterior distribution is given by 
\begin{equation*}
f(\psi|x_{1},\ldots,x_{T})\propto f(\psi)\prod_{t=1}^{T} \prod_{i=0}^{\infty}\prod_{j=0}^{\infty}P_{ij}(\psi)^{\mathbb{I}(x_t-j)\mathbb{I}(x_{t-1}-i)}
\end{equation*}
where $\psi=(\alpha,a,b,c,\beta)$ is the parameter vector $f(\psi)$ the joint prior and
\begin{equation*}
P_{ij}(\psi)= \sum_{k=0}^{\min(i,j)}d_{ijk}\binom{\frac{a+b(j-k)}{c}+j-k}{j-k}\alpha^{k}(1-\alpha)^{i-k} \beta^{j-k}(1-\beta)^{\frac{a+b(j-k)}{c}}
\end{equation*}
where $d_{ijk}=\binom{i}{k}((a/c)/((a+bx)/(c)+j-k)$.

Following the discussion above in this section, if the parameter constraint $c>0$ is not imposed, the coefficients of the Lagrangian expansion can be negative. In this case, a truncated GLK can be used, similarly to what is proposed in \cite{mccabe2020distributions} for the Katz distribution, and the inference procedure can be easily extended to include this type of distribution. The truncation can be imposed by using the following recursion for the transition probability:
\begin{equation*}
p_{i}(\psi)=p_0\prod_{j=0}^{i-1}\max\left\{0,\frac{U(\psi)+V(\psi)j}{a+j}\right\}
\end{equation*} 
where $U(\psi)=a\beta/c$, $V(\psi)=U(b+c)/(a+b)$ and
\begin{equation*}
p_0=\left(1+\sum_{j=1}^{\infty}\prod_{k=0}^{j-1} \max\left\{0,\frac{U(\psi)+V(\psi)j}{a+j}\right\}\right)^{-1}.
\end{equation*}
The probability $p_{i}$ becomes null for $i>j$ if $U(\psi)+V(\psi)j<0$ at $j$.

Since the joint posterior is not tractable, we follow a Markov Chain Monte Carlo (MCMC) framework for posterior approximation. See \cite{RobCas2013} for an introduction to MCMC methods. We overcome the difficulties in tuning the parameters of the MCMC procedure by applying the Adaptive MCMC sampler (AMCMC) proposed in \cite{Andr08}. Following a standard procedure, the following reparametrization is considered to impose constraints on the parameters of the GLK--INAR(1). Let $\eta=(\eta_1,\ldots,\eta_5)$ be the 5-dimensional parameter vector obtained by the transformation $\eta=\varphi(\psi)$ with $\eta_{1}=\log(\psi_1 /(1-\psi_1))$, $\eta_2=\log(\psi_2)$, $\eta_{3}=\log(\psi_3)$, $\eta_4 = \log(\psi_4)$, and $\eta_{5}=\log(\psi_5 /(1-\psi_5))$
and let $f(\eta|x_{1},\ldots,x_T)=f(\varphi^{-1}(\eta)|x_{1},\ldots,x_T)J(\eta)$ be the posterior of $\eta$, with $J(\eta)=\psi_1\psi_2\psi_3\psi_4\psi_5(1-\psi_1)(1-\psi_5)$ the Jacobian of the transformation $\varphi$ given above.
Given the adaptation parameters $\mu^{j}$ and $\Sigma^{(j)}$, at the $j$-th iterations, the AMCMC consists of the following three steps. First, a candidate $\eta^{\ast}$ is generated from the random walk proposal: $\eta^{\ast}=\eta^{(j-1)}+\lambda^{(j)} w^{(j)},\quad w^{(j)}\sim\mathcal{N}_{q}(\mathbf{0},\Sigma^{(j)})$.
Second, the candidate is accepted with probability $\rho^{(j)}=\rho(\eta^{(j-1)},\eta^{\ast})$, where 
\begin{equation*}
\rho(\eta^{(j-1)},\eta^{\ast}) = \min \left( 1, \frac{f(\varphi^{-1}(\eta^{\ast})|x_{1},\ldots,x_{T})J(\eta^{\ast})}{f(\varphi^{-1}(\eta^{(j-1)})|x_{1},\ldots,x_{T})J(\eta^{(j-1)})}
\right)
\end{equation*}
 and third, the adaptive parameters are updated as follows:
\begin{eqnarray*}
\mu^{(j+1)}&=&\mu^{(j)}+\gamma^{(j)}(\mu^{(j)}-\eta^{(j)})\\
\Sigma^{(j+1)}&=&\Sigma^{(j)}+\gamma^{(j)}((\mu^{(j)}-\eta^{(j)})(\mu^{(j)}-\eta^{(j)})'-\Sigma^{(j)})\\
\log \lambda^{(j+1)}&=&\log \lambda^{(j)}+\gamma^{(j)}(\rho^{(j)}-\rho^{\ast}),
\end{eqnarray*}
where $\rho^{*}$ is the target acceptance probability and $\gamma^{(j)}=j^{-a}$, $a> 0$ is the adaptive scale \cite[][, Algorithm 4]{Andr08}. Following the suggestions in \cite{Robetal1997} we set $\rho^{*} = 0.44$.

{\color{black}
The latent allocation variables in the Markov--switching specification of the GLK--INAR(1) are sampled the Forward--Filtering Backward sampling procedure (FFBS). The prediction and filtering probabilities are given by 
\begin{equation*}
	\begin{aligned}
		&\mathbb{P}(S_{t}=k|\mathcal{X}_{t-1})=\sum_{\ell=1}^K\pi_{\ell k}\mathbb{P}(S_{t-1}=\ell|\mathcal{X}_{t-1})\\
	&\mathbb{P}(S_{t}=k|\mathcal{X}_{t})\propto f(x_t|\psi_k,x_{t-1},S_t=k)\mathbb{P}(S_{t}=k|\mathcal{X}_{t-1}),
	\end{aligned}
\end{equation*}
where $\mathcal{X}_{t-1}=(x_1,\ldots,x_{t-1})'$, $f(x_t|\psi_k,x_{t-1},S_t=k)=$ $ \prod_{i=0}^{\infty} $ $\prod_{j=0}^{\infty}$ $P_{ij}(\psi_k)^{\mathbb{I}(x_t-j)\mathbb{I}(x_{t-1}-i)}$ for $k,\ell \in \{1,\ldots,K\}$. Notice that the conditioning on the parameters $\psi$ is included only in the likelihood but not in the probabilities to simplify the notation. The filtered probabilities can be smoothed by considering all the information available, i.e.   
\begin{equation*}
\mathbb{P}\left(S_{1:T}|\mathcal{X}_{T}\right)=\mathbb{P}\left(S_{T}|\mathcal{X}_T\right)\prod_{t=1}^{T-1}\mathbb{P}\left(S_{t}|S_{t+1},\mathcal{X}_t\right)
\end{equation*}	
where $\mathbb{P}\left(S_{t}|S_{t+1},\mathcal{X}_t\right)\propto \pi_{S_{t}S_{t+1}}  \mathbb{P}\left(S_{t}|\mathcal{X}_t\right)$ and $S_{1:T}=(S_1,\ldots,S_T)'$. The allocation variables are sampled directly from these smoothed probabilities.

The conditional posterior distribution of the transition probabilities of the Markov chain $S_t$ is conditionally conjugate and can be sampled directly from $
\pi_{\cdot k}|S_{1:T}\sim \mathcal{D}(d_1,\ldots,d_K),
$ where $d_k=1/K+\sum_{t=1}^T\mathbb{I}_{k}(S_t)$ for $k=1,\ldots,K$.
}

{\color{black} For the possible extensions of the GLK--INAR, such as the GLK--INRMA(p,q) and the GLK--MINAR(1), data augmentation techniques can be used to improve the efficiency of the MCMC  \citep{NeSu07,c2022bayesian}. For instance, in the case of the GLK--INRMA(p,q), conditional conjugacy of the thinning parameters can be obtained by assuming each autoregressive (moving average) component is a latent variable following a binomial distribution. In the case of GLK--MINAR(1), a similar strategy can be followed, see for instance \citet{soyer2022bayesian}. 	
}
\subsection{Simulation results}
We illustrate the Bayesian procedure's effectiveness in recovering the parameters' true value and the MCMC procedure's efficiency through some simulation experiments. We test the algorithm's efficiency in two different settings, commonly found in the data: low persistence and high persistence (see trajectories in Fig. \ref{fig:inartraj}). The true values of the parameters are: $\alpha=0.3$, $a=5.3239$, $b=0.0592$, $c=0.6$, $\beta=0.5917$ in the low persistence setting, and $\alpha=0.7$, $a=5.3239$,$b=0.0592$, $c=0.6$, $\beta=0.5917$ in the high persistence setting. For each setting, we run the Gibbs sampler for 50,000 iterations on each dataset, discard the first 10,000 draws to remove dependence on initial conditions, and apply a thinning procedure with a factor of 10 to reduce the dependence between consecutive draws. 

For illustrative purposes, in Figure \ref{fig:MCMCrawHist} in the Supplementary Material we show the MCMC posterior approximation for the parameter $\alpha$ (first row), the unconditional mean of the process (second row), and the marginal likelihood (last row), in one of our experiments for the high- and low-persistence settings. Each plot represents the true value (solid black line) and the Bayesian estimates. Posterior estimated are approximated by using 4,000 MCMC samples after thinning and burn-in removal (dashed red line).  Figures  \ref{fig:MCMCrawHigh}-\ref{fig:mcmcHistHigh} and \ref{fig:MCMCrawLow}-\ref{fig:mcmcHistLow} in the Supplementary Material  exhibit 10,000 MCMC posterior draws and the MCMC approximation of the posterior distribution for all the parameters, in the high- and low-persistence settings.

In our experiments, the acceptance rate is in the range of 40\%-53\% for both parameter settings (see Figure \ref{fig:mcmcDiagLow} in the Supplementary Material). Table \ref{Stat} in the Supplementary Material shows, for all the parameters the autocorrelation function (ACF), effective sample size (ESS), inefficiency factor (INEFF) and Geweke's convergence diagnostic (CD) before (BT subscript) and after thinning (AT subscript). The numerical standard errors are evaluated using the \textit{nse} package \citep{Geyer92, ArdBlu17, Ardiaetal18}.

The thinning procedure is effective in reducing the autocorrelation levels and in increasing the ESS. The p-values of the CD statistics indicate that the null hypothesis that two sub-samples of the MCMC draws have the same distribution is always accepted. The efficiency of the MCMC after the thinning procedure is generally improved. After thinning, on average, the inefficiency measures (5.83), the p-values of the CD statistics (0.36) and the NSE (0.02) achieved the values recommended in the literature \citep[e.g., see ][]{Robetal1997}. 

{\color{black} It is important to underline that the persistence parameter estimation and the forecast are highly sensitive to the innovation distributional assumption. An illustration is presented in the left plot of Figure \ref{fig:missp}, where the data generating process corresponds to a GLK--INAR(1) with large overdispersion (VRM=8.6). The standard model for count data is the Poisson INAR(1) model (PINAR(1)), which cannot capture overdispersion. This misspecified model entails an underestimation of the persistence parameter (medium gray histogram). The NBINAR(1) captures the overdispersion and provides reliable persistence estimates (light gray) comparable with the one of GLK--INAR (dark gray). Nevertheless, in the case of underdispersion (VRM=0.4, right plot of Figure \ref{fig:missp}), both NBINAR(1) and PINAR(1) return an estimation bias in the persistence parameter, while the INAR--GLK 
gives a good approximation of the true persistence. 
In summary, the INAR--GLK(1) model nests standard models, such as Generalized Poisson and Negative Binomial INARs, and allows for different degrees of underdispersion and overdispersion. Hence, it can be used without preliminary testing of the dispersion features of the series.
	
	\begin{figure}[t]
    \centering
		\begin{tabular}{cc}
			\includegraphics[scale=0.5,trim={3.85cm 9.5cm 4cm 9.5cm},clip]{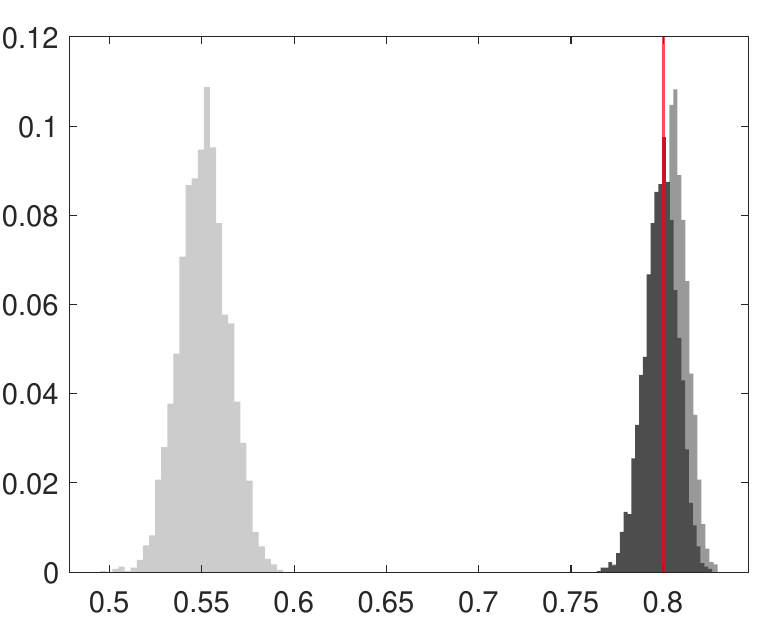}& \includegraphics[scale=0.5,trim={3.85cm 9.5cm 4cm 9.5cm},clip]{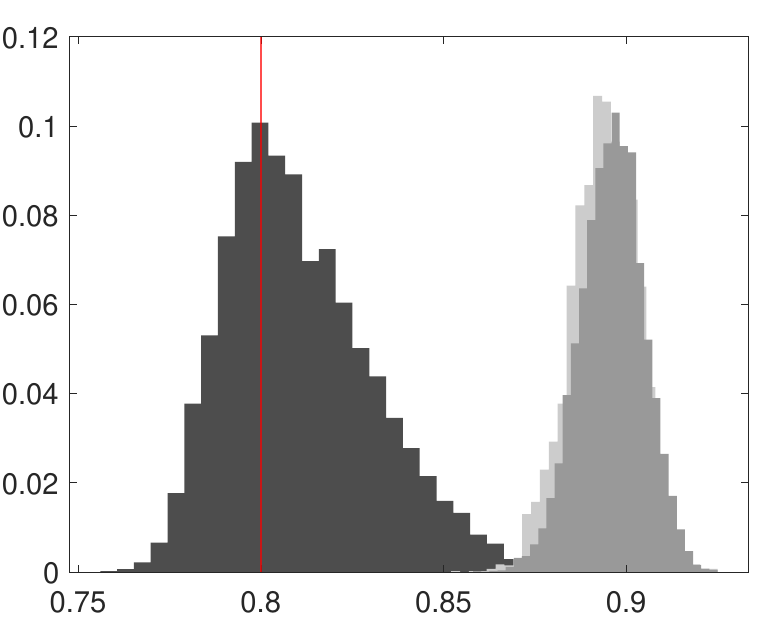}
		\end{tabular}
		\caption{\color{black} Posterior approximation of the persistence parameter under a PINAR(1) (medium gray), NBINAR(1) (light gray) and GLK--INAR(1) (dark gray) for over (left plot) and under (right) dispersion scenarios. The red line indicates the true level of persistence. \label{fig:missp}}
	\end{figure}
	
	Similarly, to exemplify the effectiveness and efficiency of the estimation procedure in different scenarios we considered: i) high and low persistence regimes with the same parameter configuration of the settings presented before and ii) a large mean regime and an zero--inflated regime where $\alpha\to 0$, $a=1$, $b\to 0$, $c=1$, $\beta\to 0$. The simulated trajectories are shown in Figure \ref{fig:twoR} (\ref{fig:zeroinf}) in the Supplementary Material together with the estimated allocations of the regimes, represented by the shaded areas, with an accuracy of 97\% (100\%)  for the two regimes (zero--inflated) scenario. Moreover, the parameters are successfully retrieved, see  in Figures \ref{fig:twoMCMCh}, \ref{fig:twoMCMCt} and \ref{fig:zeroMCMC} in the Supplementary Material. Notice that the zero--inflated parameters are not estimated but set by default to approximate the Dirac distribution. 
    
    In conclusion, the Gibbs sampler is computationally efficient and can retrieve the true parameter values of the MS--GLK--INAR in different settings, including the single--regime and the zero--inflated specifications. The MCMC for the GLK--INAR takes 0.5 minutes for a sample size of $T=260$ observations and for 30,000 MCMC iterations. This is comparable with the Negative Binomial INAR (0.4 minutes). The method is scalable and can be applied to datasets with thousands of observations. For larger-size datasets, the theoretical moment of the process can be used to devise alternative estimation procedures, such as the method of moments.  The moments of the distribution are provided in closed form in Proposition \ref{Th:Moment} in the Supplementary Material.}

\section{Application to climate change}\label{sec:climate}
\subsection{Data description}
\begin{figure}[t]
\begin{center}
\renewcommand{\arraystretch}{0.2}
\setlength{\tabcolsep}{0pt}
\begin{tabular}{cc}
\includegraphics[scale=0.47]{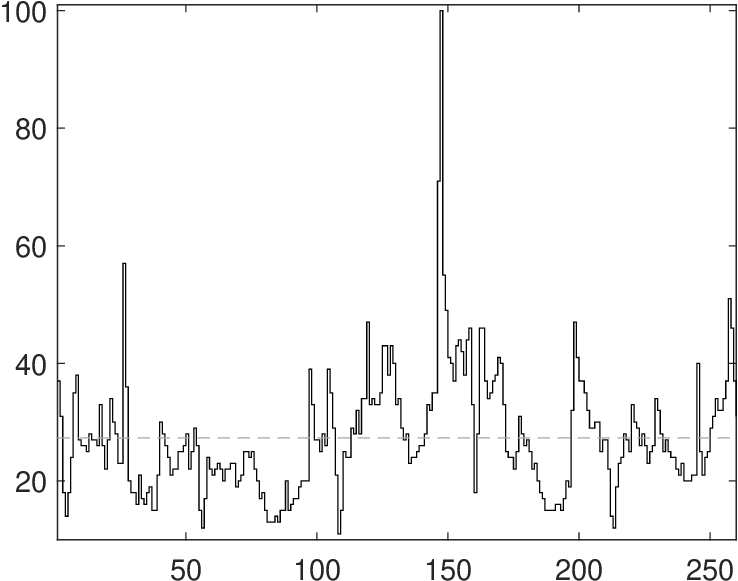}&\includegraphics[scale=0.47]{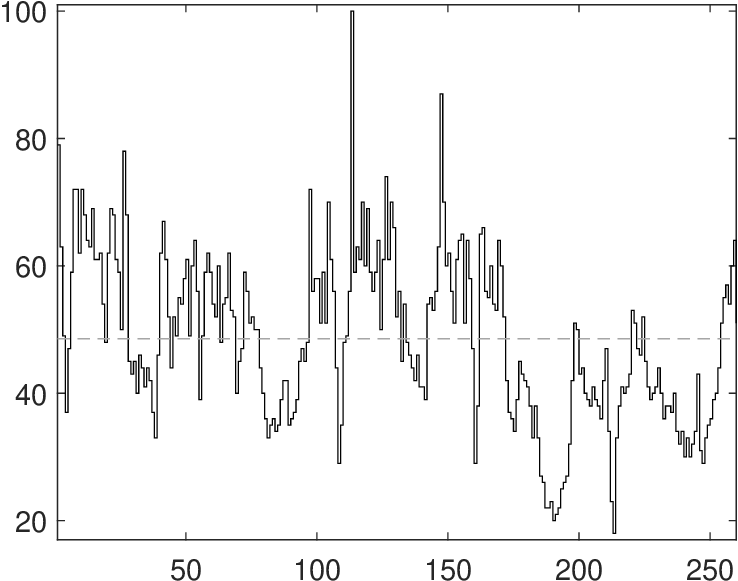}\\\includegraphics[scale=0.47]{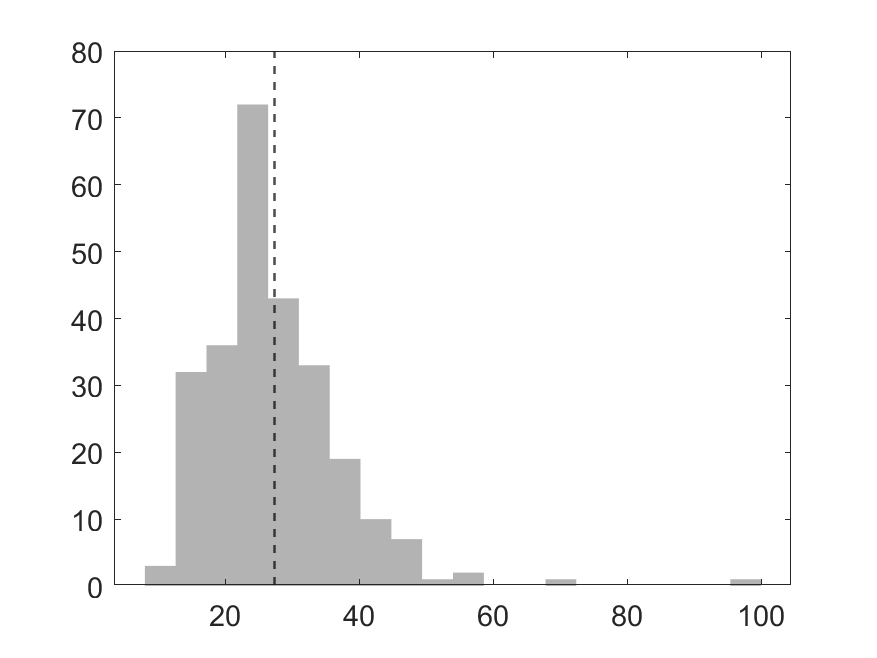}&\includegraphics[scale=0.47]{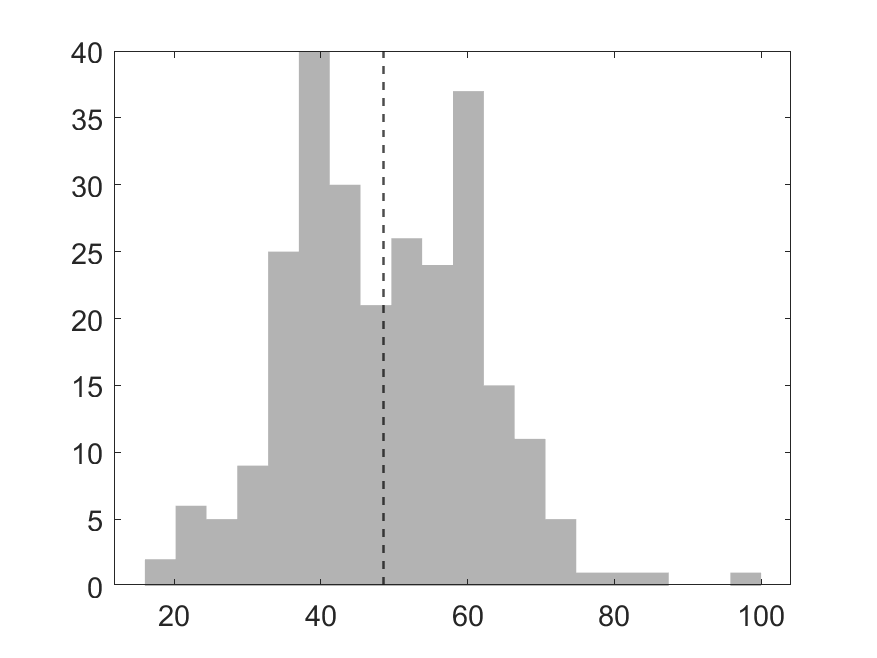}
\end{tabular}
\end{center}
\caption{Time series (top) and histograms (bottom) of the global Google search of the words ``Climate Change" (left) and ``Global Warming" (right). Weekly frequency from 4th December 2016 to 21st November 2021. Empirical mean (dashed line).}\label{fig:inartrajReal1}
\end{figure}

We used Google Trends data to measure the changes in public concern about climate change. Google Trend represents a source of big data \citep{choi2012,scott2014} which have been used in many studies, for example, \cite{AndHel21} studied domestic violence during covid-19, \cite{Yaetal21} studied influenza trends, \cite{Schetal2020} and \cite{Dinetal21} presented applications to unemployment and \cite{Yuetal19} studied oil consumption. In this study, we follow \cite{lineman2015talking} and use Google search volumes as a proxy for public concern about ``Climate Change'' (CC) and ``Global Warming'' (GW). The search volume is the traffic for the specific combination of keywords relative to all queries submitted in Google Search in the world or a given region over a defined period. The indicator ranges from 0 to 100, with 100 corresponding to the largest relative search volume during the period of interest. The search volume is sampled weekly from 4th December 2016 until 21st November 2021. We analysed the dynamics at the global and country level. Countries with an excess of zeros above 95\% in the search volume series have been excluded. The final dataset includes 65 countries of the about 200 countries provided by Google Trends. For illustration purposes, we report in the top plots of Fig. \ref{fig:inartrajReal1} the series of the world volume. The CC global volume exhibits overdispersion with $\widehat{VMR}=102/27.33=3.73$, skewness and kurtosis $\widehat{S}=2.09$ and $\widehat{K}=13.47$, respectively. The GW global volume has over--dispersion $\widehat{VMR}=170.42/48.56=3.51$, skewness $\widehat{S}=0.27$  and kurtosis $\widehat{K}=3.22$ (see also the histograms in the bottom plots). The country-specific indexes exhibit different levels of persistence and over--dispersion.

\begin{figure}[t]
\begin{center}
\renewcommand{\arraystretch}{0.2}
\setlength{\tabcolsep}{0pt}
\begin{tabular}{cc}
\multicolumn{2}{c}{\small (a) Google search dataset ``Climate Change"}\\
\includegraphics[scale=0.47]{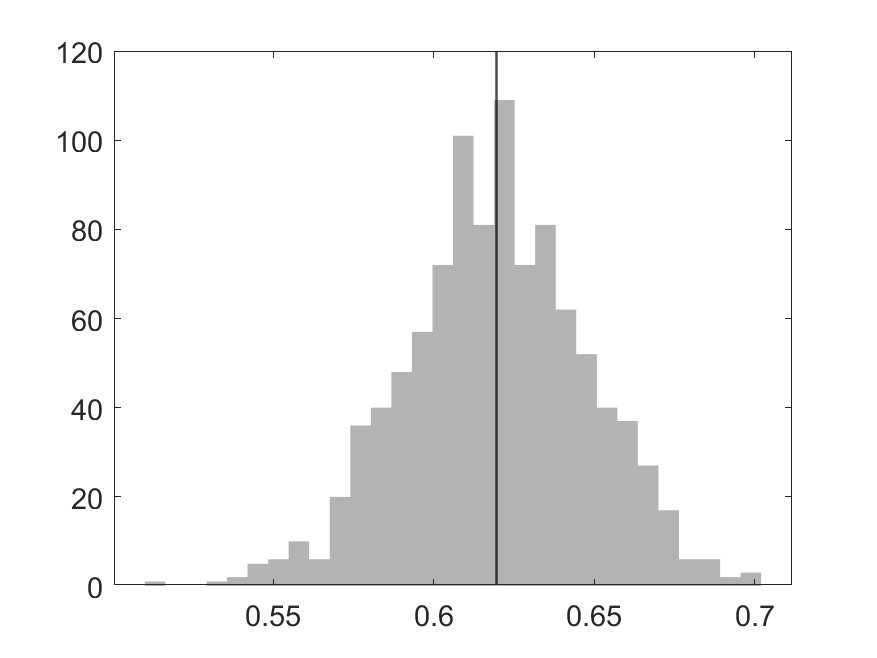}&\includegraphics[scale=0.47]{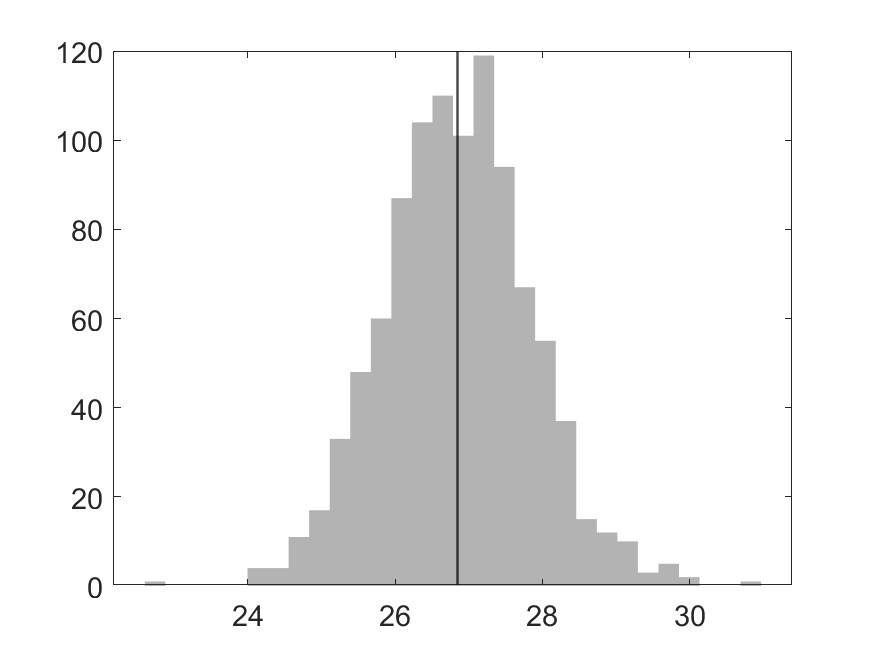}\\
\multicolumn{2}{c}{\small (b) Google search dataset ``Global Warming"}\\
\includegraphics[scale=0.47]{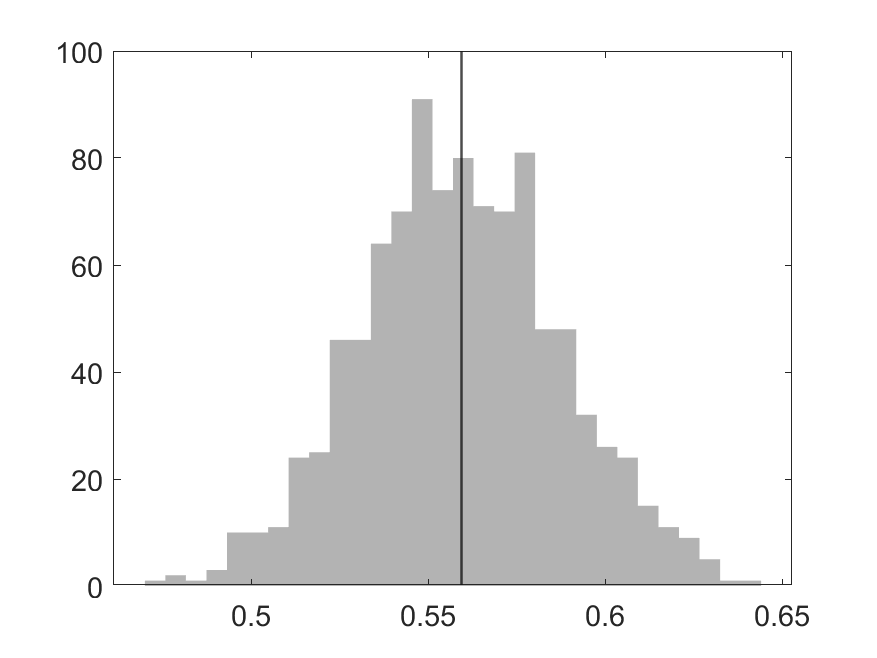}&\includegraphics[scale=0.47]{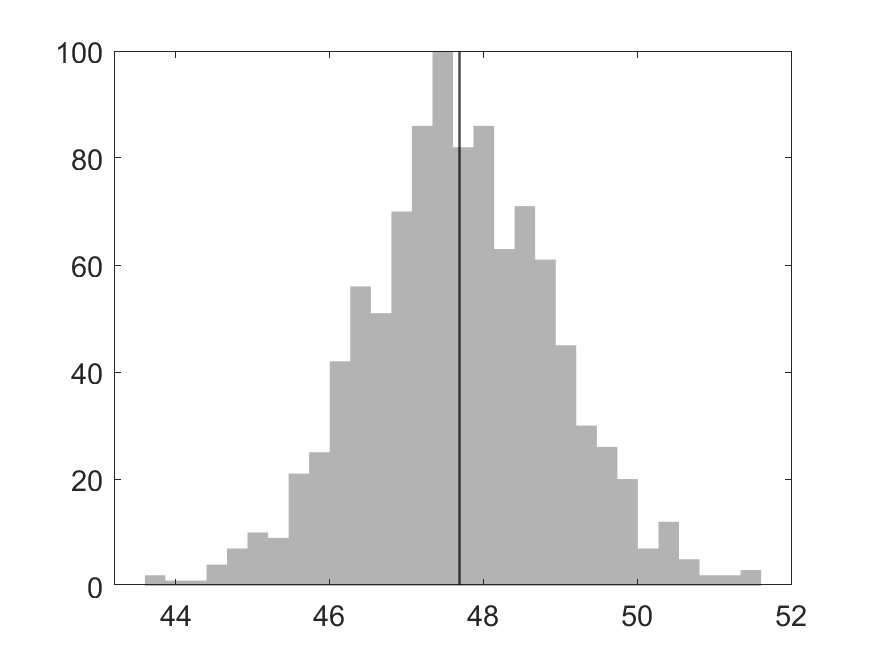}
\end{tabular}
\end{center}
\caption{Posterior approximation of the persistence parameter $\alpha$ (left) and the unconditional moment $\mu_\varepsilon/(1-\alpha)$ (right) for the global search volume.}\label{fig:inartrajReal2}
\end{figure}

\begin{figure}[t]
	\begin{center}
		\renewcommand{\arraystretch}{0.2}
		\setlength{\tabcolsep}{0pt}
		\begin{tabular}{ccc}
			\multicolumn{3}{c}{\small (a) Google search dataset ``Climate Change"}\\
			\includegraphics[scale=0.35,trim={3.5cm 9.5cm 4cm 9.5cm},clip]{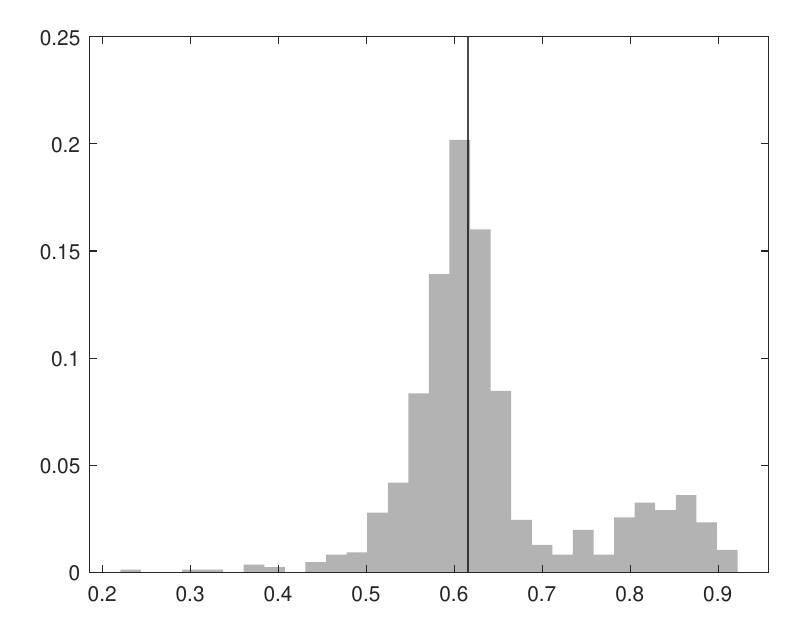}&\includegraphics[scale=0.35,trim={3.5cm 9.5cm 4cm 9.5cm},clip]{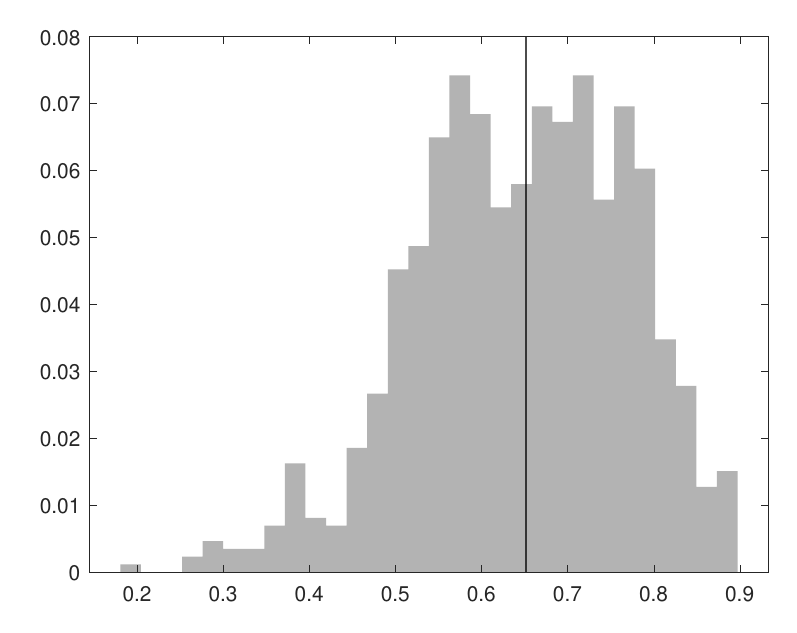} & \includegraphics[scale=0.35,trim={3.5cm 9.5cm 4cm 9.5cm},clip]{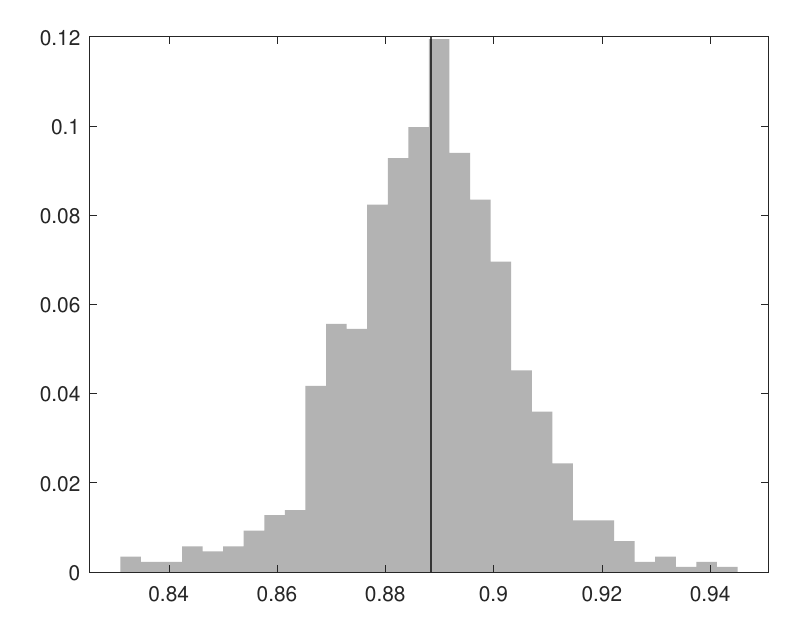}  
			\\
			\multicolumn{3}{c}{\small (b) Google search dataset ``Global Warming"}\\
			\includegraphics[scale=0.35,trim={3.5cm 9.5cm 4cm 9.5cm},clip]{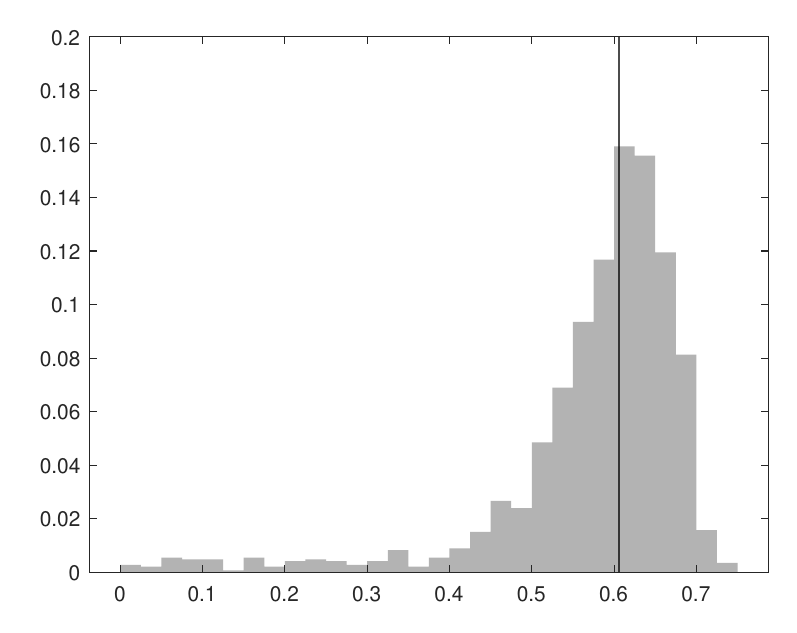}&\includegraphics[scale=0.35,trim={3.5cm 9.5cm 4cm 9.5cm},clip]{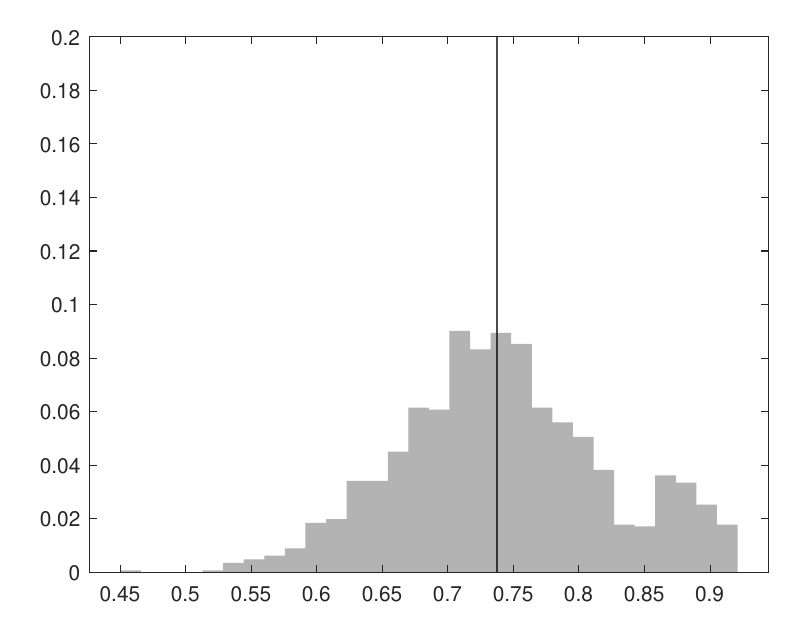} & \includegraphics[scale=0.35,trim={3.5cm 9.5cm 4cm 9.5cm},clip]{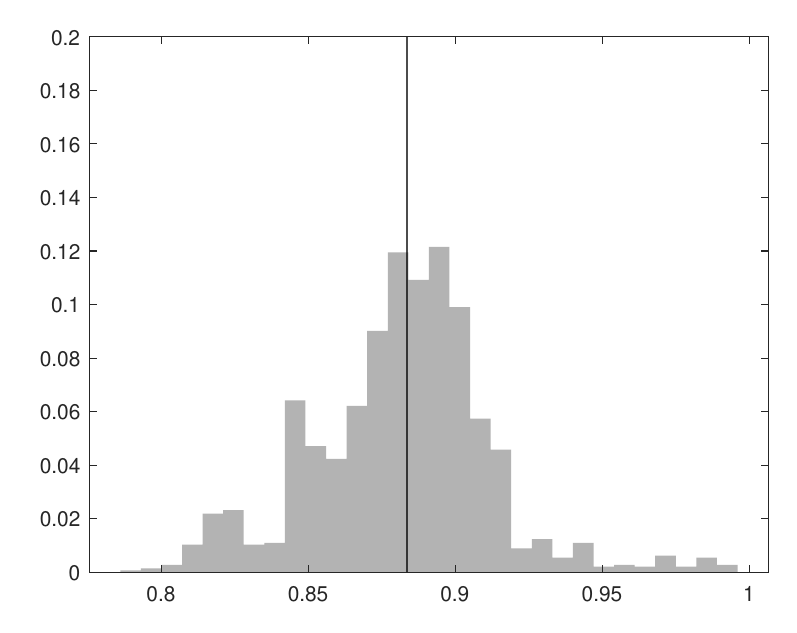}
		\end{tabular}
	\end{center}
	\caption{\color{black} Posterior approximation of the persistence parameter for a three--state Markov Switching model: high (right plot), medium (middle) and low (left) persistence, for the Climate Change (top) and Global Warming (bottom) datasets.}\label{fig:MSinartrajReal2}
\end{figure}

\subsection{Estimation results}
The posterior distribution of the autoregressive coefficient is given in Fig. \ref{fig:inartrajReal2}. The coefficient estimate and posterior credible interval (in parenthesis) are $\widehat{\alpha}=0.56$ $(0.50,0.62)$ and $\widehat{\alpha}=0.62$ $(0.56,0.67)$ for the GW and the CC dataset, respectively (see also the approximation to the posterior distribution of the parameters in Figures \ref{fig:mcmcHistGoogle} and \ref{fig:mcmcHistGoogleWarm} in the Supplementary Material). This result indicates that the public concern about climate risk is persistent over time worldwide at an aggregate level. The estimated parameter of the innovation process and their 0.95\% credible intervals (in parenthesis) are $\widehat{a}=3.53$ $(1.56,6.08)$, $\widehat{b}=0.04$ $(0.01, 0.11)$, $\widehat{c}=0.21$ $(0.05,0.47)$ and $\widehat{\beta}=0.48$ $(0.20,0.65)$ for the GW dataset and $\widehat{a}=3.26$ $(1.44,5.72)$, $\widehat{b}=0.12$ $(0.021,0.310)$, $\widehat{c}=0.26$ $(0.032,0.726)$ and $\widehat{\beta}=0.35$ $(0.067,0.623)$ for the CC one. 

\subsection{Model comparison}
The results indicate a deviation from the Negative Binomial model. Thus we apply the DIC criterion $DIC=-4\mathbb{E}(\log f(X|\psi)|y)+2\log f(X|\widehat{\psi})$ to compare GLK--INAR(1) and NB--INAR(1). The $DIC$ is computed following \citep[][]{Spiegel2002}:
\begin{equation}
\hbox{DIC}=-4\frac{1}{N}\sum_{j=1}^{N}\log f(X|\psi^{(j)})+2\log f(X|\widehat{\psi})
\end{equation}
where $f(X|\psi)$ is the likelihood of the model, $\psi^{(j)}$ $j=1,\ldots,N$ the MCMC draws after thinning and burn-in sample removal, and $\widehat{\psi}$ is the parameter estimate. The DICs for the GLK (NB) INARs fitted on the aggregate CC and GW series are $1.6743\cdot 10^3$ $(1.6862\cdot 10^3)$ and $1.8735\cdot 10^3$ $(1.8834\cdot 10^3)$, respectively.

\begin{figure}[t]
\centering
	\begin{tabular}{cc}
	 \includegraphics[scale=0.45,trim={3.5cm 10cm 4cm 10cm},clip]{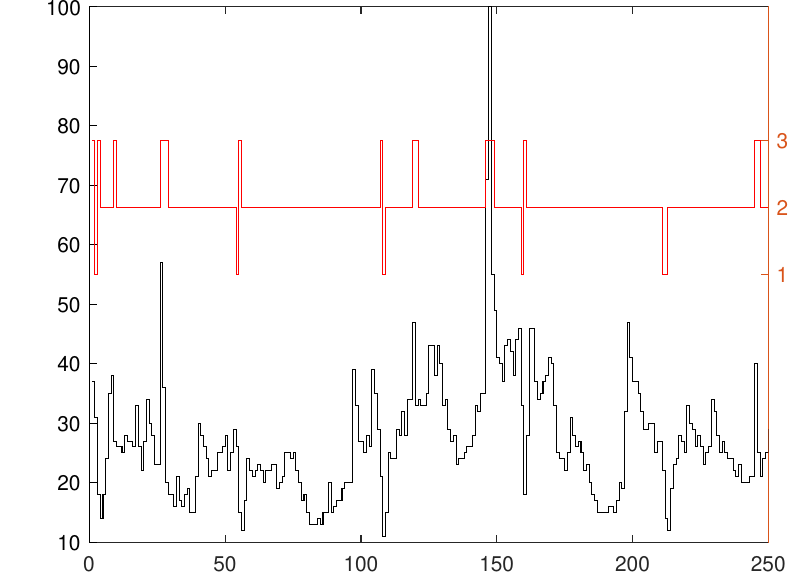} & \includegraphics[scale=0.4,trim={4cm 9.5cm 3cm 9.5cm},clip]{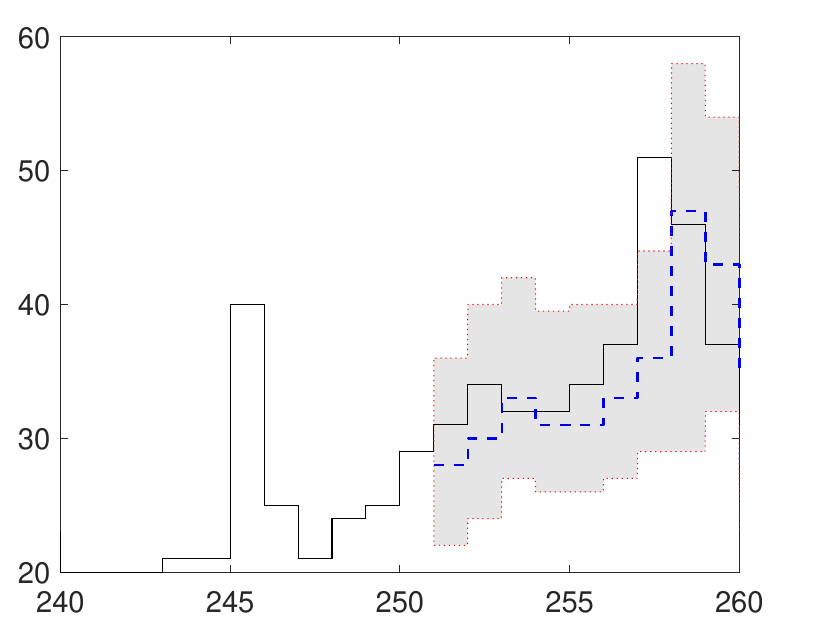}\\
	 	 \includegraphics[scale=0.45,trim={3.5cm 10cm 4cm 10cm},clip]{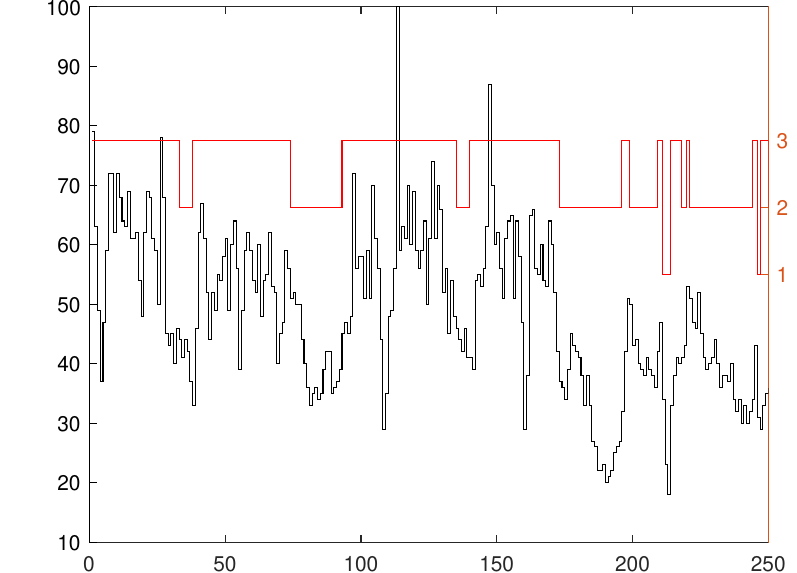} & \includegraphics[scale=0.4,trim={4cm 9.5cm 3cm 9.5cm},clip]{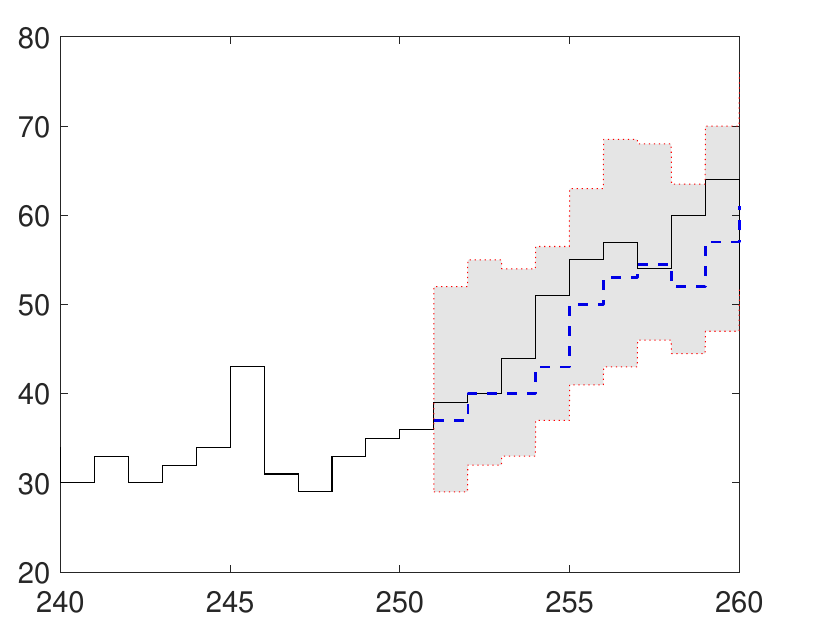} 
	\end{tabular}
\caption{\color{black} MS--GLK--INAR(1) results with three regimes (left) and one--step forecast (right) including point (dashed--blue line) and 90\% credible intervals (shaded region) for the Climate Change (top) and Global Warming (bottom) database} \label{fig:MSfor}
\end{figure}

{\color{black} Given the high kurtosis levels and the multi--modality in the empirical distribution of both series (see Figure \ref{fig:inartrajReal1}), the Markov Switching INAR is used to deal with outliers and parameter instability. We use DIC and RMSE to select the number of regimes and the model (see Table \ref{tab:DICMSEaphor} in the Supplementary Material). We find that GLK--INAR with two or three regimes present the best fit in--sample and out--sample for both the CC and GW datasets. The results with three--regimes are presented in Figure \ref{fig:MSinartrajReal2}. The three regimes identify different persistence levels: high (right plot), medium (middle) and low (left). Some of the regimes also have different unconditional mean levels (see Figure \ref{fig:MSfor} left plots). In terms of one--step--ahead forecasting, in both datasets, the model can reproduce the upward trend at the end of the sample and effectively cover the true values within their 90\% credible intervals (see Figure \ref{fig:MSfor} right plots).}

{\color{black}\subsection{Disaggregate analysis}
We run the analysis at a disaggregate level. The results are given in Figures \ref{fig:country}-\ref{fig:countryUncVMR} and Tables \ref{tab:pers1}-\ref{tab:pers2} in the Supplementary Material. Figure \ref{fig:country} provides evidence of an inverse relationship between estimated persistence $\widehat{\alpha}$ and dispersion $\widehat{VMR}$ cross countries (reference lines in the left plot). There is evidence of this inverse relationship in both the CC (blue dots) and GW (red dots) datasets. The plot on the right indicates an inverse (direct) relationship between the estimated unconditional mean $\widehat{\mu_\varepsilon}/(1-\widehat{\alpha})$ and the dispersion index $\widehat{VMR}$ for the GW (CC). In the same picture, we indicate the parameter estimates for the world volume of searches (stars). 
}

\begin{figure}[p]
\centering
\renewcommand{\arraystretch}{0.7}
\setlength{\tabcolsep}{0pt}
\begin{tabular}{cc}
\includegraphics[scale=0.47]{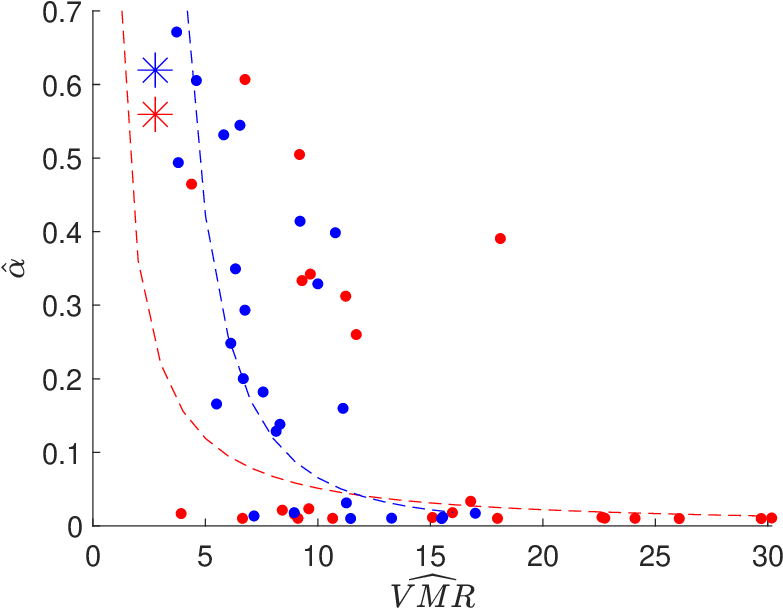}&\includegraphics[scale=0.47]{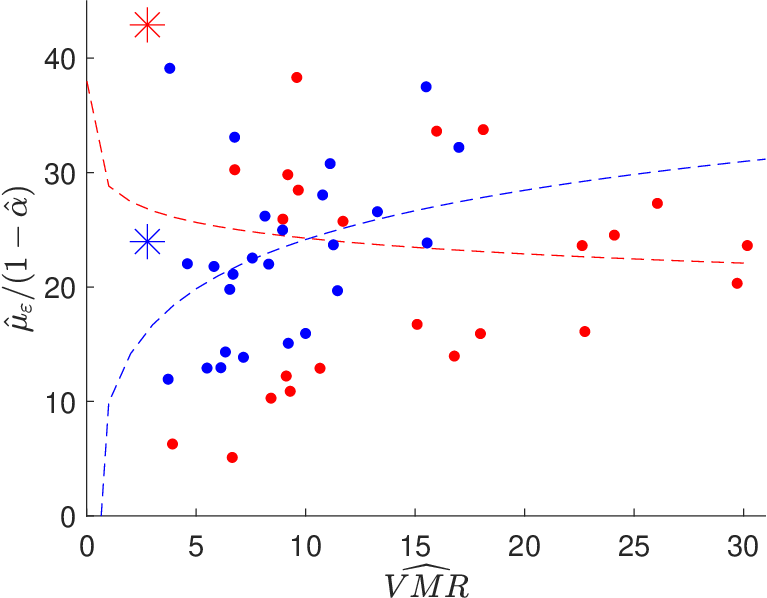}\\
\end{tabular}
\caption{Persistence-dispersion ($\widehat{\alpha}$ and $\widehat{VMR}$, left) and unconditional mean and dispersion ($\widehat{\mu_\varepsilon}/(1-\widehat{\alpha})$ and $\widehat{VMR}$, right) scatter plots for all countries in the ``Climate Change" (\textcolor{blue}{$\bullet$}) and ``Global Warming" (\textcolor{red}{$\bullet$}) datasets. Only countries with less than 21\% of zeros are reported. Stars indicate the parameters of the world's volume of searches. ``*" indicates the parameter estimates for the aggregated search volume.}\label{fig:country}
\end{figure}

\begin{figure}[p]
\centering
\renewcommand{\arraystretch}{0.7}
\setlength{\tabcolsep}{12pt}
\begin{tabular}{cc}
\includegraphics[scale=0.47]{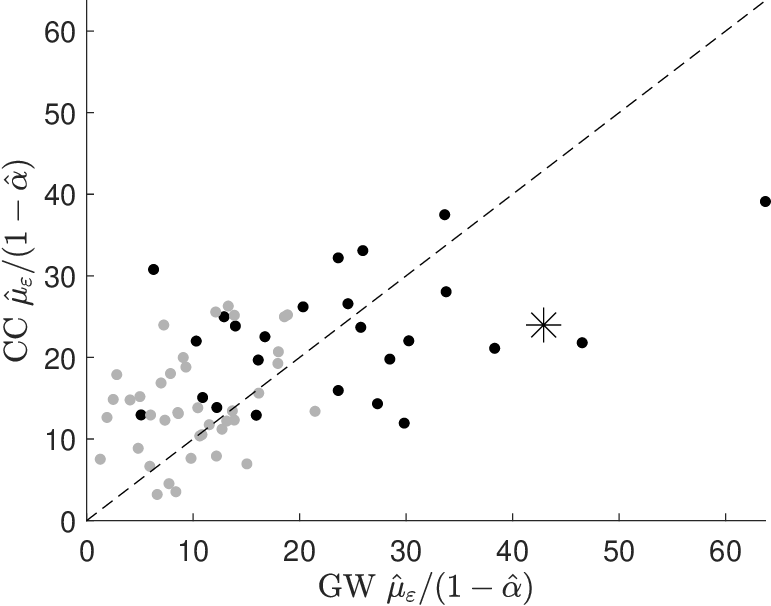}&\includegraphics[scale=0.47]{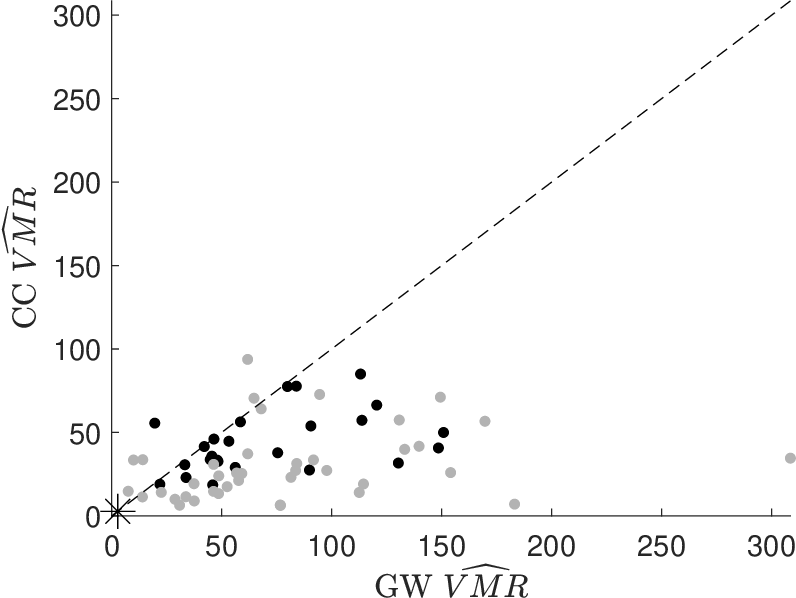}\\
\end{tabular}
\caption{Unconditional mean (left) and dispersion index (right) of the GW (horizontal) and CC (vertical) for countries with more than 21\% of zeros (\textcolor{gray}{$\bullet$}) less than 21\% ($\bullet$, values rescaled by five for visualization purposes) in the number of searches. In each plot, the 45$^{\circ}$ reference line.}\label{fig:countryUncVMR}
\end{figure}

The terms ``Climate Change" and ``Global Warming" are used interchangeably. Nevertheless, they describe different phenomena and can be used to determine the public's level of understanding about these two parallel concepts \cite{lineman2015talking}. We investigate the relationships in the search volumes through the lens of our GLK--INAR(1) model. The left plot in Fig. \ref{fig:countryUncVMR} shows the unconditional mean of the search volumes for the two concepts in all countries (dots). In public attention, the two concepts are connected
in the long run. We find a positive association for both countries with large (percentage of zeros $<$ 21\%) and low search volumes (percentage of zeros $>$ 21\%). There is an asymmetric effect in the overdispersion (right plot), and in all countries, the GW search volume has a larger VMR than the CC volume. This can be explained by the larger variability induced by the changes in the use of the GW term in official communications.

Comparing the coefficients across the rows of Tables \ref{tab:pers1}-\ref{tab:pers2} in the Supplementary Material, we find evidence of two types of series, one with high persistence and the other with low persistence. Moreover, for each country, the level of persistence is similar across the two datasets (compare columns of Tables \ref{tab:pers1}-\ref{tab:pers2} in the Supplementary Material).

Tables \ref{tab:pers1}-\ref{tab:pers2} in the Supplementary Material report the marginal likelihood of the GLK--INAR(1) and Lagrangian Katz INAR(1) in columns GLK and LK, respectively. We find evidence of a better fitting of the GLK--INAR(1) for some countries and variables, e.g. CC searches in India and CC and GW searches in South Africa. To get further insights into the results, we study the relationship between the dynamic and dispersion properties of the series and the actual level of climate risk of the countries. We consider the Global Climate Risk Index (CRI), which ranks countries and regions following the impacts of extreme weather events (such as storms, hurricanes, floods, heatwaves, etc.). The lower the index value, the larger the climate risk is. Following the values of the CRI for 2021, based on the events recorded from 2000 to 2019, our dataset includes some of the countries most exposed to climate risk, such as Japan, Philippines, Germany, South Africa, India, Sri Lanka and Canada  \citep[see]{eckstein2021}. 

\begin{figure}[t]
\begin{center}
\renewcommand{\arraystretch}{0.7}
\setlength{\tabcolsep}{0pt}
\begin{tabular}{cc}
\includegraphics[scale=0.47]{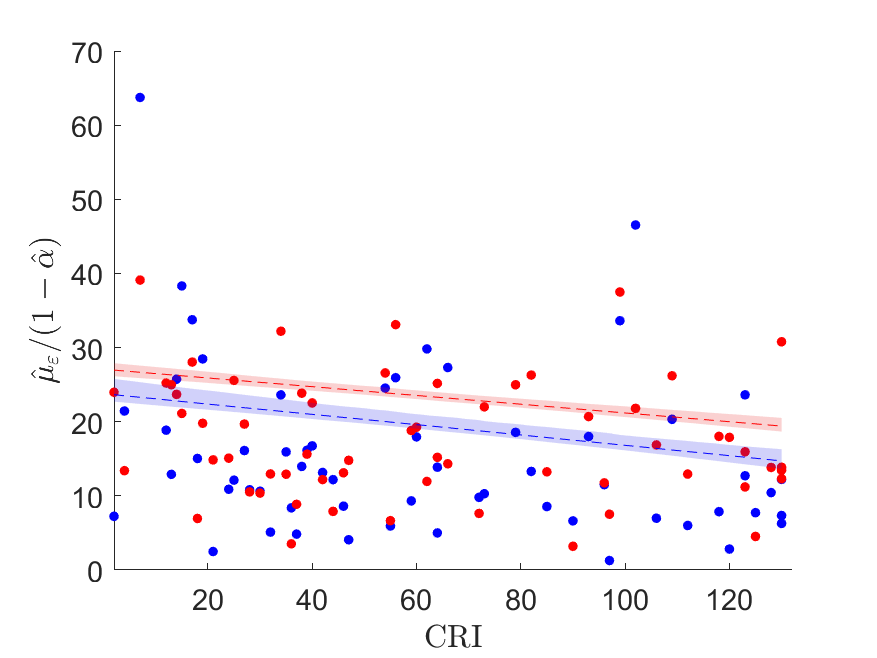}&\includegraphics[scale=0.47]{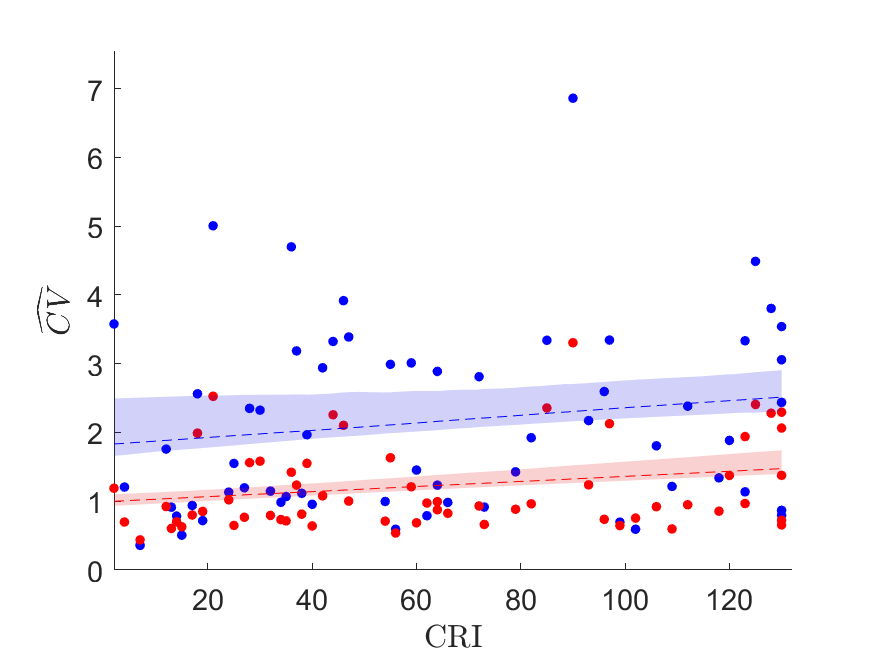}\\
\end{tabular}
\end{center}
\caption{Climate Risk Index and unconditional mean scatter plot (CRI-$\mu_\varepsilon/(1-\alpha)$, left) and Climate Risk Index and dispersion scatter plot (CRI-$CV$, right) scatter plots for all countries in the ``Climate Change" (\textcolor{blue}{$\bullet$}) and ``Global Warming" (\textcolor{red}{$\bullet$}) datasets. Dashed lines represent the linear regression estimated on the data.}\label{fig:CRI}
\end{figure}

The left plot in Fig. \ref{fig:CRI} shows the unconditional mean against the CRI. There is evidence of a positive relationship between the public interest in climate-related topics and the actual level of climatic risk. The lower the CRI level, the larger the Google search volumes are (see dashed lines). For example, India has a high risk (CRI equal to 7) and a very high long-run level of public attention.

The right plot reports the coefficient of variation against the CRI for all countries in the ``Climate Change" (blue) and ``Global Warming" (red) datasets. The dashed lines represent linear regressions estimated on the data. There is evidence of a negative relationship between the dispersion of public concern and climatic risk; in countries with more significant risk levels, the Google search volumes are less over--dispersed.

{\color{black}
To deal with the excess of zeros, which are very frequent in more than 62\% of the series, we apply the MS--GLK--INAR(1) with two states, where the first state represents a prolonged absence of searches on Google and the second a persistent search activity. The MS--GLK--INAR(1) performs better than MS--NBINAR(1) in 119 out of the 130 CC and GW time series following the DIC (see Table \ref{tab:MSallcou2} and \ref{tab:MSallcou1} in the Supplementary Material). Accounting for the excess of zeros allows for improving the estimation of the persistence and provides an estimate of the probability $\hat{\pi}_{11}$ to stay in an inactive search regime.  The findings on the persistence parameter discussed in this section for the GLK--INAR(1) are confirmed by the MS--GLK--INAR. Furthermore, there is evidence of a positive relationship between the probability $\hat{\pi}_{11}$ and the CRI, consistent with the results on the Google search persistence.}

\section{Conclusion}
\textcolor{black}{A novel integer--valued autoregressive process is proposed with Generalized Lagrangian Katz innovations (GLK--INAR). Theoretical properties of the model, such as stationarity, moments, and semi--self--decomposability, are provided. To deal with parameter instability and excess of zeroes, we also propose a Markov--Switching GLK--INAR. A Bayesian approach to inference and an efficient Gibbs sampling procedure have been proposed, which naturally account for uncertainty when forecasting. The modelling framework is applied to a Google Trend dataset measuring the public concern about climate change in 65 countries. The greater flexibility of the GLK--INAR allows for a superior fitting compared to the standard INAR models and a better comprehension of the heterogeneity in public perception. More specifically, new evidence is provided about the long-run level of public attention, its persistence and dispersion in countries with low and high levels of climate risk. The Markov-switching GLK-INAR identified regimes with the absence of searches and changes in the dynamic features of the series.}

\section*{Acknowledgment}
The authors acknowledge support from: the MUR– PRIN project ‘Discrete random structures for Bayesian learning and prediction’ under g.a. n. 2022CLTYP4 and the Next Generation EU -- ‘GRINS– Growing Resilient, INclusive and Sustainable’ project (PE0000018), National Recovery and Resilience Plan (NRRP)-- PE9 -- Mission 4, C2, Intervention 1.3. The views and opinions expressed are only those of the authors and do not necessarily reflect those of the European Union or the European Commission. Neither the European Union nor the European Commission can be held responsible for them. 

\bibliography{katz}

\begin{thebibliography}{70}
\expandafter\ifx\csname natexlab\endcsname\relax\def\natexlab#1{#1}\fi
\providecommand{\url}[1]{\texttt{#1}}
\providecommand{\href}[2]{#2}
\providecommand{\path}[1]{#1}
\providecommand{\DOIprefix}{doi:}
\providecommand{\ArXivprefix}{arXiv:}
\providecommand{\URLprefix}{URL: }
\providecommand{\Pubmedprefix}{pmid:}
\providecommand{\doi}[1]{\href{http://dx.doi.org/#1}{\path{#1}}}
\providecommand{\Pubmed}[1]{\href{pmid:#1}{\path{#1}}}
\providecommand{\bibinfo}[2]{#2}
\ifx\xfnm\relax \def\xfnm[#1]{\unskip,\space#1}\fi
\bibitem[{Afrifa-Yamoah \& Mueller(2022)}]{Afrifa2022}
\bibinfo{author}{Afrifa-Yamoah, E.}, \& \bibinfo{author}{Mueller, U.} (\bibinfo{year}{2022}).
\newblock \bibinfo{title}{Modeling digital camera monitoring count data with intermittent zeros for short-term prediction}.
\newblock {\it \bibinfo{journal}{Heliyon}\/},  {\it \bibinfo{volume}{8}\/}, \bibinfo{pages}{e08774}.
\bibitem[{Aknouche et~al.(2021)Aknouche, Almohaimeed \& Dimitrakopoulos}]{AknoucheAlmohaimeedDimitrakopoulos2021}
\bibinfo{author}{Aknouche, A.}, \bibinfo{author}{Almohaimeed, B.~S.}, \& \bibinfo{author}{Dimitrakopoulos, S.} (\bibinfo{year}{2021}).
\newblock \bibinfo{title}{Forecasting transaction counts with integer-valued garch models}.
\newblock {\it \bibinfo{journal}{Studies in Nonlinear Dynamics \& Econometrics}\/},  {\it \bibinfo{volume}{26}\/}, \bibinfo{pages}{529--539}.
\bibitem[{Al-Osh \& Alzaid(1987)}]{AlOAlz1987}
\bibinfo{author}{Al-Osh, M.}, \& \bibinfo{author}{Alzaid, A.~A.} (\bibinfo{year}{1987}).
\newblock \bibinfo{title}{First-order integer-valued autoregressive ({INAR} (1)) process}.
\newblock {\it \bibinfo{journal}{Journal of Time Series Analysis}\/},  {\it \bibinfo{volume}{8}\/}, \bibinfo{pages}{261--275}.
\bibitem[{Al-Osh \& Aly(1992)}]{AlOAly1992}
\bibinfo{author}{Al-Osh, M.~A.}, \& \bibinfo{author}{Aly, E.-E.~A.} (\bibinfo{year}{1992}).
\newblock \bibinfo{title}{First order autoregressive time series with negative binomial and geometric marginals}.
\newblock {\it \bibinfo{journal}{Communications in Statistics - Theory and Methods}\/},  {\it \bibinfo{volume}{21}\/}, \bibinfo{pages}{2483--2492}.
\bibitem[{Alzaid \& Al-Osh(1988)}]{AlOAlz1988}
\bibinfo{author}{Alzaid, A.}, \& \bibinfo{author}{Al-Osh, M.} (\bibinfo{year}{1988}).
\newblock \bibinfo{title}{First-order integer-valued autoregressive ({INAR} {$(1)$}) process: distributional and regression properties}.
\newblock {\it \bibinfo{journal}{Statist. Neerlandica}\/},  {\it \bibinfo{volume}{42}\/}, \bibinfo{pages}{53--61}.
\bibitem[{Alzaid \& Al-Osh(1993)}]{AlzAlO1993}
\bibinfo{author}{Alzaid, A.}, \& \bibinfo{author}{Al-Osh, M.} (\bibinfo{year}{1993}).
\newblock \bibinfo{title}{Generalized {P}oisson {{ARMA}} processes}.
\newblock {\it \bibinfo{journal}{Annals of the Institute of Statistical Mathematics}\/},  {\it \bibinfo{volume}{45}\/}, \bibinfo{pages}{223--232}.
\bibitem[{Alzaid \& Omair(2014)}]{AlzOma2014}
\bibinfo{author}{Alzaid, A.~A.}, \& \bibinfo{author}{Omair, M.~A.} (\bibinfo{year}{2014}).
\newblock \bibinfo{title}{{P}oisson difference integer valued autoregressive model of order one}.
\newblock {\it \bibinfo{journal}{Bulletin of the Malaysian Mathematical Sciences Society}\/},  {\it \bibinfo{volume}{37}\/}, \bibinfo{pages}{465--485}.
\bibitem[{Anderberg et~al.(2021)Anderberg, Rainer \& Siuda}]{AndHel21}
\bibinfo{author}{Anderberg, D.}, \bibinfo{author}{Rainer, H.}, \& \bibinfo{author}{Siuda, F.} (\bibinfo{year}{2021}).
\newblock \bibinfo{title}{Quantifying domestic violence in times of crisis: An internet search activity-based measure for the covid-19 pandemic}.
\newblock {\it \bibinfo{journal}{Journal of the Royal Statistical Society: Series A (Statistics in Society)}\/}, .
\bibitem[{Andersson \& Karlis(2014)}]{AndKar2014}
\bibinfo{author}{Andersson, J.}, \& \bibinfo{author}{Karlis, D.} (\bibinfo{year}{2014}).
\newblock \bibinfo{title}{A parametric time series model with covariates for integers in {Z}}.
\newblock {\it \bibinfo{journal}{Statistical Modelling}\/},  {\it \bibinfo{volume}{14}\/}, \bibinfo{pages}{135--156}.
\bibitem[{Andrieu \& Thoms(2008)}]{Andr08}
\bibinfo{author}{Andrieu, C.}, \& \bibinfo{author}{Thoms, J.} (\bibinfo{year}{2008}).
\newblock \bibinfo{title}{A tutorial on adaptive {MCMC}}.
\newblock {\it \bibinfo{journal}{Statistics and computing}\/},  {\it \bibinfo{volume}{18}\/}, \bibinfo{pages}{343--373}.
\bibitem[{Ardia \& Bluteau(2017)}]{ArdBlu17}
\bibinfo{author}{Ardia, D.}, \& \bibinfo{author}{Bluteau, K.} (\bibinfo{year}{2017}).
\newblock \bibinfo{title}{nse: Computation of numerical standard errors in r}.
\newblock {\it \bibinfo{journal}{Journal of Open Source Software}\/},  {\it \bibinfo{volume}{2}\/}, \bibinfo{pages}{172}.
\bibitem[{Ardia et~al.(2018)Ardia, Bluteau \& Hoogerheide}]{Ardiaetal18}
\bibinfo{author}{Ardia, D.}, \bibinfo{author}{Bluteau, K.}, \& \bibinfo{author}{Hoogerheide, L.~F.} (\bibinfo{year}{2018}).
\newblock \bibinfo{title}{Methods for computing numerical standard errors: {R}eview and application to value-at-risk estimation}.
\newblock {\it \bibinfo{journal}{Journal of Time Series Econometrics}\/},  {\it \bibinfo{volume}{10}\/}.
\bibitem[{Battaglini et~al.(2009)Battaglini, Barbeau, Bindi \& Badeck}]{bat2009}
\bibinfo{author}{Battaglini, A.}, \bibinfo{author}{Barbeau, G.}, \bibinfo{author}{Bindi, M.}, \& \bibinfo{author}{Badeck, F.-W.} (\bibinfo{year}{2009}).
\newblock \bibinfo{title}{European winegrowers’ perceptions of climate change impact and options for adaptation}.
\newblock {\it \bibinfo{journal}{Regional Environmental Change}\/},  {\it \bibinfo{volume}{9}\/}, \bibinfo{pages}{61–73}.
\bibitem[{Berry \& West(2020)}]{Berri2020}
\bibinfo{author}{Berry, L.~R.}, \& \bibinfo{author}{West, M.} (\bibinfo{year}{2020}).
\newblock \bibinfo{title}{Bayesian forecasting of many count-valued time series}.
\newblock {\it \bibinfo{journal}{Journal of Business \& Economic Statistics}\/},  {\it \bibinfo{volume}{38}\/}, \bibinfo{pages}{872--887}.
\bibitem[{Bourguignon et~al.(2016)Bourguignon, Vasconcellos, Reisen \& Ispány}]{Bouetal16}
\bibinfo{author}{Bourguignon, M.}, \bibinfo{author}{Vasconcellos, K.~L.}, \bibinfo{author}{Reisen, V.~A.}, \& \bibinfo{author}{Ispány, M.} (\bibinfo{year}{2016}).
\newblock \bibinfo{title}{A {P}oisson {INAR(1)} process with a seasonal structure}.
\newblock {\it \bibinfo{journal}{Journal of Statistical Computation and Simulation}\/},  {\it \bibinfo{volume}{86}\/}, \bibinfo{pages}{373--387}.
\bibitem[{Bouzar(2008)}]{bouzar2008semi}
\bibinfo{author}{Bouzar, N.} (\bibinfo{year}{2008}).
\newblock \bibinfo{title}{Semi-self-decomposable distributions on $\mathbf{Z}_{+}$}.
\newblock {\it \bibinfo{journal}{Annals of the Institute of Statistical Mathematics}\/},  {\it \bibinfo{volume}{60}\/}, \bibinfo{pages}{901--917}.
\bibitem[{Chen \& Lee(2016)}]{Chen2016}
\bibinfo{author}{Chen, C.~W.}, \& \bibinfo{author}{Lee, S.} (\bibinfo{year}{2016}).
\newblock \bibinfo{title}{Generalized {P}oisson autoregressive models for time series of counts}.
\newblock {\it \bibinfo{journal}{Computational Statistics \& Data Analysis}\/},  {\it \bibinfo{volume}{99}\/}, \bibinfo{pages}{51--67}.
\bibitem[{Chen \& Lee(2017)}]{Chen17}
\bibinfo{author}{Chen, C.~W.}, \& \bibinfo{author}{Lee, S.} (\bibinfo{year}{2017}).
\newblock \bibinfo{title}{Bayesian causality test for integer-valued time series models with applications to climate and crime data}.
\newblock {\it \bibinfo{journal}{Journal of the Royal Statistical Society: Series C (Applied Statistics)}\/},  {\it \bibinfo{volume}{66}\/}, \bibinfo{pages}{797--814}.
\bibitem[{Choi \& Varian(2012)}]{choi2012}
\bibinfo{author}{Choi, H.}, \& \bibinfo{author}{Varian, H.} (\bibinfo{year}{2012}).
\newblock \bibinfo{title}{Predicting the present with {G}oogle {T}rends}.
\newblock {\it \bibinfo{journal}{Economic Record}\/},  {\it \bibinfo{volume}{88}\/}, \bibinfo{pages}{2--9}.
\bibitem[{Consul \& Famoye(2006)}]{ConFam2006}
\bibinfo{author}{Consul, P.~C.}, \& \bibinfo{author}{Famoye, F.} (\bibinfo{year}{2006}).
\newblock {\it \bibinfo{title}{{L}agrangian probability distributions}\/}.
\newblock \bibinfo{publisher}{Springer}.
\bibitem[{Cunha et~al.(2018)Cunha, Vasconcellos \& Bourguignon}]{CunVasBou2018}
\bibinfo{author}{Cunha, E. T.~d.}, \bibinfo{author}{Vasconcellos, K.~L.}, \& \bibinfo{author}{Bourguignon, M.} (\bibinfo{year}{2018}).
\newblock \bibinfo{title}{A skew integer-valued time-series process with generalized {P}oisson difference marginal distribution}.
\newblock {\it \bibinfo{journal}{Journal of Statistical Theory and Practice}\/},  {\it \bibinfo{volume}{12}\/}, \bibinfo{pages}{718--743}.
\bibitem[{Diafouka et~al.(2022)Diafouka, Louzayadio, Malouata, Ngabassaka \& Bidounga}]{diafouka2022bivariate}
\bibinfo{author}{Diafouka, M.~K.}, \bibinfo{author}{Louzayadio, C.~G.}, \bibinfo{author}{Malouata, R.~O.}, \bibinfo{author}{Ngabassaka, N.~R.}, \& \bibinfo{author}{Bidounga, R.} (\bibinfo{year}{2022}).
\newblock \bibinfo{title}{On a bivariate katz’s distribution}.
\newblock {\it \bibinfo{journal}{Advances in Mathematics: Scientific Journal}\/},  {\it \bibinfo{volume}{11}\/}, \bibinfo{pages}{955--968}.
\bibitem[{Douwes-Schultz \& Schmidt(2022)}]{douwes2022zero}
\bibinfo{author}{Douwes-Schultz, D.}, \& \bibinfo{author}{Schmidt, A.~M.} (\bibinfo{year}{2022}).
\newblock \bibinfo{title}{Zero-state coupled markov switching count models for spatio-temporal infectious disease spread}.
\newblock {\it \bibinfo{journal}{Journal of the Royal Statistical Society Series C: Applied Statistics}\/},  {\it \bibinfo{volume}{71}\/}, \bibinfo{pages}{589--612}.
\bibitem[{Drovandi et~al.(2016)Drovandi, Pettitt \& McCutchan}]{Droetal16}
\bibinfo{author}{Drovandi, C.~C.}, \bibinfo{author}{Pettitt, A.~N.}, \& \bibinfo{author}{McCutchan, R.~A.} (\bibinfo{year}{2016}).
\newblock \bibinfo{title}{{Exact and Approximate Bayesian Inference for Low Integer-Valued Time Series Models with Intractable Likelihoods}}.
\newblock {\it \bibinfo{journal}{Bayesian Analysis}\/},  {\it \bibinfo{volume}{11}\/}, \bibinfo{pages}{325 -- 352}.
\bibitem[{Eckstein et~al.(2021)Eckstein, K{\"u}nzel \& Sch{\"a}fer}]{eckstein2021}
\bibinfo{author}{Eckstein, D.}, \bibinfo{author}{K{\"u}nzel, V.}, \& \bibinfo{author}{Sch{\"a}fer, L.} (\bibinfo{year}{2021}).
\newblock \bibinfo{title}{Global climate risk index 2021. {W}ho suffers most from extreme weather events? {W}eather-related loss events in 2019 and 2000-2019}.
\newblock {\it \bibinfo{journal}{Bonn: Germanwatch}\/},  {\it \bibinfo{volume}{2021}\/}.
\bibitem[{Fahad \& Wang(2018)}]{Fahad2018}
\bibinfo{author}{Fahad, S.}, \& \bibinfo{author}{Wang, J.} (\bibinfo{year}{2018}).
\newblock \bibinfo{title}{Farmers’ risk perception, vulnerability, and adaptation to climate change in rural pakistan}.
\newblock {\it \bibinfo{journal}{Land Use Policy}\/},  {\it \bibinfo{volume}{79}\/}, \bibinfo{pages}{301--309}.
\bibitem[{Freeland \& McCabe(2004)}]{FreCab2004}
\bibinfo{author}{Freeland, R.}, \& \bibinfo{author}{McCabe, B.~P.} (\bibinfo{year}{2004}).
\newblock \bibinfo{title}{Analysis of low count time series data by {P}oisson autoregression}.
\newblock {\it \bibinfo{journal}{Journal of Time Series Analysis}\/},  {\it \bibinfo{volume}{25}\/}, \bibinfo{pages}{701--722}.
\bibitem[{Freeland(2010)}]{Fre2010}
\bibinfo{author}{Freeland, R.~K.} (\bibinfo{year}{2010}).
\newblock \bibinfo{title}{True integer value time series}.
\newblock {\it \bibinfo{journal}{AStA Advances in Statistical Analysis}\/},  {\it \bibinfo{volume}{94}\/}, \bibinfo{pages}{217--229}.
\bibitem[{Fried et~al.(2015)Fried, Agueusop, Bornkamp, Fokianos, Fruth \& Ickstadt}]{fried2015retrospective}
\bibinfo{author}{Fried, R.}, \bibinfo{author}{Agueusop, I.}, \bibinfo{author}{Bornkamp, B.}, \bibinfo{author}{Fokianos, K.}, \bibinfo{author}{Fruth, J.}, \& \bibinfo{author}{Ickstadt, K.} (\bibinfo{year}{2015}).
\newblock \bibinfo{title}{Retrospective {B}ayesian outlier detection in {INGARCH} series}.
\newblock {\it \bibinfo{journal}{Statistics and Computing}\/},  {\it \bibinfo{volume}{25}\/}, \bibinfo{pages}{365--374}.
\bibitem[{Frondel et~al.(2017)Frondel, Simora \& Sommer}]{Frondel2017}
\bibinfo{author}{Frondel, M.}, \bibinfo{author}{Simora, M.}, \& \bibinfo{author}{Sommer, S.} (\bibinfo{year}{2017}).
\newblock \bibinfo{title}{Risk perception of climate change: {E}mpirical evidence for {G}ermany}.
\newblock {\it \bibinfo{journal}{Ecological Economics}\/},  {\it \bibinfo{volume}{137}\/}, \bibinfo{pages}{173--183}.
\bibitem[{Garay et~al.(2020{\natexlab{a}})Garay, Medina, Cabral \& Lin}]{Garayetal20}
\bibinfo{author}{Garay, A.~M.}, \bibinfo{author}{Medina, F.~L.}, \bibinfo{author}{Cabral, C.~R.}, \& \bibinfo{author}{Lin, T.-I.} (\bibinfo{year}{2020}{\natexlab{a}}).
\newblock \bibinfo{title}{Bayesian analysis of the p-order integer-valued ar process with zero-inflated poisson innovations}.
\newblock {\it \bibinfo{journal}{Journal of Statistical Computation and Simulation}\/},  {\it \bibinfo{volume}{90}\/}, \bibinfo{pages}{1943--1964}.
\bibitem[{Garay et~al.(2020{\natexlab{b}})Garay, Medina, Cabral \& Lin}]{Garetal20}
\bibinfo{author}{Garay, A.~M.}, \bibinfo{author}{Medina, F.~L.}, \bibinfo{author}{Cabral, C. R.~B.}, \& \bibinfo{author}{Lin, T.-I.} (\bibinfo{year}{2020}{\natexlab{b}}).
\newblock \bibinfo{title}{Bayesian analysis of the p-order integer-valued ar process with zero-inflated poisson innovations}.
\newblock {\it \bibinfo{journal}{Journal of Statistical Computation and Simulation}\/},  {\it \bibinfo{volume}{90}\/}, \bibinfo{pages}{1943--1964}.
\bibitem[{Geyer(1992)}]{Geyer92}
\bibinfo{author}{Geyer, C.~J.} (\bibinfo{year}{1992}).
\newblock \bibinfo{title}{Practical {M}arkov chain {M}onte {C}arlo}.
\newblock {\it \bibinfo{journal}{Statistical Science}\/},  (pp. \bibinfo{pages}{473--483}).
\bibitem[{Janardan(1998)}]{jan98}
\bibinfo{author}{Janardan, K.} (\bibinfo{year}{1998}).
\newblock \bibinfo{title}{Generalized {P}olya {E}ggenberger family of distributions and its relation to {L}agrangian {K}atz family}.
\newblock {\it \bibinfo{journal}{Communications in Statistics-Theory and Methods}\/},  {\it \bibinfo{volume}{27}\/}, \bibinfo{pages}{2423--2442}.
\bibitem[{Janardan(1999)}]{jan99}
\bibinfo{author}{Janardan, K.} (\bibinfo{year}{1999}).
\newblock \bibinfo{title}{Estimation of parameters of the {GPED}}.
\newblock {\it \bibinfo{journal}{Communications in Statistics-Theory and Methods}\/},  {\it \bibinfo{volume}{28}\/}, \bibinfo{pages}{2167--2179}.
\bibitem[{Jin-Guan \& Yuan(1991)}]{DuLi1991}
\bibinfo{author}{Jin-Guan, D.}, \& \bibinfo{author}{Yuan, L.} (\bibinfo{year}{1991}).
\newblock \bibinfo{title}{The integer-valued autoregressive ({INAR} (p)) model}.
\newblock {\it \bibinfo{journal}{Journal of {T}ime {S}eries {A}nalysis}\/},  {\it \bibinfo{volume}{12}\/}, \bibinfo{pages}{129--142}.
\bibitem[{Katz(1965)}]{katz1965}
\bibinfo{author}{Katz, L.} (\bibinfo{year}{1965}).
\newblock \bibinfo{title}{Unified treatment of a broad class of discrete probability distributions}.
\newblock {\it \bibinfo{journal}{Classical and contagious discrete distributions}\/},  {\it \bibinfo{volume}{1}\/}, \bibinfo{pages}{175--182}.
\bibitem[{Kim \& Lee(2017)}]{KimLee2017}
\bibinfo{author}{Kim, H.}, \& \bibinfo{author}{Lee, S.} (\bibinfo{year}{2017}).
\newblock \bibinfo{title}{On first-order integer-valued autoregressive process with {K}atz family innovations}.
\newblock {\it \bibinfo{journal}{Journal of Statistical Computation and Simulation}\/},  {\it \bibinfo{volume}{87}\/}, \bibinfo{pages}{546--562}.
\bibitem[{Kim \& Park(2008)}]{KimPar2008}
\bibinfo{author}{Kim, H.-Y.}, \& \bibinfo{author}{Park, Y.} (\bibinfo{year}{2008}).
\newblock \bibinfo{title}{A non-stationary integer-valued autoregressive model}.
\newblock {\it \bibinfo{journal}{Statistical {P}apers}\/},  {\it \bibinfo{volume}{49}\/}, \bibinfo{pages}{485}.
\bibitem[{Liesenfeld et~al.(2006)Liesenfeld, Nolte \& Pohlmeier}]{LieMolPoh2006}
\bibinfo{author}{Liesenfeld, R.}, \bibinfo{author}{Nolte, I.}, \& \bibinfo{author}{Pohlmeier, W.} (\bibinfo{year}{2006}).
\newblock \bibinfo{title}{Modelling financial transaction price movements: {A} dynamic integer count data model}.
\newblock {\it \bibinfo{journal}{Empirical Economics}\/},  {\it \bibinfo{volume}{30}\/}, \bibinfo{pages}{795--825}.
\bibitem[{Lineman et~al.(2015)Lineman, Do, Kim \& Joo}]{lineman2015talking}
\bibinfo{author}{Lineman, M.}, \bibinfo{author}{Do, Y.}, \bibinfo{author}{Kim, J.~Y.}, \& \bibinfo{author}{Joo, G.-J.} (\bibinfo{year}{2015}).
\newblock \bibinfo{title}{Talking about climate change and global warming}.
\newblock {\it \bibinfo{journal}{PloS one}\/},  {\it \bibinfo{volume}{10}\/}, \bibinfo{pages}{e0138996}.
\bibitem[{Maiti et~al.(2015)Maiti, Biswas \& Das}]{maiti2015time}
\bibinfo{author}{Maiti, R.}, \bibinfo{author}{Biswas, A.}, \& \bibinfo{author}{Das, S.} (\bibinfo{year}{2015}).
\newblock \bibinfo{title}{Time series of zero-inflated counts and their coherent forecasting}.
\newblock {\it \bibinfo{journal}{Journal of Forecasting}\/},  {\it \bibinfo{volume}{34}\/}, \bibinfo{pages}{694--707}.
\bibitem[{Malyshkina et~al.(2009)Malyshkina, Mannering \& Tarko}]{malyshkina2009markov}
\bibinfo{author}{Malyshkina, N.~V.}, \bibinfo{author}{Mannering, F.~L.}, \& \bibinfo{author}{Tarko, A.~P.} (\bibinfo{year}{2009}).
\newblock \bibinfo{title}{Markov switching negative binomial models: an application to vehicle accident frequencies}.
\newblock {\it \bibinfo{journal}{Accident Analysis \& Prevention}\/},  {\it \bibinfo{volume}{41}\/}, \bibinfo{pages}{217--226}.
\bibitem[{Marques et~al.(2022)Marques, Graziadei \& Lopes}]{c2022bayesian}
\bibinfo{author}{Marques, P.}, \bibinfo{author}{Graziadei, H.}, \& \bibinfo{author}{Lopes, H.~F.} (\bibinfo{year}{2022}).
\newblock \bibinfo{title}{Bayesian generalizations of the integer-valued autoregressive model}.
\newblock {\it \bibinfo{journal}{Journal of Applied Statistics}\/},  {\it \bibinfo{volume}{49}\/}, \bibinfo{pages}{336--356}.
\bibitem[{McCabe \& Martin(2005)}]{CabMar05}
\bibinfo{author}{McCabe, B.}, \& \bibinfo{author}{Martin, G.} (\bibinfo{year}{2005}).
\newblock \bibinfo{title}{Bayesian predictions of low count time series}.
\newblock {\it \bibinfo{journal}{International Journal of Forecasting}\/},  {\it \bibinfo{volume}{21}\/}, \bibinfo{pages}{315--330}.
\bibitem[{McCabe \& Skeels(2020)}]{mccabe2020distributions}
\bibinfo{author}{McCabe, B.~P.}, \& \bibinfo{author}{Skeels, C.~L.} (\bibinfo{year}{2020}).
\newblock \bibinfo{title}{Distributions you can count on... {B}ut what’s the point?}
\newblock {\it \bibinfo{journal}{Econometrics}\/},  {\it \bibinfo{volume}{8}\/}, \bibinfo{pages}{9}.
\bibitem[{McKenzie(1985)}]{McK1985}
\bibinfo{author}{McKenzie, E.} (\bibinfo{year}{1985}).
\newblock \bibinfo{title}{Some simple models for discrete variate time series}.
\newblock {\it \bibinfo{journal}{Water Resources Bulletin}\/},  {\it \bibinfo{volume}{21}\/}, \bibinfo{pages}{645--650}.
\bibitem[{McKenzie(1986)}]{McK1986}
\bibinfo{author}{McKenzie, E.} (\bibinfo{year}{1986}).
\newblock \bibinfo{title}{Autoregressive moving-average processes with negative-binomial and geometric marginal distributions}.
\newblock {\it \bibinfo{journal}{Advances in Applied Probability}\/},  {\it \bibinfo{volume}{18}\/}, \bibinfo{pages}{679?705}.
\bibitem[{McKenzie(2003)}]{mckenzie2003ch}
\bibinfo{author}{McKenzie, E.} (\bibinfo{year}{2003}).
\newblock \bibinfo{title}{Discrete variate time series}.
\newblock In \bibinfo{editor}{D.~N. Shanbhag}, \& \bibinfo{editor}{C.~R. Rao} (Eds.), {\it \bibinfo{booktitle}{Stochastic Processes: Modelling and Simulation}\/} (pp. \bibinfo{pages}{573--606}).
\newblock \bibinfo{publisher}{Elsevier}.
\bibitem[{Neal \& Subba~Rao(2007)}]{NeSu07}
\bibinfo{author}{Neal, P.}, \& \bibinfo{author}{Subba~Rao, T.} (\bibinfo{year}{2007}).
\newblock \bibinfo{title}{{MCMC} for integer-valued {ARMA} processes}.
\newblock {\it \bibinfo{journal}{Journal of Time Series Analysis}\/},  {\it \bibinfo{volume}{28}\/}, \bibinfo{pages}{92--110}.
\bibitem[{Pedeli \& Karlis(2011)}]{PedKar2011}
\bibinfo{author}{Pedeli, X.}, \& \bibinfo{author}{Karlis, D.} (\bibinfo{year}{2011}).
\newblock \bibinfo{title}{A bivariate {INAR}(1) process with application}.
\newblock {\it \bibinfo{journal}{Statistical Modelling}\/},  {\it \bibinfo{volume}{11}\/}, \bibinfo{pages}{325--349}.
\bibitem[{Robert \& Casella(2013)}]{RobCas2013}
\bibinfo{author}{Robert, C.}, \& \bibinfo{author}{Casella, G.} (\bibinfo{year}{2013}).
\newblock {\it \bibinfo{title}{Monte Carlo statistical methods}\/}.
\newblock \bibinfo{publisher}{Springer Science \& Business Media}.
\bibitem[{Roberts et~al.(1997)Roberts, Gelman, Gilks et~al.}]{Robetal1997}
\bibinfo{author}{Roberts, G.~O.}, \bibinfo{author}{Gelman, A.}, \bibinfo{author}{Gilks, W.~R.} et~al. (\bibinfo{year}{1997}).
\newblock \bibinfo{title}{Weak convergence and optimal scaling of random walk metropolis algorithms}.
\newblock {\it \bibinfo{journal}{The Annals of Applied Probability}\/},  {\it \bibinfo{volume}{7}\/}, \bibinfo{pages}{110--120}.
\bibitem[{Schiavoni et~al.(2021)Schiavoni, Palm, Smeekes \& van~den Brakel}]{Schetal2020}
\bibinfo{author}{Schiavoni, C.}, \bibinfo{author}{Palm, F.}, \bibinfo{author}{Smeekes, S.}, \& \bibinfo{author}{van~den Brakel, J.} (\bibinfo{year}{2021}).
\newblock \bibinfo{title}{A dynamic factor model approach to incorporate big data in state space models for official statistics}.
\newblock {\it \bibinfo{journal}{Journal of the Royal Statistical Society: Series A (Statistics in Society)}\/},  {\it \bibinfo{volume}{184}\/}, \bibinfo{pages}{324--353}.
\bibitem[{Schweer \& Wei\ss(2014)}]{Schweer2014}
\bibinfo{author}{Schweer, S.}, \& \bibinfo{author}{Wei\ss, C.~H.} (\bibinfo{year}{2014}).
\newblock \bibinfo{title}{Compound {P}oisson {INAR}(1) processes: stochastic properties and testing for overdispersion}.
\newblock {\it \bibinfo{journal}{Comput. Statist. Data Anal.}\/},  {\it \bibinfo{volume}{77}\/}, \bibinfo{pages}{267--284}.
\bibitem[{Scott \& Varian(2014)}]{scott2014}
\bibinfo{author}{Scott, S.~L.}, \& \bibinfo{author}{Varian, H.~R.} (\bibinfo{year}{2014}).
\newblock \bibinfo{title}{Predicting the present with {B}ayesian structural time series}.
\newblock {\it \bibinfo{journal}{International Journal of Mathematical Modelling and Numerical Optimisation}\/},  {\it \bibinfo{volume}{5}\/}, \bibinfo{pages}{4--23}.
\bibitem[{Scotto et~al.(2015)Scotto, Wei{\ss} \& Gouveia}]{ScoWeiGou2015}
\bibinfo{author}{Scotto, M.~G.}, \bibinfo{author}{Wei{\ss}, C.~H.}, \& \bibinfo{author}{Gouveia, S.} (\bibinfo{year}{2015}).
\newblock \bibinfo{title}{Thinning-based models in the analysis of integer-valued time series: {A} review}.
\newblock {\it \bibinfo{journal}{Statistical Modelling}\/},  {\it \bibinfo{volume}{15}\/}, \bibinfo{pages}{590--618}.
\bibitem[{Shahtahmassebi \& Moyeed(2016)}]{ShaMoy2016}
\bibinfo{author}{Shahtahmassebi, G.}, \& \bibinfo{author}{Moyeed, R.} (\bibinfo{year}{2016}).
\newblock \bibinfo{title}{An application of the generalized {P}oisson difference distribution to the {B}ayesian modelling of football scores}.
\newblock {\it \bibinfo{journal}{Statistica Neerlandica}\/},  {\it \bibinfo{volume}{70}\/}, \bibinfo{pages}{260--273}.
\bibitem[{Shang \& Zhang(2018)}]{ShaZha18}
\bibinfo{author}{Shang, H.}, \& \bibinfo{author}{Zhang, B.} (\bibinfo{year}{2018}).
\newblock \bibinfo{title}{Outliers detection in {INAR} (1) time series}.
\newblock {\it \bibinfo{journal}{Journal of Physics: Conference Series}\/},  {\it \bibinfo{volume}{1053}\/}, \bibinfo{pages}{012094}.
\bibitem[{Soyer \& Zhang(2022)}]{soyer2022bayesian}
\bibinfo{author}{Soyer, R.}, \& \bibinfo{author}{Zhang, D.} (\bibinfo{year}{2022}).
\newblock \bibinfo{title}{Bayesian modeling of multivariate time series of counts}.
\newblock {\it \bibinfo{journal}{Wiley Interdisciplinary Reviews: Computational Statistics}\/},  {\it \bibinfo{volume}{14}\/}, \bibinfo{pages}{e1559}.
\bibitem[{Spiegelhalter et~al.(2002)Spiegelhalter, Best, Carlin \& Van Der~Linde}]{Spiegel2002}
\bibinfo{author}{Spiegelhalter, D.~J.}, \bibinfo{author}{Best, N.~G.}, \bibinfo{author}{Carlin, B.~P.}, \& \bibinfo{author}{Van Der~Linde, A.} (\bibinfo{year}{2002}).
\newblock \bibinfo{title}{Bayesian measures of model complexity and fit}.
\newblock {\it \bibinfo{journal}{Journal of the Royal Statistical Society: Series B (Statistical Methodology)}\/},  {\it \bibinfo{volume}{64}\/}, \bibinfo{pages}{583--639}.
\bibitem[{Steutel \& van Harn(1979)}]{Ste79}
\bibinfo{author}{Steutel, F.~W.}, \& \bibinfo{author}{van Harn, K.} (\bibinfo{year}{1979}).
\newblock \bibinfo{title}{Discrete analogues of self-decomposability and stability}.
\newblock {\it \bibinfo{journal}{The Annals of Probability}\/},  (pp. \bibinfo{pages}{893--899}).
\bibitem[{Ullah et~al.(2018)Ullah, Rashid, Liu \& Hussain}]{Ullah2018}
\bibinfo{author}{Ullah, H.}, \bibinfo{author}{Rashid, A.}, \bibinfo{author}{Liu, G.}, \& \bibinfo{author}{Hussain, M.} (\bibinfo{year}{2018}).
\newblock \bibinfo{title}{Perceptions of mountainous people on climate change, livelihood practices and climatic shocks: {A} case study of {S}wat {D}istrict, {P}akistan}.
\newblock {\it \bibinfo{journal}{Urban Climate}\/},  {\it \bibinfo{volume}{26}\/}, \bibinfo{pages}{244--257}.
\bibitem[{Wei{\ss}(2008)}]{weiss2008combined}
\bibinfo{author}{Wei{\ss}, C.~H.} (\bibinfo{year}{2008}).
\newblock \bibinfo{title}{The combined {INAR} (p) models for time series of counts}.
\newblock {\it \bibinfo{journal}{Statistics \& probability letters}\/},  {\it \bibinfo{volume}{78}\/}, \bibinfo{pages}{1817--1822}.
\bibitem[{Wei\ss(2013)}]{weiss2013}
\bibinfo{author}{Wei\ss, C.~H.} (\bibinfo{year}{2013}).
\newblock \bibinfo{title}{Integer-valued autoregressive models for counts showing underdispersion}.
\newblock {\it \bibinfo{journal}{J. Appl. Stat.}\/},  {\it \bibinfo{volume}{40}\/}, \bibinfo{pages}{1931--1948}.
\bibitem[{Wei{\ss} \& Kim(2013)}]{WeiKim2013}
\bibinfo{author}{Wei{\ss}, C.~H.}, \& \bibinfo{author}{Kim, H.-Y.} (\bibinfo{year}{2013}).
\newblock \bibinfo{title}{Parameter estimation for binomial {AR(1)} models with applications in finance and industry}.
\newblock {\it \bibinfo{journal}{Statistical Papers}\/},  {\it \bibinfo{volume}{54}\/}, \bibinfo{pages}{563--590}.
\bibitem[{Yang et~al.(2021)Yang, Ning \& Kou}]{Yaetal21}
\bibinfo{author}{Yang, S.}, \bibinfo{author}{Ning, S.}, \& \bibinfo{author}{Kou, S.~C.} (\bibinfo{year}{2021}).
\newblock \bibinfo{title}{Use internet search data to accurately track state level influenza epidemics}.
\newblock {\it \bibinfo{journal}{Scientific Reports}\/},  {\it \bibinfo{volume}{11}\/}, \bibinfo{pages}{4023}.
\bibitem[{Yi et~al.(2021)Yi, Ning, Chang \& Kou}]{Dinetal21}
\bibinfo{author}{Yi, D.}, \bibinfo{author}{Ning, S.}, \bibinfo{author}{Chang, C.-J.}, \& \bibinfo{author}{Kou, S.~C.} (\bibinfo{year}{2021}).
\newblock \bibinfo{title}{Forecasting unemployment using internet search data via prism}.
\newblock {\it \bibinfo{journal}{Journal of the American Statistical Association}\/},  {\it \bibinfo{volume}{116}\/}, \bibinfo{pages}{1662--1673}.
\bibitem[{Yu et~al.(2019)Yu, Zhao, Tang \& Yang}]{Yuetal19}
\bibinfo{author}{Yu, L.}, \bibinfo{author}{Zhao, Y.}, \bibinfo{author}{Tang, L.}, \& \bibinfo{author}{Yang, Z.} (\bibinfo{year}{2019}).
\newblock \bibinfo{title}{Online big data-driven oil consumption forecasting with {G}oogle trends}.
\newblock {\it \bibinfo{journal}{International Journal of Forecasting}\/},  {\it \bibinfo{volume}{35}\/}, \bibinfo{pages}{213--223}.
\bibitem[{Ziegler(2017)}]{Zieg2017}
\bibinfo{author}{Ziegler, A.} (\bibinfo{year}{2017}).
\newblock \bibinfo{title}{Political orientation, environmental values, and climate change beliefs and attitudes: An empirical cross country analysis}.
\newblock {\it \bibinfo{journal}{Energy Economics}\/},  {\it \bibinfo{volume}{63}\/}, \bibinfo{pages}{144--153}.

\end{thebibliography}

\newpage

\begin{center}
\Large First--order integer--valued autoregressive processes with Generalized Katz innovations

Supplement 
\end{center}

This supplement consists of four Appendices. Appendix A provides some properties of the GLK distributions. Appendix B provides proof of the paper's results. Appendix C contains the simulation results, while Appendix D includes more details on the empirical application.

\renewcommand\thefigure{A.\arabic{figure}}
\setcounter{figure}{0}
\renewcommand\theequation{A.\arabic{equation}}
\setcounter{equation}{0}
\renewcommand\thetable{A.\arabic{table}}
\setcounter{table}{0}

\appendix
\section{Properties of the GLK distributions}\label{App:Prop}
Studying the moments allows for a better understanding of the flexibility of the GLK distribution. The following are four moments relevant to our analysis.
\begin{prop}\label{Th:Moment}
Let $X\sim \mathcal{GLK}(a,b,c,\beta)$, define $\mu_k'=\mathbb{E}((X-\mathbb{E}(X))^k)$ and $\mu_k=\mathbb{E}(X^k)$ then
\begin{eqnarray*}
\mu_1&=& \frac{a\theta}{\kappa},\quad\quad \mu_2' = \frac{(1-\beta)a\theta}{\kappa^{3}},\quad\quad \\
\mu_3' &=& \frac{a\theta(1-2\beta)(1-\beta)}{\kappa^{4}} + \frac{3a\theta^{2} (1-\beta)^{2}(b+c)}{\kappa^{5}}\\
\mu_4' &=& a\theta(1-\beta)(1+2\theta b-(b+c)\beta\theta)\left( \frac{1-\beta-\beta^{2}}{\kappa^{6}}\right.\\
&\,&\left. + \frac{5a\theta (1-\beta)(b+c)}{\kappa^{7}} \right)+3(\mu_{2}')^{2},
\end{eqnarray*}
where $\kappa=1 -\beta - b \beta/c>0$ and $\theta=\beta/c$.
\end{prop}
For a proof, see \cite{jan98} Theorems 1--3.
\begin{figure}[t]
\begin{center}
\renewcommand{\arraystretch}{1}
\setlength{\tabcolsep}{5pt}
\begin{tabular}{cc}
\small (a) Mean ($\mu_1$) & \small (b) Index of Dispersion ($VMR$)\vspace{7pt}\\
\includegraphics[scale=0.37]{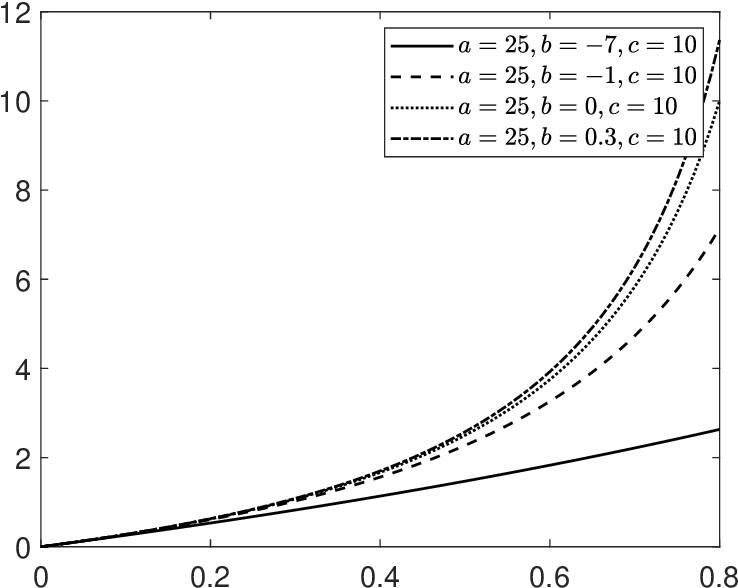}&
\includegraphics[scale=0.37]{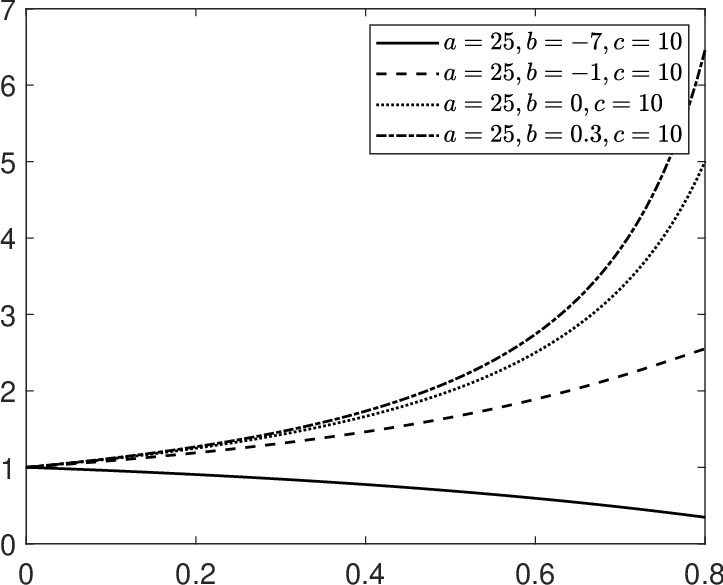}\\
\small (c) Skewness ($S$) & \small (d) Kurtosis ($K$)\vspace{7pt}\\
\includegraphics[scale=0.37]{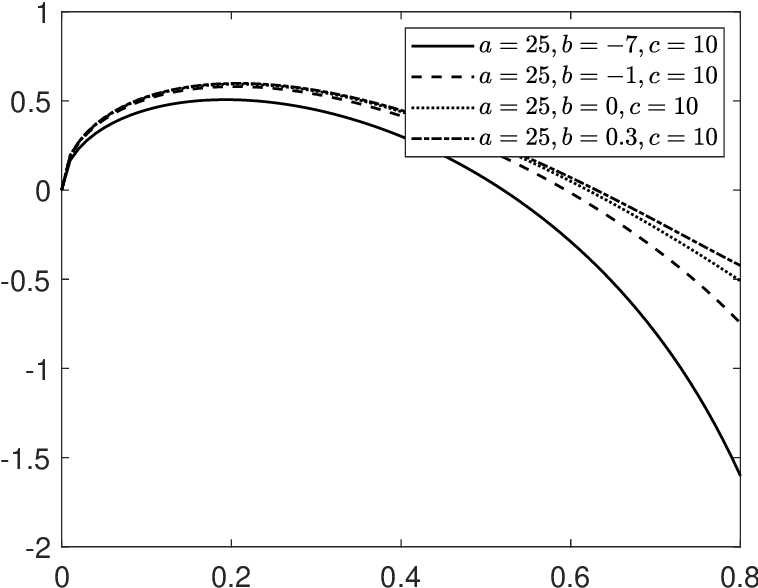}&
\includegraphics[scale=0.37]{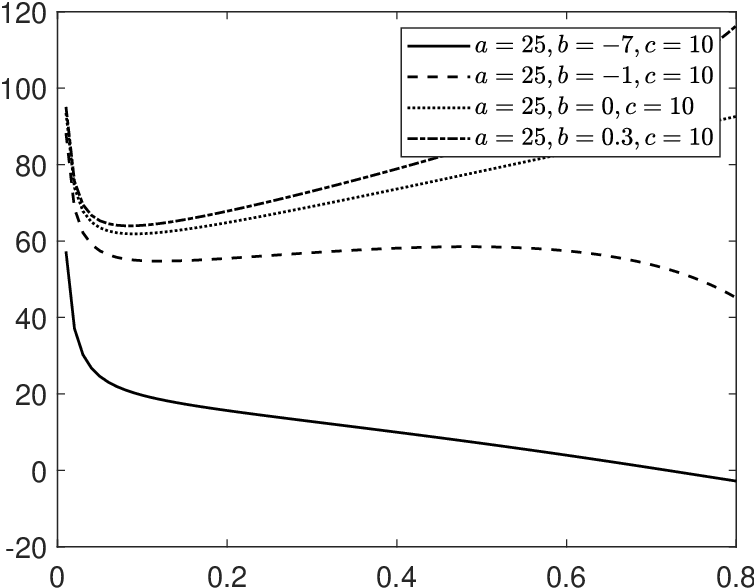}
\end{tabular}
\end{center}
\caption{GLK moments when increasing the value of $\beta$ (horizontal axis) for different values of $b$ (lines).}\label{fig:moments}
\end{figure}

The skewness and the kurtosis of the distribution are
\begin{eqnarray*}
S &=& \frac{(1-2\beta)\kappa^{1/2}}{((1-\beta)a\theta)^{1/2}} + \frac{3((1-\beta)\theta)^{1/2} (b+c)}{a\kappa^{1/2}},\\
K &=& (1+2b\theta -(b+c)c\theta^{2})\left(\frac{(1-\beta -\beta^{2})}{a\theta (1-\beta)} + \frac{5(b+c)}{\kappa}\right) +3,
\end{eqnarray*}
respectively. For a given value of $\theta$, there is negative skewness if $\beta<(1+\xi)/(2\kappa+\xi)$ with $\xi=3\theta(b+c)a^{-1/2}$ and positive otherwise. 
%

Figure \ref{fig:moments} illustrates the effect of the parameter values on the mean, dispersion index, skewness and kurtosis. Increasing the value of $\beta$ (horizontal axis) the $\mathcal{GLK}(a,b,c,\beta)$ distribution allows for different types of dispersion (panel b), for both negative and positive skewness (panel c) and various degrees of excess of kurtosis (panel d).

\pagebreak

\renewcommand\thefigure{B.\arabic{figure}}
\setcounter{figure}{0}
\renewcommand\theequation{B.\arabic{equation}}
\setcounter{equation}{0}
\renewcommand\thetable{B.\arabic{table}}
\setcounter{table}{0}

\section{Details of some statements of  Section \ref{sec:INARtheory}}\label{App:ThProof}
\subsection{Connection with generalized Lagrangian distributions}

As stated in  Remark \ref{rem:GLKpmf}
 it is possible to derive the Generalized Lagrangian Katz distribution 
 as a "generalized Lagrangian distribution". 
Let $f (z)$ and $g(z)$  be two analytic functions of $z$, which are infinitely differentiable
in $[-1,1]$ with $g(0)\not =0$. 
Following \cite[][p. 10-11]{ConFam2006} the general Lagrangian expansion of  $f$ is
\begin{equation}
\frac{f(z)}{1-zg^{\prime}/g(z)}=\sum_{j=0}^{\infty}\frac{u^j}{j!}\left\vert\partial^{j}(g^{j}(z)f(z))\right\vert_{z=0},
\end{equation}
where $u$ satisfies $z=ug(z)$. The definition of Lagrangian distribution given in \cite{jan98} uses a slightly different expansion, which is obtained from the one given above by replacing $f(z)$ with $f(z)(1-z g^{\prime}(z)/g(z)))$. 
By applying iteratively the derivative $\partial$ to the product of functions, we obtain the coefficient in the $j$-th term of the expansion
\begin{eqnarray*}
&&\hspace{-5pt}\frac{1}{j!}\vert\partial^{j}\left(g(z)^{j}f(z)(1-z g^{\prime}(z)/g(z))\right)\vert_{z=0}\nonumber\\
&&=\frac{1}{j!}\vert\partial^{j-1}(g(z)^{j}f^{\prime}(z)\nonumber\\
&&\hspace{-10pt}+(j-1) g^{\prime}(z)g(z)^{j-1}f(z)-z\partial g^{j-1}(z)g^{\prime}(z)f(z))\vert_{z=0}=\ldots \nonumber\\
&&\hspace{-10pt}=\frac{1}{j!}\vert\partial^{j-1}(g(z)^{j}f^{\prime}(z))\vert_{z=0}+\vert\partial^{j-\ell}\left((j-\ell) \partial^{\ell-1} (g^{\prime}(z)g(z)^{j-1}f(z))\right.\nonumber\\
&&\hspace{-10pt}\left. -z \partial^{\ell}(g^{j-1}(z)g^{\prime}(z)f(z))\right)\vert_{z=0}=\frac{1}{j!}\vert\partial^{j-1}(g(z)^{j}f^{\prime}(z))\vert_{z=0},
\end{eqnarray*}
where we set $\ell=j$ to get the result  and the following equivalent Lagrangian expansion used in \cite{jan98}
\begin{eqnarray}\label{lagr1}
&&\frac{f(z)}{1-zg^{\prime}/g(z)}=\sum_{j=0}^{\infty}\frac{u^j}{j!}\left\vert\partial^{j}(g^{j}(z)f(z))\right\vert_{z=0}\\
&&\Leftrightarrow f(z)=\sum_{j=0}^{\infty}\frac{u^j}{j!}\left\vert\partial^{j-1}(g^{j}(z) f^{\prime}(z))\right\vert_{z=0}
\end{eqnarray}
In particular, if $f(1)=g(1)=1$, the function 
 $u \mapsto f(z(u))$ defines the pgf of  the 
 "generalized Lagrangian distribution"
$
p_j=\frac{1}{j!} \left\vert\partial^{j-1}(g^{j}(z) f^{\prime}(z))\right\vert_{z=0}
$
provided that $p_j\geq0$ for $j=0,1,\dots$.
Assuming  $f(z) = \left( \frac{1-\beta}{1-\beta z} \right)^{a/c}$ and $g(z) = \left( \frac{1-\beta}{1-\beta z} \right)^{b/c}$,
 the expressions in \eqref{pgf1} and \eqref{pmf} follows after some algebra as detailed in the following. The  expansion coefficients become 
\begin{equation*}
f'(z)=\frac{a}{c} \left( \frac{1-\beta}{1-\beta z}\right)^{\frac{a}{c} + 1} \frac{\beta}{1-\beta},
\end{equation*}
\begin{equation*}
g^{k}(z)f'(z) = \frac{a}{c} \left( \frac{1-\beta}{1-\beta z}\right)^{\frac{a}{c}+k\frac{b}{c}+1} \frac{\beta}{1-\beta}.
\end{equation*}
Hence 
\[
p_0 =f(0)= \left({1-\beta}\right)^{\frac{a}{c}}
\quad 
p_1=g^{1}(0)f'(0) =\frac{a}{c} \left({1-\beta}\right)^{\frac{a}{c}+\frac{b}{c}} {\beta}.
\]
while, for $k \geq 2$, the $k$-th coefficient of the Lagrangian expansion in Eq. \ref{lagr1} is 
\begin{eqnarray*}
&&\hspace{-25pt}p_k=\frac{1}{k!}\vert\partial^{k-1}(g(z))^{k}f'(z)\vert_{z=0} =\\
&&=\frac{1}{k!}\partial^{k-2}\vert(\frac{a}{c}\frac{\beta^2}{(1-\beta)^2}\xi_k\left( \frac{1-\beta}{1-\beta z}\right)^{\xi_k+1})\vert_{z=0}\\
&&\hspace{-25pt}=\frac{1}{k!}\vert\partial^{k-3}(\frac{a}{c}\frac{\beta^3}{(1-\beta)^{3}}(\xi_k(\xi_k+1))\left( \frac{1-\beta}{1-\beta z}\right)^{\xi_k+2})\vert_{z=0}\\
&&\hspace{-25pt}=...\\
&&\hspace{-25pt}=\frac{1}{k!}\beta^{k}\frac{a}{c}(1-\beta)^{\frac{a}{c}+k\frac{b}{c}} \prod_{m=0}^{k-2}\left( \xi_k +m\right) \\
&&\hspace{-25pt}=\frac{1}{k!}\beta^{k}\frac{a}{c}(1-\beta)^{\frac{a}{c}+k\frac{b}{c}} \prod_{m=1}^{k-1}\left( \frac{a}{c}+k\frac{b}{c} +m\right) \\
&&\hspace{-25pt}=\frac{1}{k!}\beta^{k}\frac{a}{c}(1-\beta)^{\frac{a}{c}+k\frac{b}{c}} \left( \frac{a}{c}+k\frac{b}{c} +1\right)_{{k-1}\uparrow} \\
&&\hspace{-25pt}=\frac{1}{k!}\beta^{k}\frac{a}{c} \frac{1}{(\frac{a}{c}+k\frac{b}{c}+k) }(1-\beta)^{\frac{a}{c}+k\frac{b}{c}} \left( \frac{a}{c}+k\frac{b}{c} +1\right)_{{k}\uparrow} \\
\end{eqnarray*}
where  
 $\xi_k=\frac{a}{c}+k\frac{b}{c}+1$ and $(x)_{{k} \uparrow} = x(x+1) \ldots (x+k-1)$ is the rising factorial.

 We now discuss conditions for which $p_k \geq 0$ for all $k \geq 0$. 
  \begin{itemize}
  \item 
 If $a>0,b>0,c>0$ one has $p_k>0$ for every $k \geq 1$. 
 \item 
If  $-c<b<0$, $a/c,b/c\in\mathbb{N}$ and $(c-a)/(c+b) \leq (a+c)/|b|$,  then
for $k < k^*=(a+c)/|b|$ one has 
\[
\frac{a+bk}{c} + 1 > 0
\]
and hence also 
\[
 \prod_{m=1}^{k-1}\left( \frac{a}{c}+k\frac{b}{c} +m\right)>0
\]
proving that $p_k>0$. For $k \geq k^*\geq (c-a)/(c+b)$ one has that 
$m_k=(|b|k-a)/c$ is an integer with $1 \leq m_k \leq k-1$ and hence the product 
$\prod_{m=1}^{k-1}\left( \frac{a}{c}+k\frac{b}{c} +m\right)=0$ since 
for $m=m_k$ one has  $\frac{a}{c}+k\frac{b}{c} +m=0$. This shows that $p_k=0$ for every $k \geq k^*$. 
  \end{itemize}

\subsection{Generalized Poisson as limit}\label{Appendix:LKlimit}

One obtains a Lagrangian Katz distribution by replacing $c$ by $\beta$. The LK is one of the few distributions which admit more pgfs. Let us consider the following definition of pgf for a $\mathcal{LK}(a,b,\beta)$ 
\begin{equation}
G(u,a,b,\beta)=\left(\frac{1-\beta z}{1-\beta}\right)^{-\frac{a}{\beta}},\quad \hbox{with}\,\,z(a,b,\beta)=u\left(\frac{1-\beta z}{1-\beta}\right)^{-\frac{b}{\beta}}.
\end{equation}
given in \citep[][p. 241]{ConFam2006}. Defining $n=(1-\beta z)/(\beta(z-1))$ and $1/\beta=n(z-1)+z$ the limiting pgf becomes
\begin{equation}
\underset{\beta \rightarrow 0^{+}}{\lim} G(u;a,b,\beta)=\underset{n \rightarrow +\infty}{\lim}\left(1+\frac{1}{n}\right)^{\frac{n(z-1)+z}{a}}=e^{a(z-1)},
\end{equation}
with
\begin{equation}
\underset{\beta \rightarrow 0^{+}}{\lim} z(a,b,\beta)=\underset{n \rightarrow +\infty}{\lim}\left(1+\frac{1}{n}\right)^{\frac{n(z-1)+z}{b}}=e^{b(z-1)}
\end{equation}
which is the pgf of the Generalized Poisson  given in \citep[][pp. 166]{ConFam2006}.

\subsection{Proof of the results in Theorem \ref{Th:stationaryMomentsCor}}\label{AppendixThmMoments}
\textit{(i)} Under stationarity assumption one has $\mu_X=E(X_s)$ for all $s\in \mathbb{Z}$, thus $\mu_X=\alpha \mu_X+\mathbb{E}(\varepsilon_t)$ which implies $\mu_X=\mu_{\varepsilon}/(1-\alpha)$.

\noindent\textit{(ii)} Let $\mu_X^{(2)}=E(X_s^2)$ for all $s\in\mathbb{Z}$, then $\mathbb{E}(X_t^2)=\mathbb{E}((\alpha \circ X_{t-1})^2)+\mathbb{E}(\varepsilon_t^2)+\mathbb{E}(2(\alpha \circ X_{t-1})\varepsilon_t)=$ $\mathbb{V}((\alpha \circ X_{t-1})^2)+(\mathbb{E}(\alpha \circ X_{t-1}))^2+\mathbb{E}(\varepsilon_t^2)+\mathbb{E}(2(\alpha \circ X_{t-1})\varepsilon_t)$. By the law of iterated expectation
\[
\begin{split}
\mu_X^{(2)}=& \alpha^2 (\mu_X^{(2)}-\mu_X^{2})+\alpha(1-\alpha)\mu_X+\alpha^2\mu_X^2 \\
& +\mu_{\varepsilon}^{(2)}+\alpha \mu_X \mu_{\varepsilon}\\
\end{split}
\]
and hence 
\[
 \mu_X^{(2)}=\frac{1}{1-\alpha^2}\left(\alpha\mu_{\varepsilon}+\mu_{\varepsilon}^{(2)}+\frac{2\alpha}{1-\alpha} \mu_{\varepsilon}^2\right)
\]

\noindent\textit{(iii)} 
One has 
\[
\begin{split}
& \mathbb{E}(X_{t}X_{t-k})=\mathbb{E}((\alpha\circ X_{t-1}+\varepsilon_{t})X_{t-k})=
\\
& \mathbb{E}(\mathbb{E}((\alpha\circ X_{t-1})X_{t-k}|X_{t-k},X_{t-1}))+\mathbb{E}(\varepsilon_{t})\mathbb{E}(X_{t-k})\\
& =\alpha \mathbb{E}(X_{t-1}X_{t-k})+\mu_{\varepsilon}\mu_X.\\
\end{split}
\]

\noindent\textit{(iv)}
Let us denote with $(x)_m=x(x-1)\ldots(x-m+1)$ the falling factorial and with $\mu^{(\underline{k})}=\mathbb{E}((X)_k)$ the $m$-order falling factorial moment of a random variable $X$. The following two results will be used. The relationships between non-central moments and falling factorial moments are
\begin{eqnarray}\label{fallfact}
\mathbb{E}(X_t^m)&=&\sum_{k=0}^{m}S(m,k)\mathbb{E}((X_t)_k)\\
\mathbb{E}((X_t)_m)&=&\sum_{k=0}^{m}s(m,k)\mathbb{E}(X_t^k)
\end{eqnarray}
where $s(m,k)$ and $S(m,k)$ are the Stirling numbers of the I and II kind, respectively \citep[e.g., see][ p. 18]{ConFam2006}. 
Let $X$ and $Y$ be two random variables then
\begin{equation}
\mathbb{E}((X+Y)_m)=\sum_{k=0}^{m}{m \choose k}\mathbb{E}((X)_k)\mathbb{E}((X)_{m-k})
\end{equation}
which can be proved by induction. Let $\alpha\circ X$ a binomial thinning with $X$ a discrete random variable, then 
\begin{equation}
\begin{split}
\mathbb{E}((\alpha\circ X)_k)& =\mathbb{E}((\sum_{j=1}^{X}B_j)_k)=\mathbb{E}\left(\sum_{|\kappa|=k}\prod_{j=1}^{X}B_{j}^{\kappa_j}\right)
\\ & 
=\mathbb{E}\left({X\choose k} k!\alpha^{k}\right)=\alpha^{k}\mathbb{E}((X)_k) \\
\end{split}
\end{equation}
where $|\kappa|=\kappa_1+\ldots+\kappa_X$. Using the results given above and stationarity (i.e. $\mathbb{E}((X_t)_m)=\mu_{X}^{(\underline{m})}$ one obtains 
\begin{eqnarray}
&&\mathbb{E}((X_t)_m)=\mathbb{E}((\alpha\circ X_{t-1}+\varepsilon_t)_m)\\
&&=\sum_{k=0}^{m}{m \choose k}\mathbb{E}\left(\left(\alpha\circ X_{t-1}\right)_k\right)\mathbb{E}((\varepsilon_t)_{m-k})\\
&&=\sum_{k=0}^{m}{m \choose k}\alpha^{k}\mathbb{E}((X_{t-1})_k)\mu_{\varepsilon}^{(\underline{m-k})}
\end{eqnarray}
which implies the $m$-order falling factorial moment of a INAR(1) is
\begin{eqnarray}
&&\mu_{X}^{(\underline{m})}=\frac{1}{1-\alpha^m}\sum_{k=0}^{m-1}{m \choose k}\alpha^{k}\mu_{X}^{(\underline{k})}\mu_{\varepsilon}^{(\underline{m-k})}\\
&&=\frac{1}{1-\alpha^m}\sum_{k=0}^{m-1}\sum_{l=0}^{m-k}{m \choose k}s(m-k,l)\alpha^{k}\mu_{X}^{(\underline{k})}\mu_{\varepsilon}^{(l)}
\end{eqnarray}
and the $m$-order moment is
\begin{equation}
\mu_{X}^{(m)}=\sum_{i=0}^{m}S(m,i)\frac{1}{1-\alpha^i}\sum_{k=0}^{i-1}\sum_{l=0}^{i-k}{i \choose k}s(i-k,l)\alpha^{k}\mu_{X}^{(\underline{k})}\mu_{\varepsilon}^{(l)}
\end{equation}

\vfill

\newpage

\pagebreak

\renewcommand\thefigure{C.\arabic{figure}}
\setcounter{figure}{0}
\renewcommand\theequation{C.\arabic{equation}}
\setcounter{equation}{0}
\renewcommand\thetable{C.\arabic{table}}
\setcounter{table}{0}

\section{Further simulation results}\label{App:EmpSimul}
\vspace{-10pt}
\begin{figure}[h!]
\begin{center}
\renewcommand{\arraystretch}{0.9}
\setlength{\tabcolsep}{0pt}
\begin{tabular}{cc}
\includegraphics[scale=0.37]{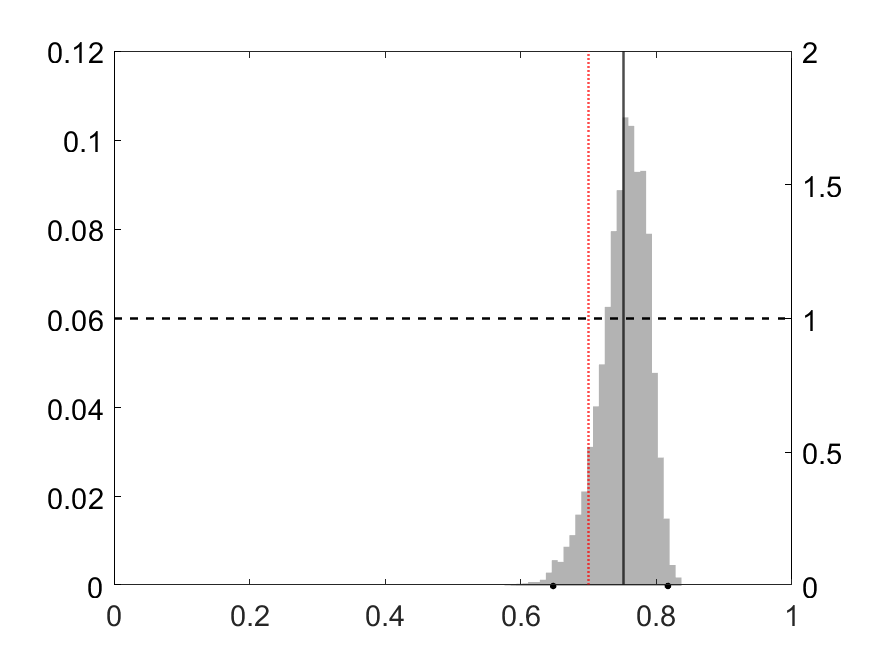}&
\includegraphics[scale=0.37]{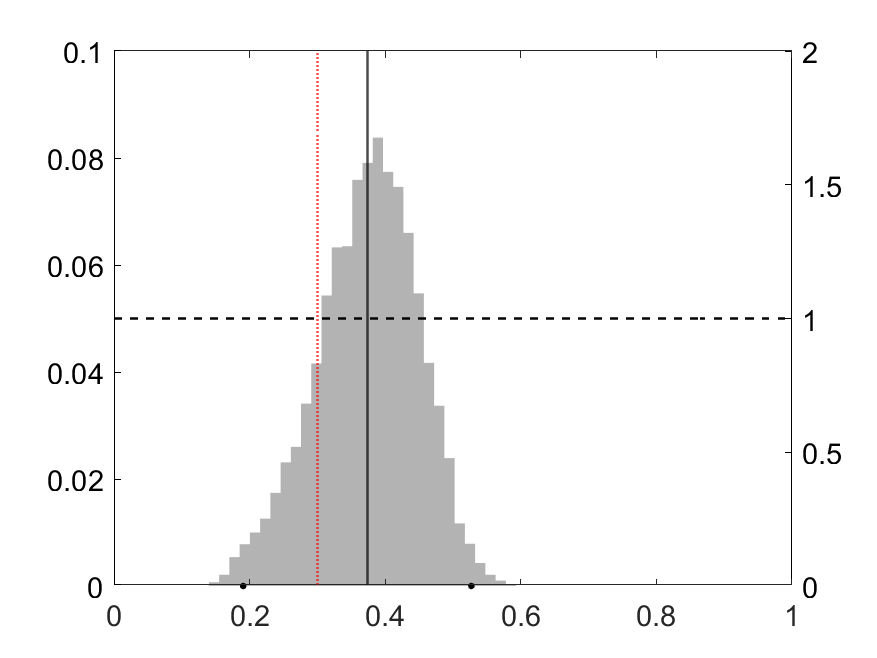}\\
\includegraphics[scale=0.37]{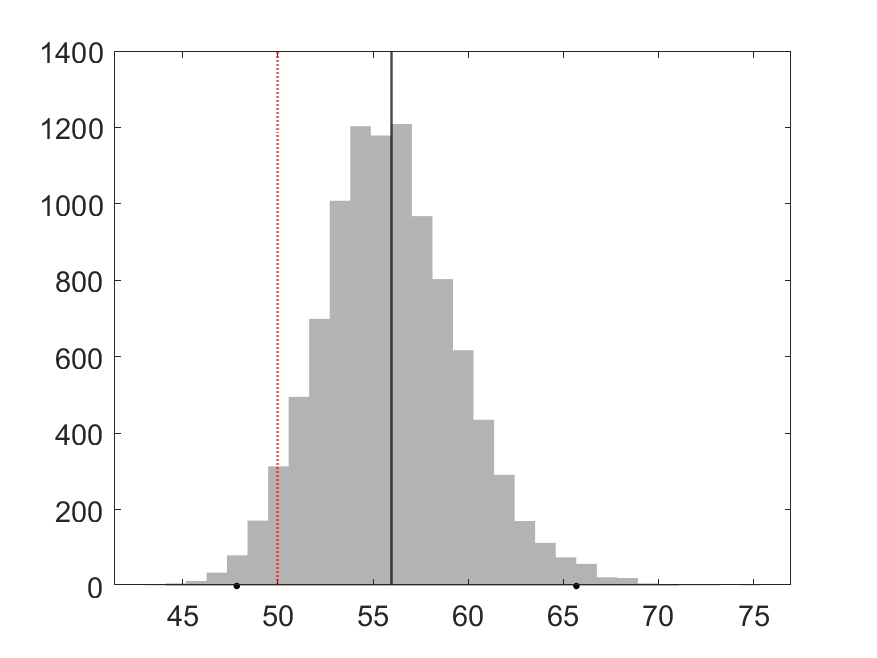}&
\includegraphics[scale=0.37]{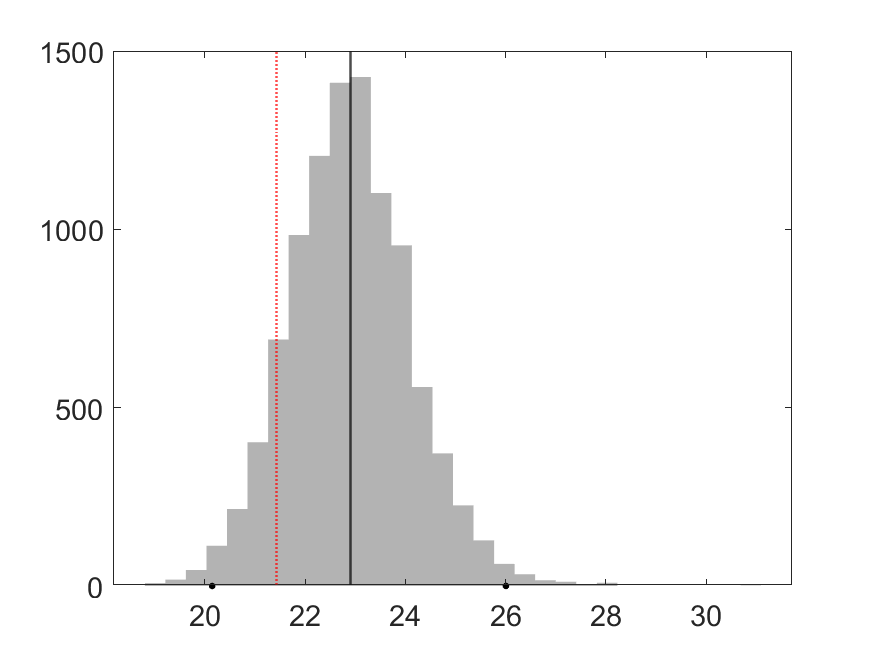}\\
\includegraphics[scale=0.37]{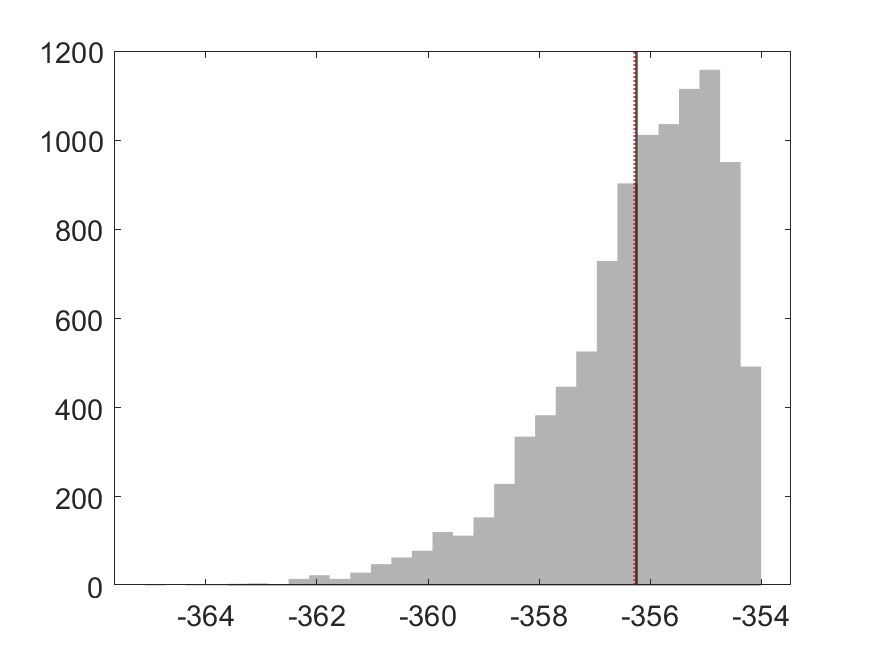}&
\includegraphics[scale=0.37]{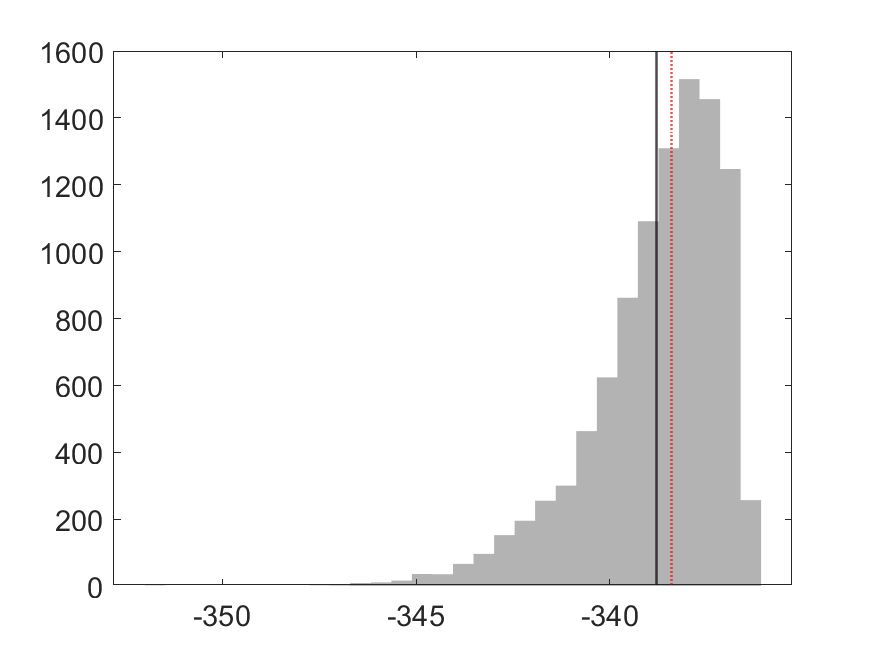}\\
\end{tabular}
\end{center}
\caption{MCMC approximation of the posterior distribution (histogram) of the parameters $\alpha$ (top), the unconditional mean $\mu_\varepsilon/(1-\alpha)$ (middle) and the marginal likelihood (bottom) of the GLK--INAR(1) in the high--persistence (left) and low--persistence (right) setting. In all plots, the true parameter value (red dashed) and the estimated one (black solid).}\label{fig:MCMCrawHist}
\end{figure}

\begin{sidewaystable}[p]
\centering
\caption{Autocorrelation function (ACF), effective sample size (ESS), inefficiency factor (INEFF), numerical standard errors (NSE) and Geweke’s convergence diagnostic (CD) of the posterior MCMC samples for the two settings: low persistence and high persistence. We ran the proposed MCMC algorithm for 50,000 iterations and evaluate the statistics before (subscript BT) and after (subscript AT) removing the first 10,000 burn-in samples, and applying a thinning procedure with a factor of 10. In parenthesis the p-values of the Geweke's convergence diagnostic.}
\label{Stat}
\begin{small}
\begin{tabular}{c|c|c|c|c|c|c|c|c|c|c}
&\multicolumn{5}{|c}{Low persistence}&\multicolumn{5}{|c}{High persistence}\\
&\multicolumn{5}{|c}{\makecell{$\alpha=0.3$, $a=5.3239$,\\ $b=0.0592$, $c=0.6$, $\beta=0.5917$}}&\multicolumn{5}{|c}{\makecell{$\alpha=0.7$, $a=5.3239$,\\ $b=0.0592$, $c=0.6$, $\beta=0.5917$}}\\
\hline
&$\alpha$&$a$&$b$&$c$&$\beta$&$\alpha$&$a$&$b$&$c$&$\beta$\\
\hline
$ACF(1)_{BT}$ & 0.93 & 0.92 & 0.93 & 0.92 & 0.94 & 0.94 & 0.92 & 0.93 & 0.92 & 0.94\\
$ACF(5)_{BT}$ & 0.71 & 0.68 & 0.72 & 0.66 & 0.74 & 0.72 & 0.68 & 0.73 & 0.65 & 0.74\\
$ACF(10)_{BT}$ & 0.52 & 0.47 & 0.54 & 0.45 & 0.56 & 0.53 & 0.48 & 0.55 & 0.44 & 0.55\\
$ACF(1)_{AT}$ & 0.54 & 0.47 & 0.55 & 0.44 & 0.58 & 0.53 & 0.48 & 0.56 & 0.45 & 0.57\\
$ACF(5)_{AT}$ & 0.08 & 0.06 & 0.13 & 0.08 & 0.16 & 0.06 & 0.05 & 0.11 & 0.02 & 0.08\\
$ACF(10)_{AT}$ & -0.01 & 0.02 & 0.02 & 0.04 & 0.06 & 0.01 & -0.003 & -0.004 & 0.06 & 0.03\\
\hline
$ESS_{BT}$ & 0.06 & 0.06 & 0.06 & 0.06 & 0.06 & 0.06 & 0.06 & 0.06 & 0.06 & 0.06\\
$ESS_{AT}$ & 0.17 & 0.18 & 0.15 & 0.18 & 0.14 & 0.19 & 0.20 & 0.15 & 0.20 & 0.16\\
\hline
$INEFF_{BT}$ & 17.05 & 16.43 & 17.20 & 16.12 & 17.61 & 17.30 & 16.51 & 17.46 & 16.00 & 17.57\\
$INEFF_{AT}$ & 5.84 & 5.51 & 6.45 & 5.54 & 6.97 & 5.35 & 5.07 & 6.58 & 4.87 & 6.09\\
\hline
$NSE_{BT}$ & 0.002 & 0.05 & 0.002 & 0.006 & 0.002 & 0.001 & 0.06 & 0.004 & 0.01 & 0.004\\
$NSE_{AT}$ & 0.002 & 0.09 & 0.003 & 0.01 & 0.004 & 0.002 & 0.09 & 0.003 & 0.01 & 0.004\\
\hline
$CD_{BT}$ & 1.12 & -2.08 & -0.42 & 0.31 & 1.72 & -0.48 & -0.92 & -0.13 & -1.48 & -0.83\\
 & (0.26)& (0.04) & (0.68) & (0.76) & (0.09) & (0.63) & (0.36) & (0.89) & (0.14) & (0.41)\\
$CD_{AT}$ & -1.047 & 0.57 & -1.32 & 0.68 & 1.12 & -1.05 & 0.57 & -1.32 & 0.68 & 1.12\\
 & (0.30) & (0.57) & (0.19) & (0.50) & (0.26) & (0.30) & (0.57) & (0.19) & (0.50) & (0.26)\\
\bottomrule
\end{tabular}
\end{small}

\end{sidewaystable}

\begin{figure}[h!]
\begin{center}
\renewcommand{\arraystretch}{0.8}
\setlength{\tabcolsep}{0pt}
\begin{tabular}{cc}
\multicolumn{2}{c}{High-persistence setting}\\
$\alpha$& $a$\vspace{5pt}\\
\includegraphics[scale=0.37]{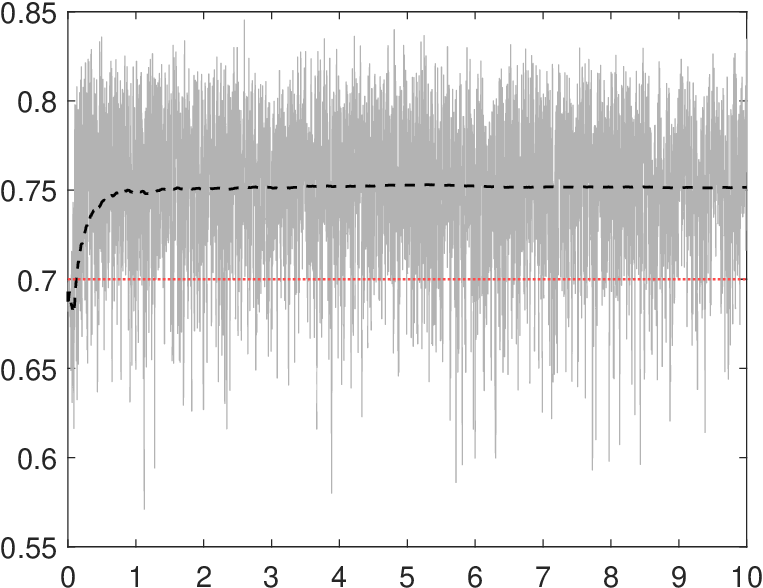}&
\includegraphics[scale=0.37]{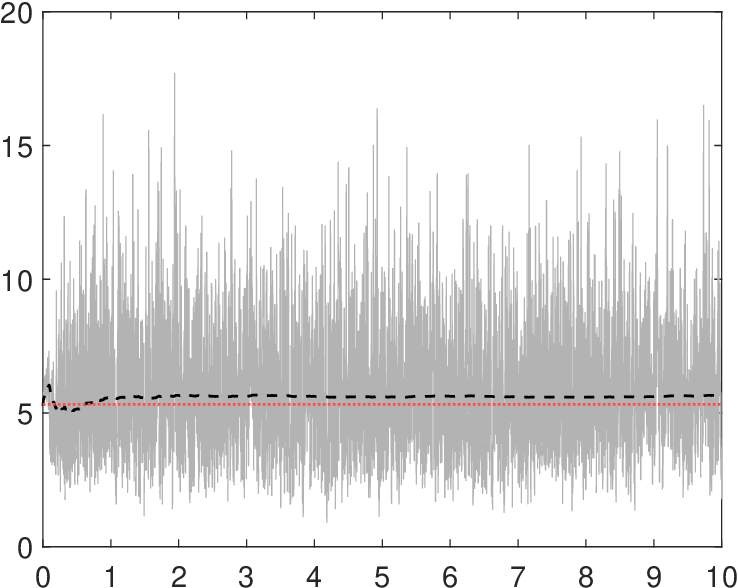}\\
$b$& $c$\vspace{5pt}\\
\includegraphics[scale=0.37]{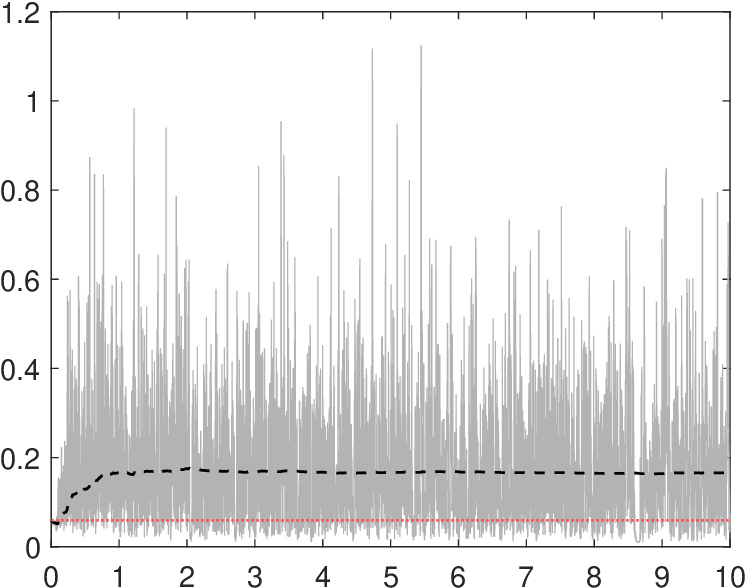}&
\includegraphics[scale=0.37]{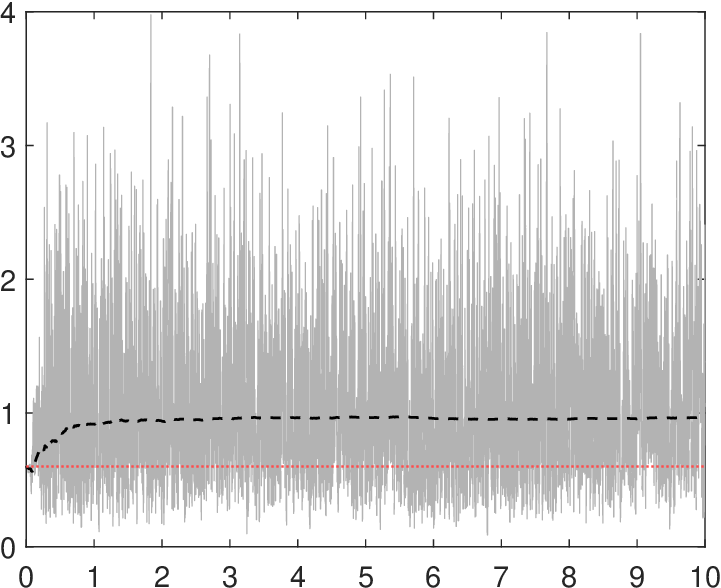}\\
$\beta$& $a\beta/(c-c\beta-\beta b)$\vspace{5pt}\\
\includegraphics[scale=0.37]{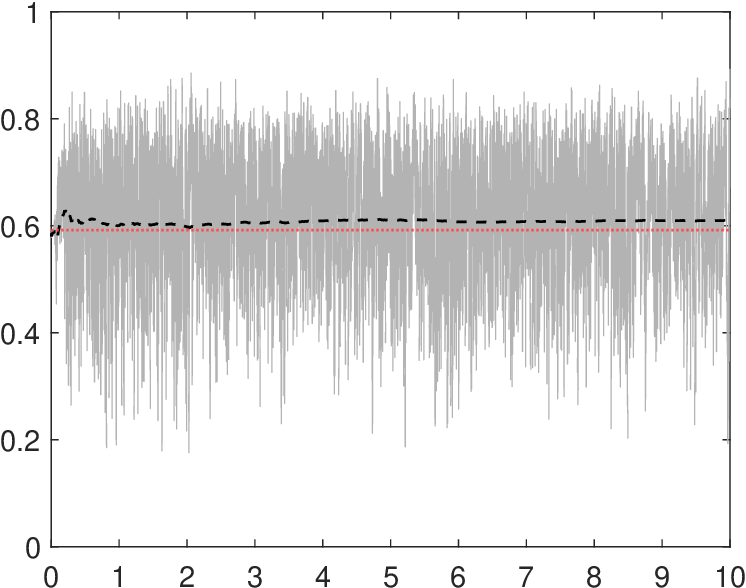}&
\includegraphics[scale=0.37]{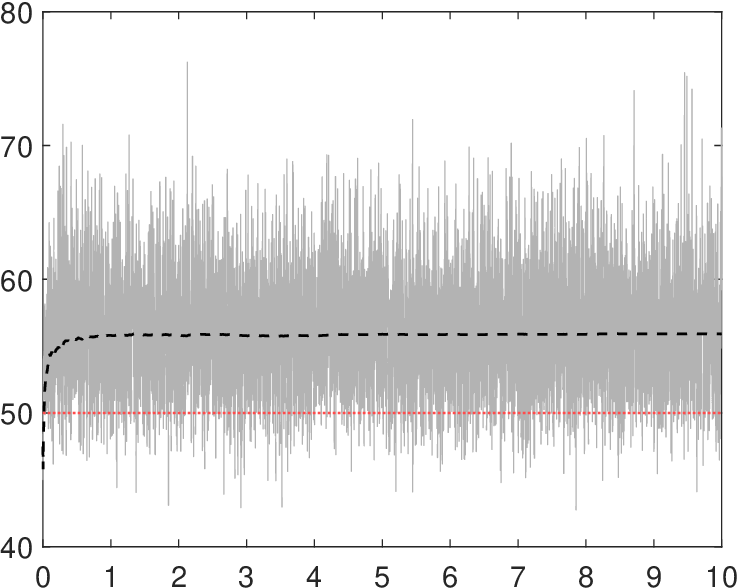}\\
\end{tabular}
\end{center}
\caption{MCMC output for the parameters of the GLK--INAR(1). In all plots, the MCMC draws (gray solid), the progressive MCMC average (dashed black) over the iterations (horizontal axis in thousands), and the true value of the parameter (horizontal red dashed).}\label{fig:MCMCrawHigh}
\end{figure}

\begin{figure}[p]
\begin{center}
\renewcommand{\arraystretch}{0.8}
\setlength{\tabcolsep}{0pt}
\begin{tabular}{cc}
\multicolumn{2}{c}{High--persistence setting}\\
$\alpha$& $a$\vspace{5pt}\\
\includegraphics[scale=0.37]{Figures/INARMCMCthetaHist1Alpha7.eps}&
\includegraphics[scale=0.37]{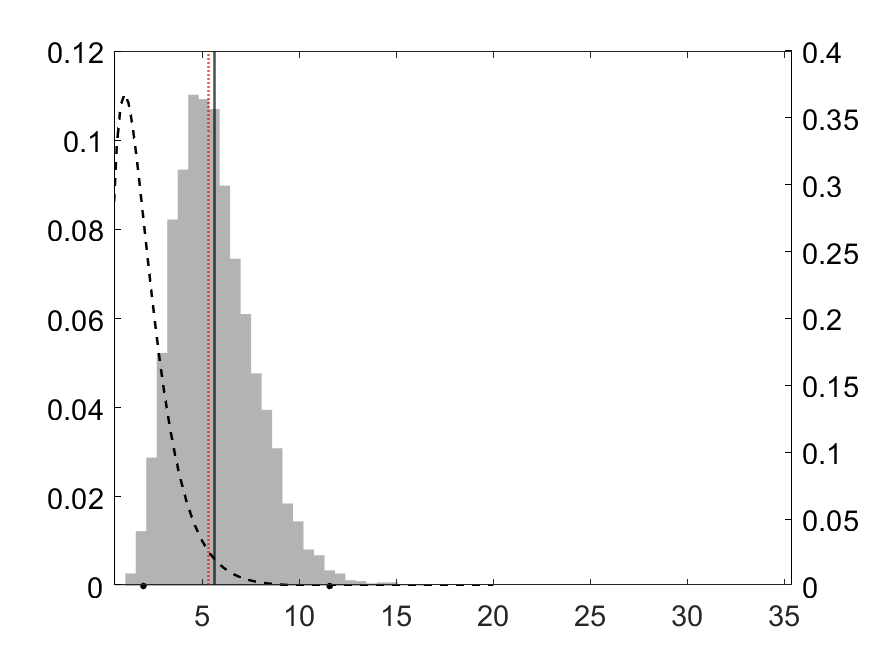}\\
$b$& $c$\vspace{5pt}\\
\includegraphics[scale=0.37]{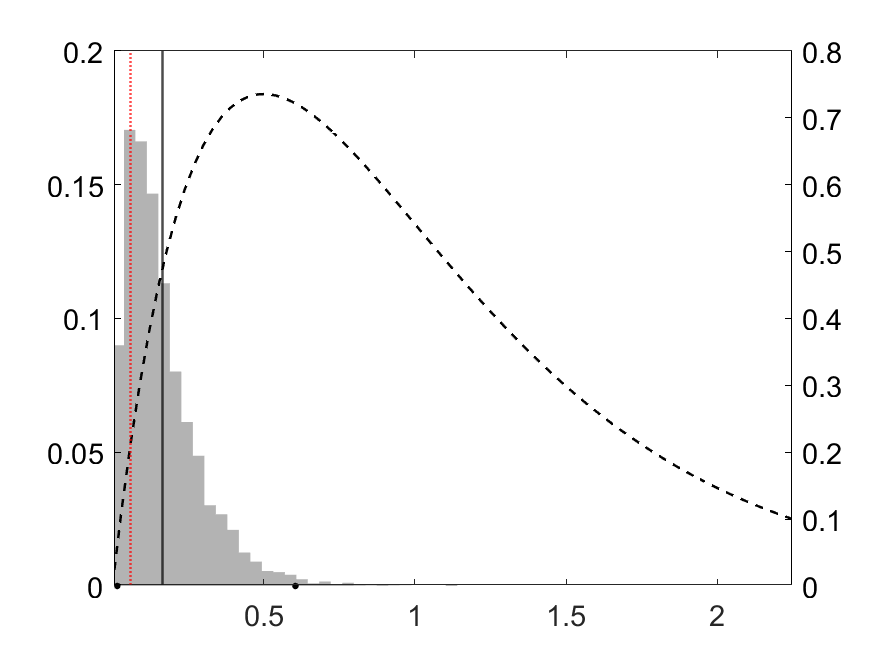}&
\includegraphics[scale=0.37]{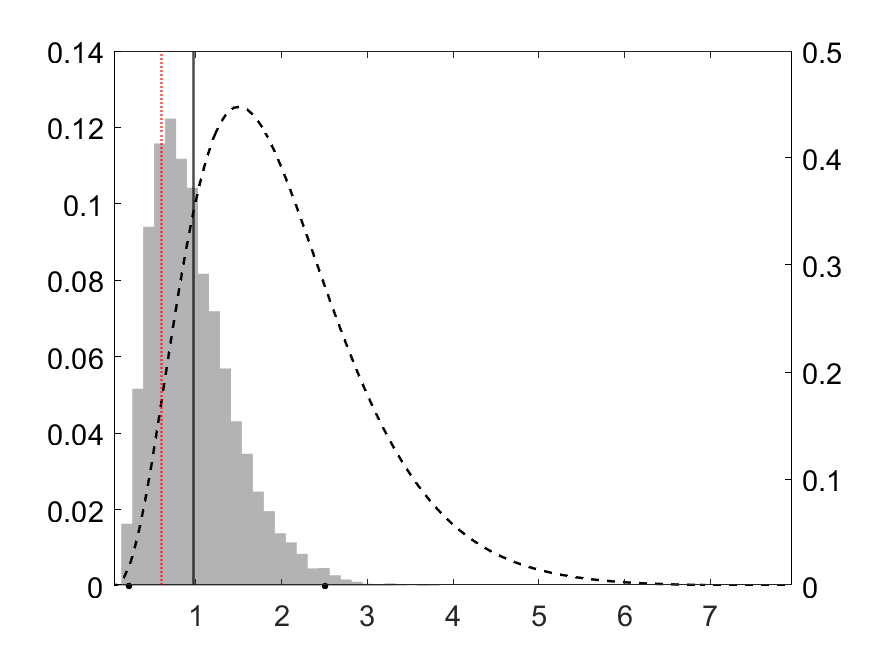}\\
$\beta$& $a\beta/(c-c\beta-\beta b)$\vspace{5pt}\\
\includegraphics[scale=0.37]{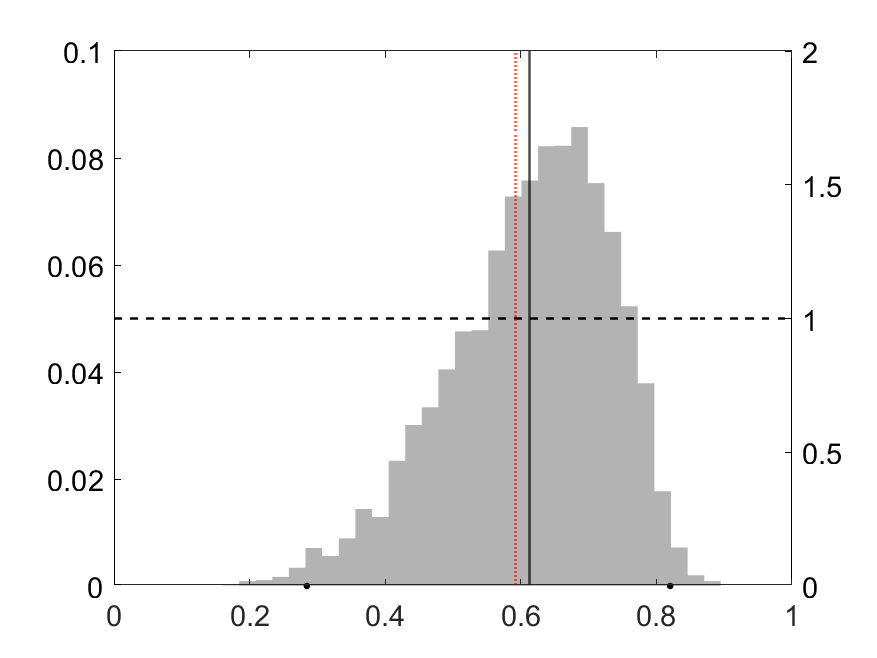}&
\includegraphics[scale=0.37]{Figures/INARMCMCthetaHistMomentAlpha7.eps}\\
\end{tabular}
\end{center}
\caption{MCMC approximation of the posterior distribution (histogram) of the parameters. In all plots, the estimated value (vertical black solid), the true value (vertical red dotted) and the prior density (dashed).}\label{fig:mcmcHistHigh}
\end{figure}

\begin{figure}[p]
\begin{center}
\renewcommand{\arraystretch}{0.8}
\setlength{\tabcolsep}{5pt}
\begin{tabular}{cc}
\multicolumn{2}{c}{Low--persistence setting}\\
$\alpha$& $a$\vspace{5pt}\\
\includegraphics[scale=0.37]{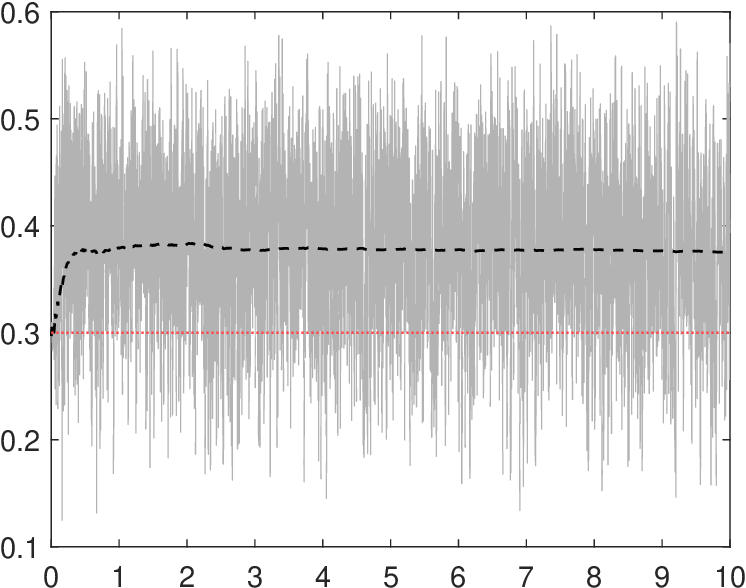}&
\includegraphics[scale=0.37]{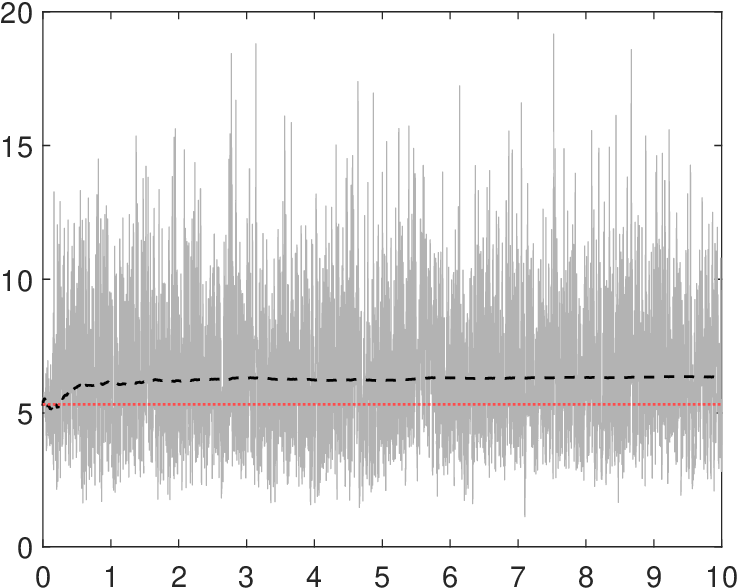}\\
$b$& $c$\vspace{5pt}\\
\includegraphics[scale=0.37]{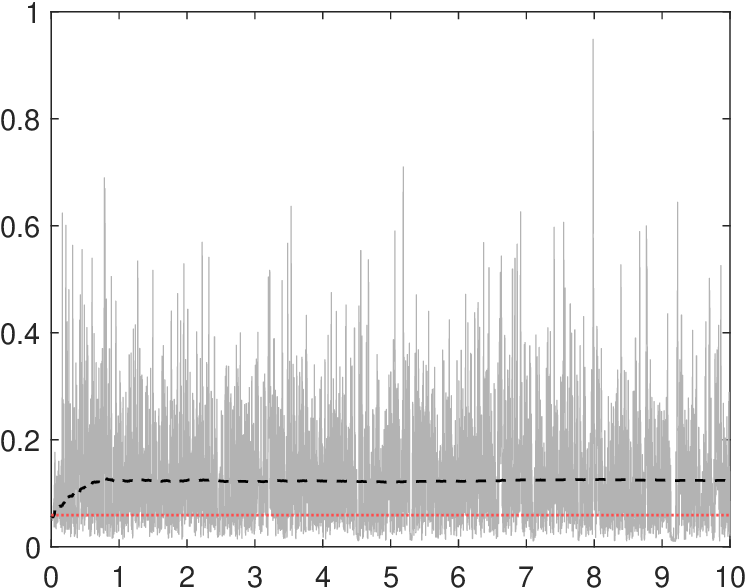}&
\includegraphics[scale=0.37]{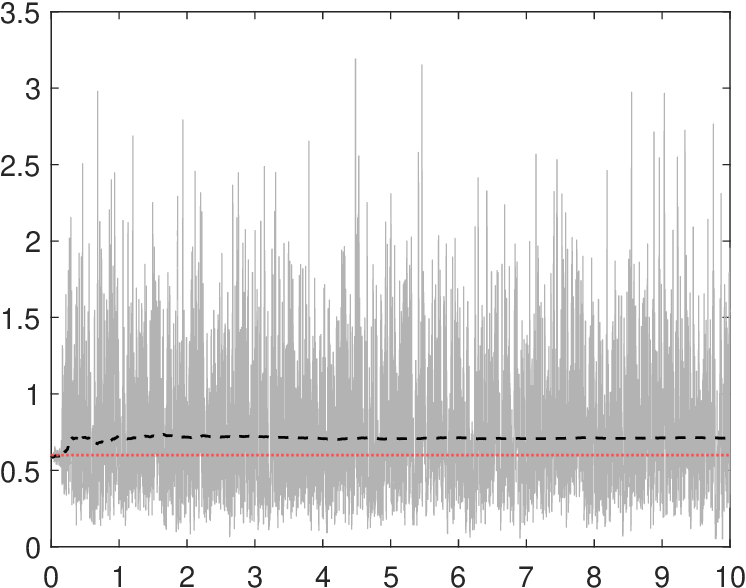}\\
$\beta$& $a\beta/(c-c\beta-\beta b)$\vspace{5pt}\\
\includegraphics[scale=0.37]{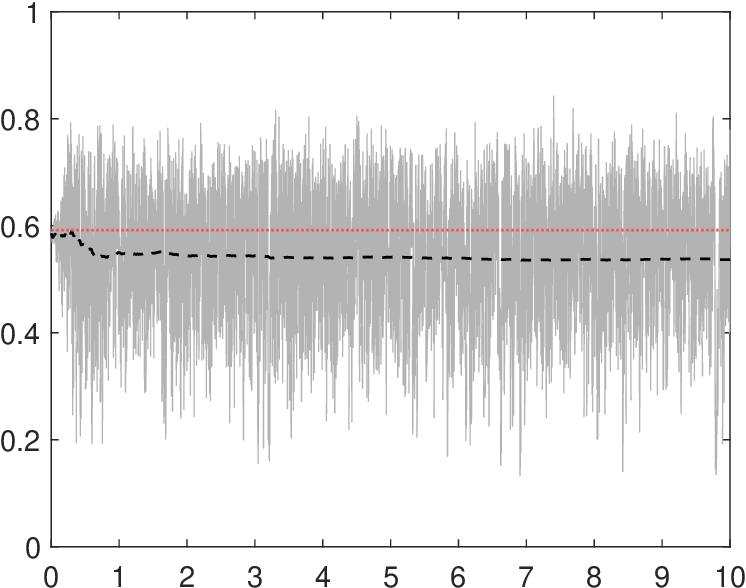}&
\includegraphics[scale=0.37]{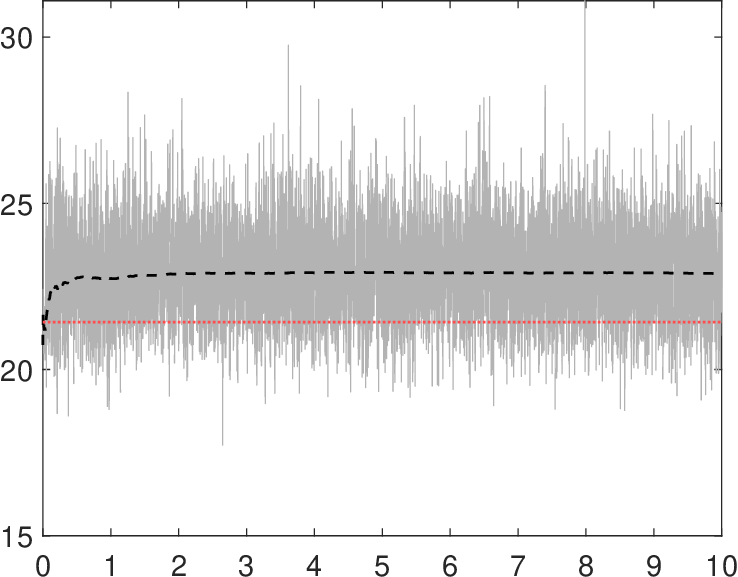}\\
\end{tabular}
\end{center}
\caption{MCMC output for the parameters of the GLK--INAR(1). In all plots, the MCMC draws (gray solid), the progressive MCMC average (dashed black) over the iterations (horizontal axis in thousands), and the true value of the parameter (horizontal red dashed).}\label{fig:MCMCrawLow}
\end{figure}

\begin{figure}[p]
\begin{center}
\renewcommand{\arraystretch}{0.8}
\setlength{\tabcolsep}{0pt}
\begin{tabular}{cc}
\multicolumn{2}{c}{Low-persistence setting}\\
$\alpha$& $a$\vspace{5pt}\\
\includegraphics[scale=0.37]{Figures/INARMCMCthetaHist1Alpha3.eps}&
\includegraphics[scale=0.37]{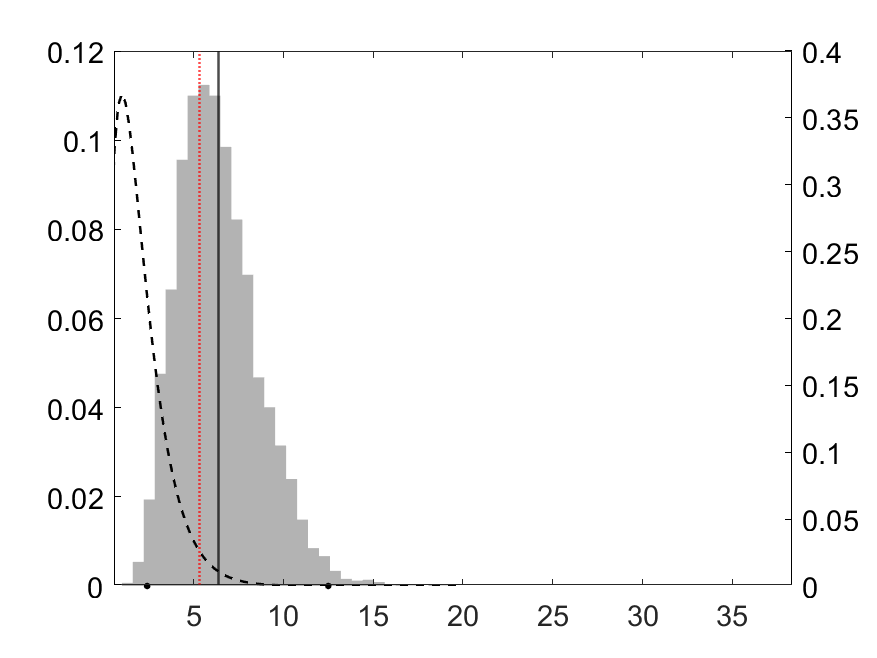}\\
$b$& $c$\vspace{5pt}\\
\includegraphics[scale=0.37]{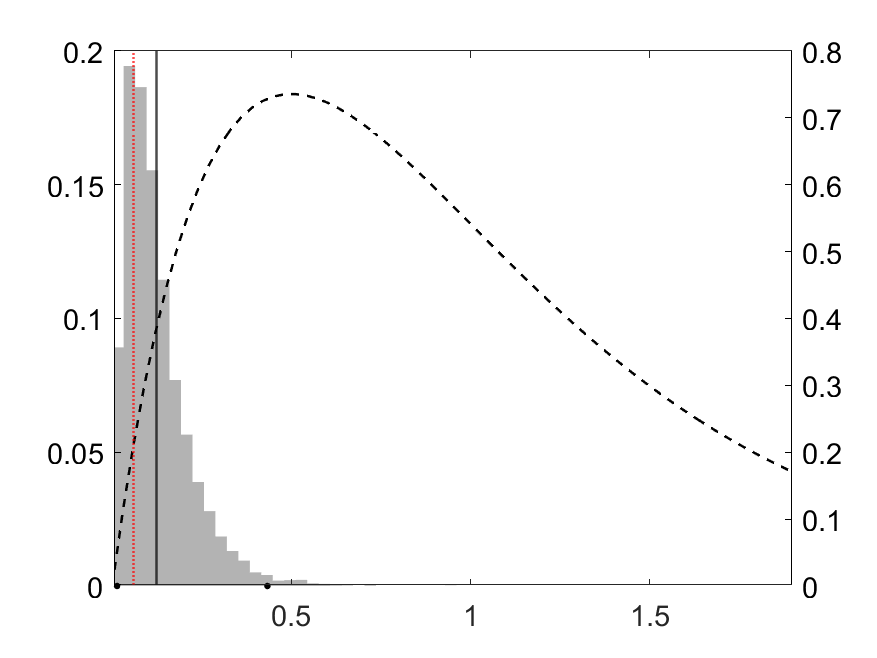}&
\includegraphics[scale=0.37]{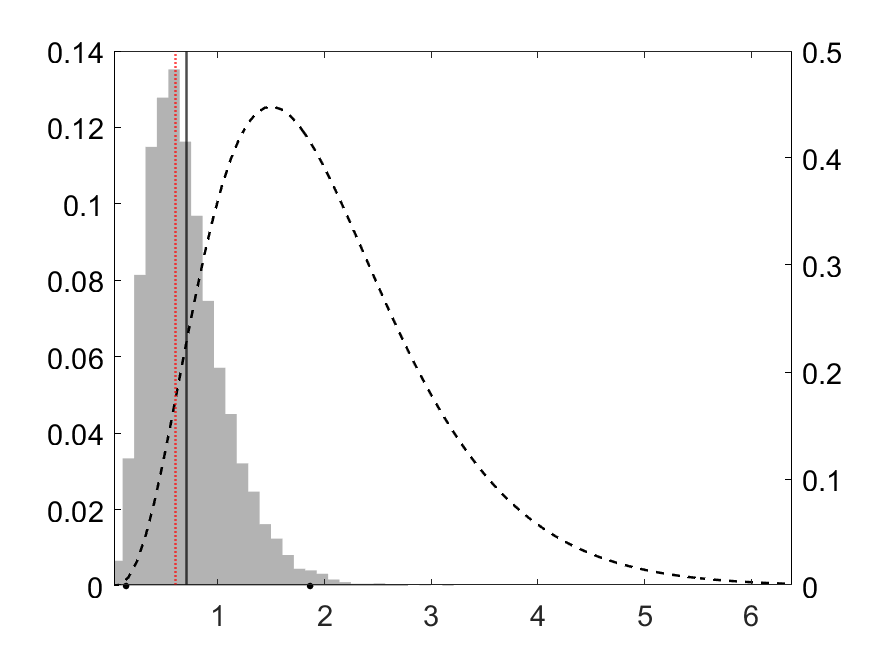}\\
$\beta$& $a\beta/(c-c\beta-\beta b)$\vspace{5pt}\\
\includegraphics[scale=0.37]{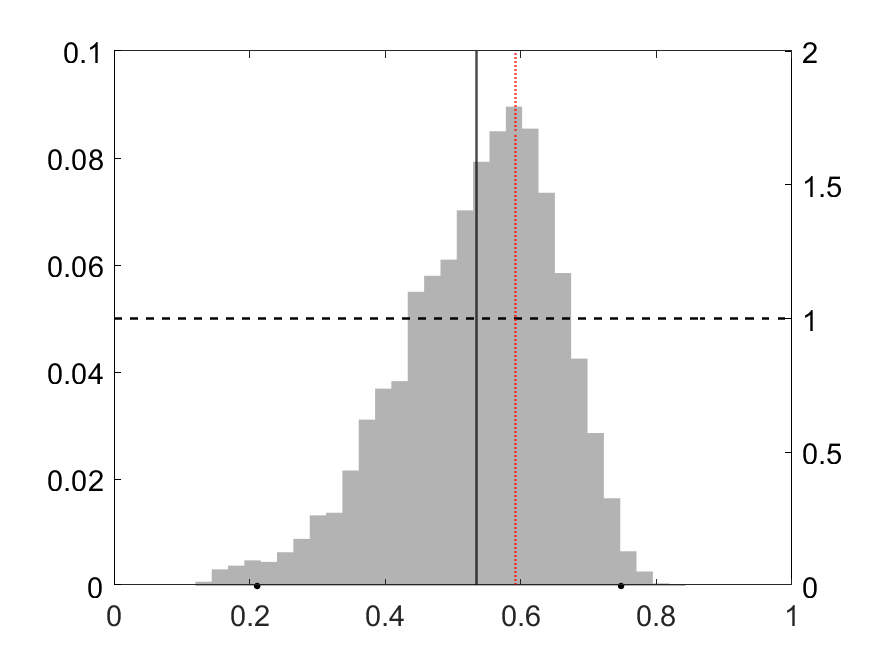}&
\includegraphics[scale=0.37]{Figures/INARMCMCthetaHistMomentAlpha3.eps}\\
\end{tabular}
\end{center}
\caption{MCMC approximation of the posterior distribution (histogram) of the parameters. In all plots, the estimated value (vertical black solid), the true value (vertical red dotted) and the prior density (dashed).}\label{fig:mcmcHistLow}
\end{figure}

\begin{figure}[p]
\begin{center}
\renewcommand{\arraystretch}{3}
\setlength{\tabcolsep}{10pt}
\begin{tabular}{cc}
\multicolumn{2}{c}{High--persistence setting}\\
\includegraphics[scale=0.37]{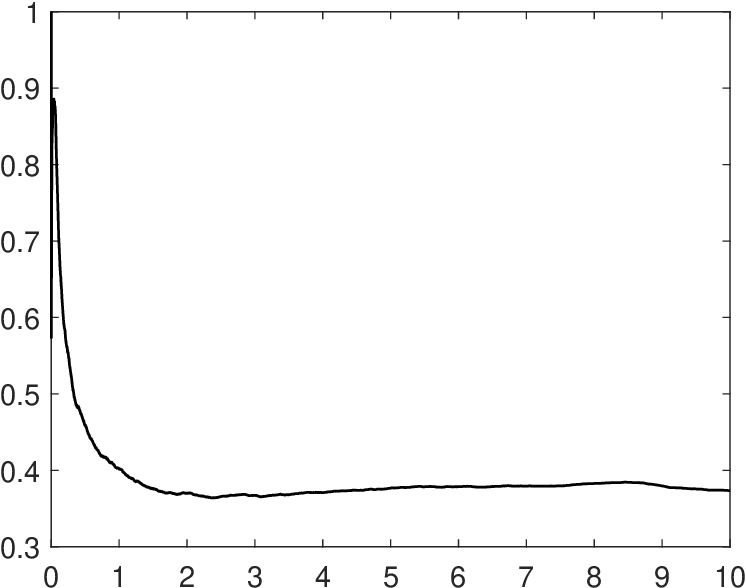}&
\includegraphics[scale=0.37]{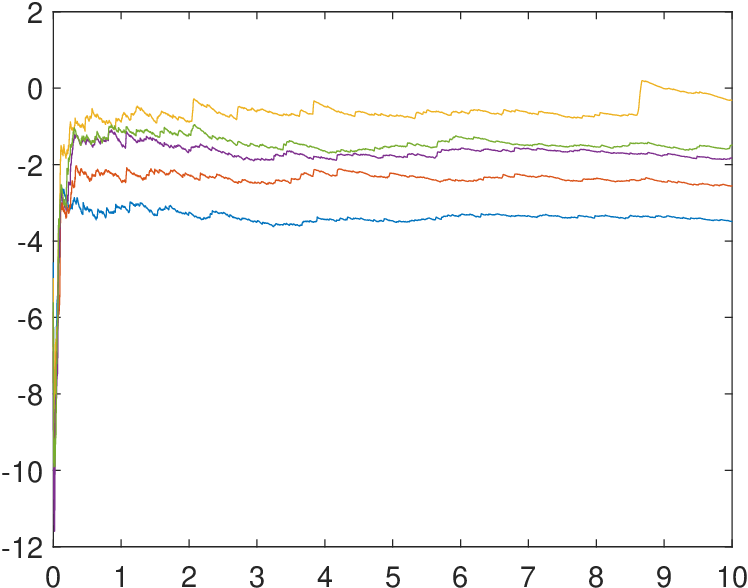}\\
\multicolumn{2}{c}{Low--persistence setting}\\
\includegraphics[scale=0.37]{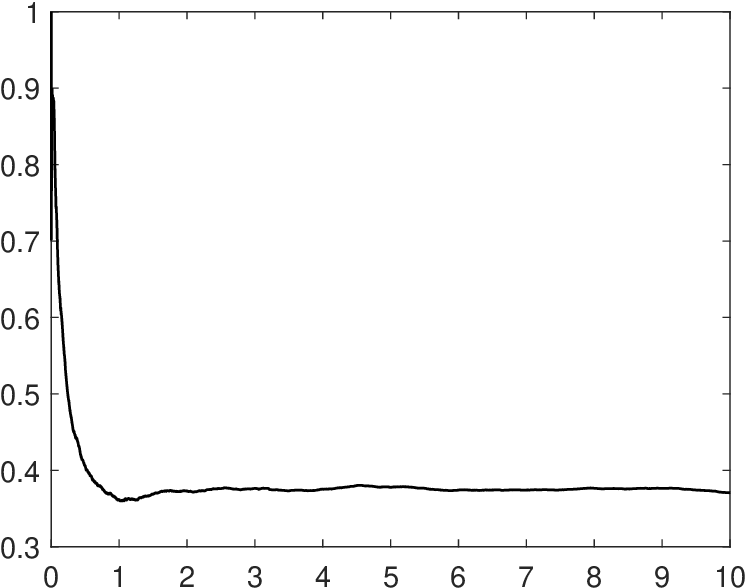}&
\includegraphics[scale=0.37]{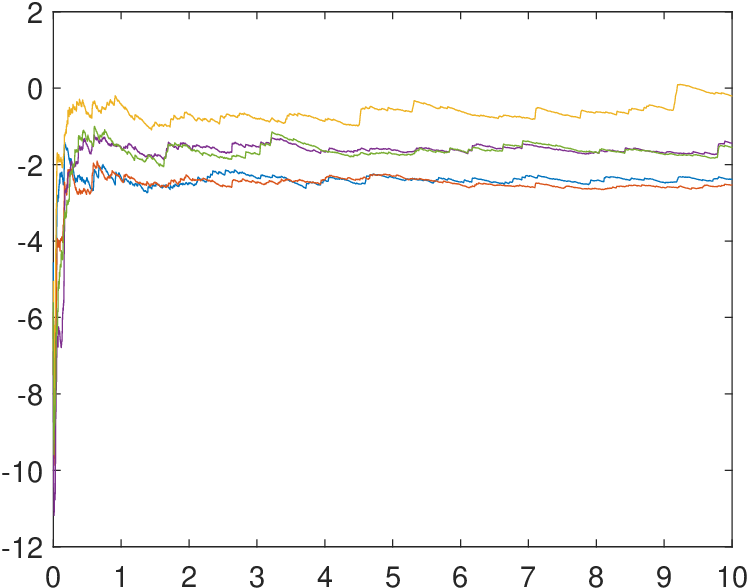}\end{tabular}
\end{center}
\caption{MCMC acceptance rate (left) and adaptive log-scales (right) over the iterations (horizontal axis in thousands).}\label{fig:mcmcDiagLow}
\end{figure}

{\color{black} 
	\begin{figure}[h!]
		\centering
		\includegraphics[scale=0.9,trim={3cm 12.5cm 3cm 10cm},clip]{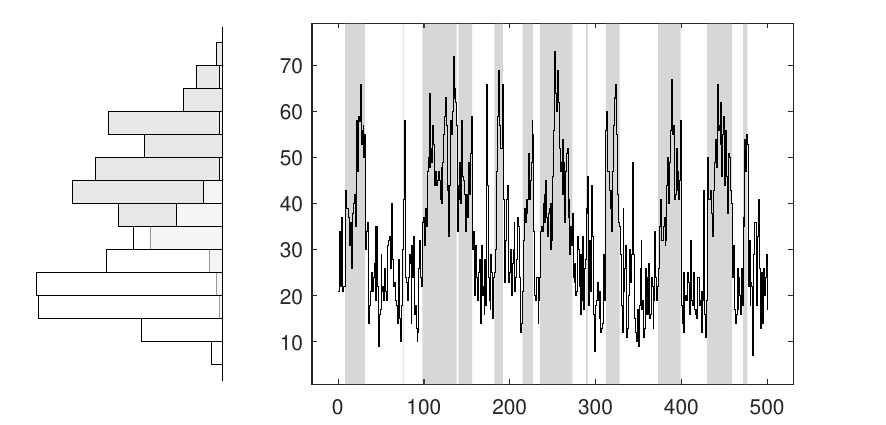}
		\caption{\color{black}Trajectory of the MS--GLK--INAR(1) (right subplot) with two regimes high (gray) and low (white) persistence and unconditional mean with their corresponding histogram (left subplot) and the estimated allocation variable in shaded rectangles.\label{fig:twoR}}
	\end{figure}

	\begin{figure}[h!]
		\centering
		\includegraphics[scale=0.9,trim={3cm 12.5cm 3cm 10cm},clip]{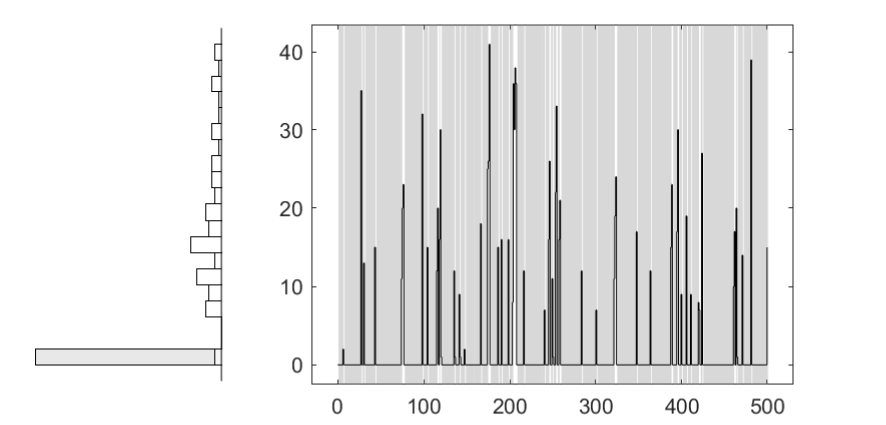}
		\caption{\color{black}Trajectory of the MS--GLK--INAR(1) (right subplot) with three regimes: inflated--zero (dark gray), high (gray) and low (white) persistence and unconditional mean with their corresponding histogram (left subplot) and the estimated allocation variable in shaded rectangles.\label{fig:zeroinf}}
	\end{figure}

	\begin{figure}[h!]
		\centering
		\begin{tblr}{Q[h,0em]cc}
			& a) Low persistence regime & b) High persistence regime\\
			{\vspace{0.8cm}\\ $\alpha$ } &\includegraphics[scale=0.3,trim={3cm 9cm 4cm 9.5cm},clip]{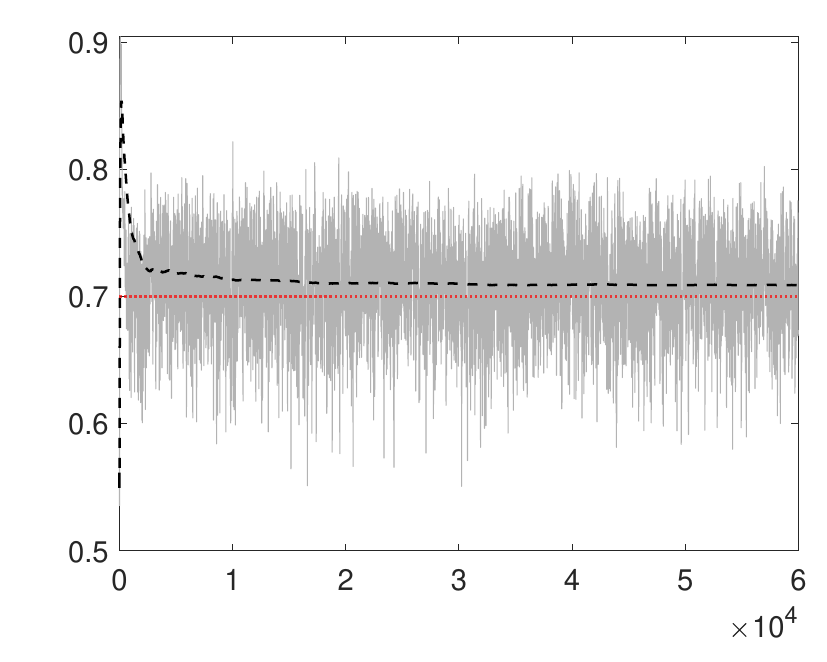} &  \includegraphics[scale=0.3,trim={3cm 9cm 4cm 9.5cm},clip]{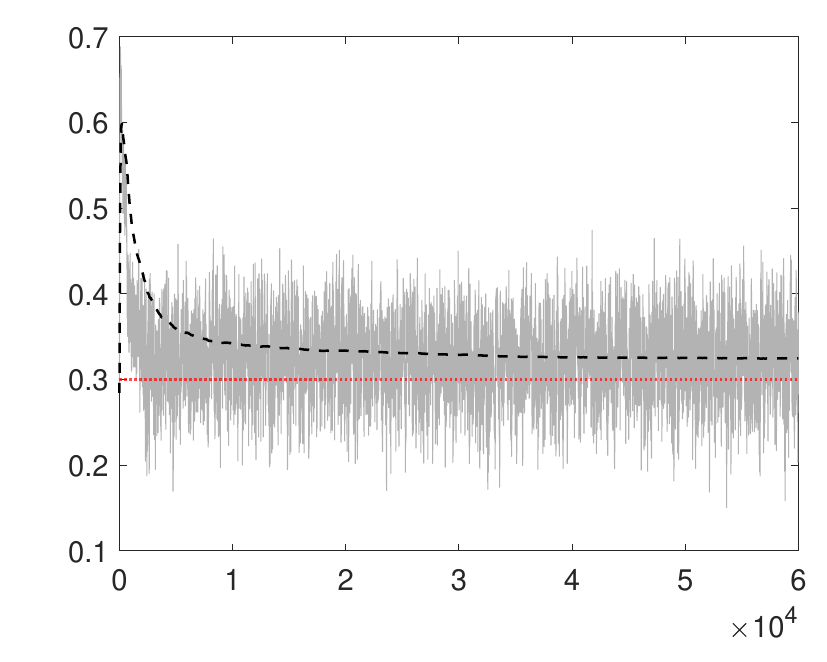}\\ 
			{\vspace{0.8cm}\\$a$} &\includegraphics[scale=0.3,trim={3cm 9cm 4cm 9.5cm},clip]{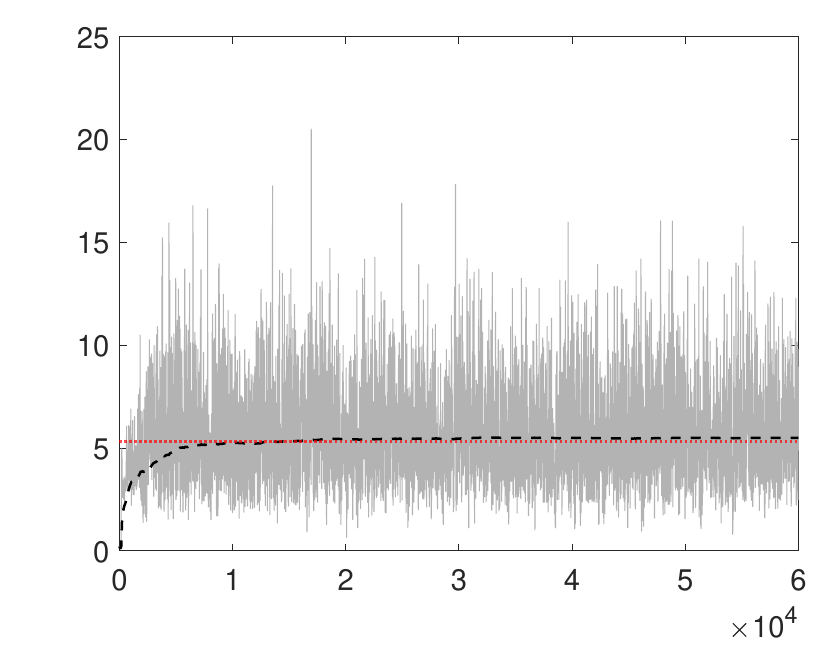} & \includegraphics[scale=0.3,trim={3cm 9cm 4cm 9.5cm},clip]{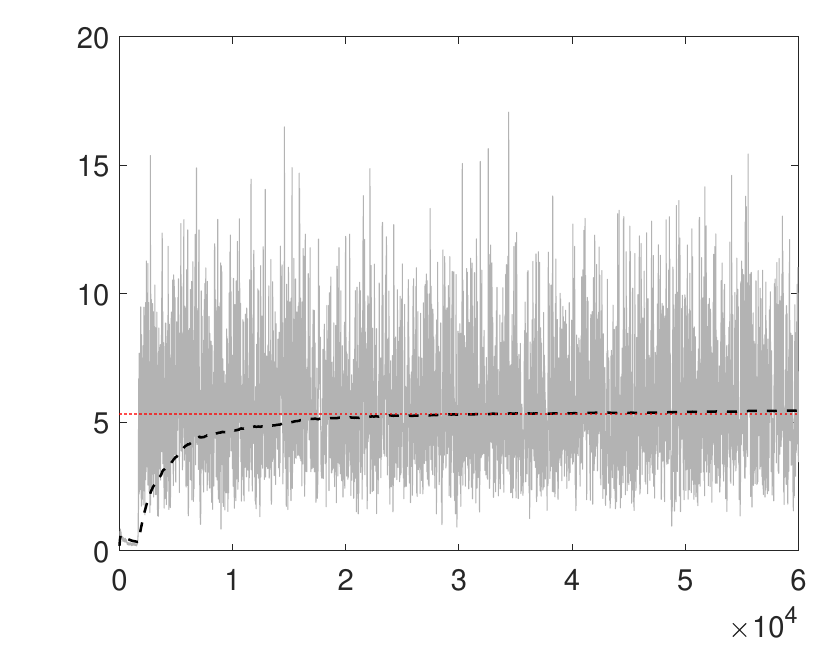}\\ 
			{\vspace{0.8cm}\\$b$}& \includegraphics[scale=0.3,trim={3cm 9cm 4cm 9.5cm},clip]{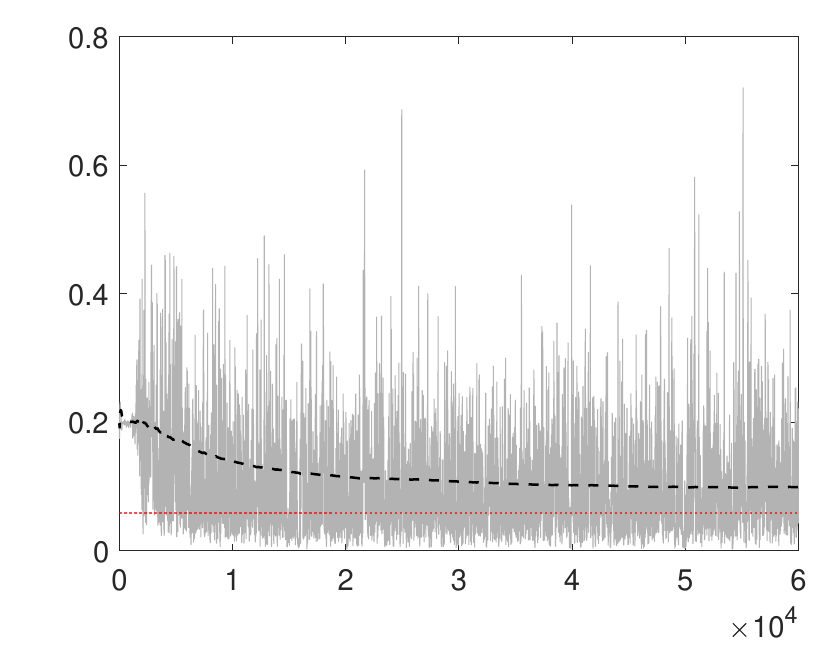} & \includegraphics[scale=0.3,trim={3cm 9cm 4cm 9.5cm},clip]{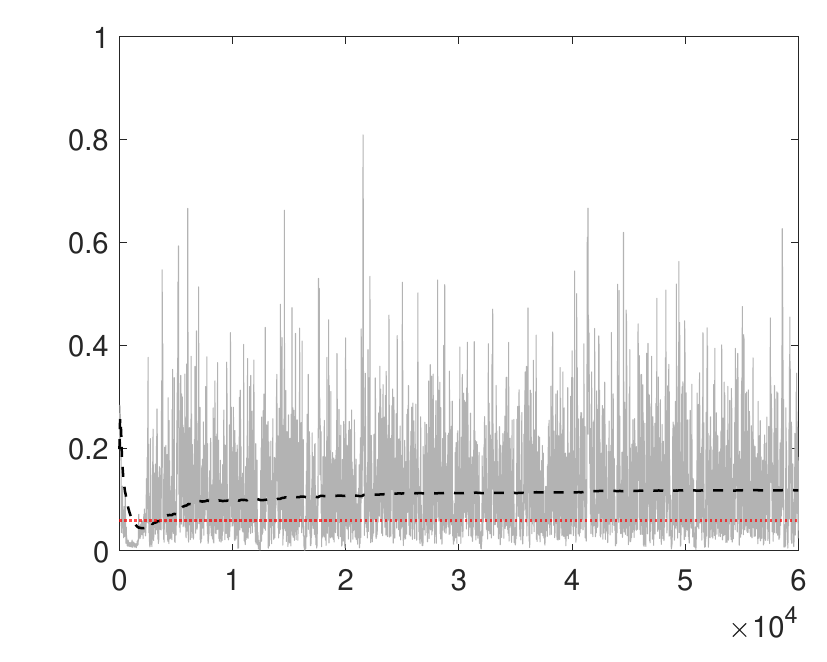} \\ 
			{\vspace{0.8cm}\\$c$} & \includegraphics[scale=0.3,trim={3cm 9cm 4cm 9.5cm},clip]{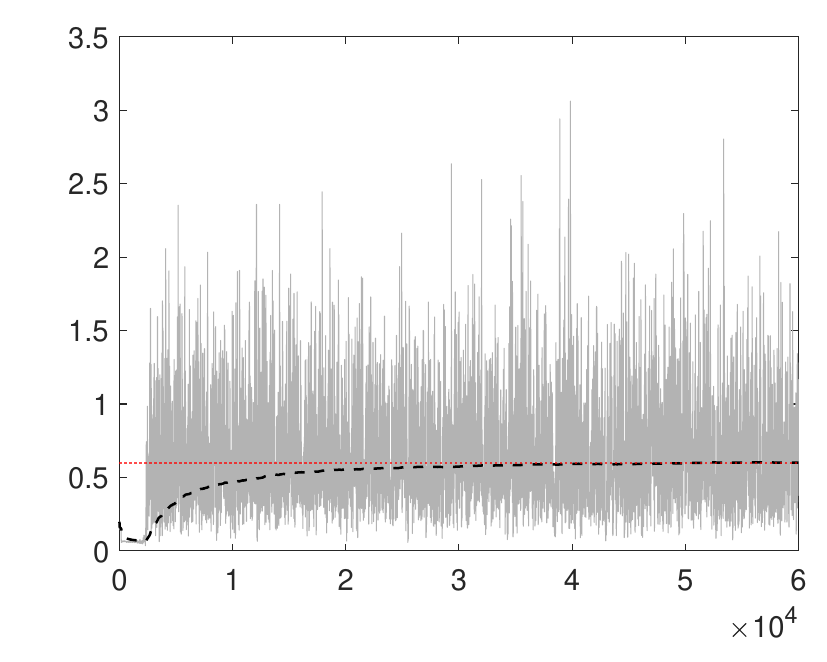} & \includegraphics[scale=0.3,trim={3cm 9cm 4cm 9.5cm},clip]{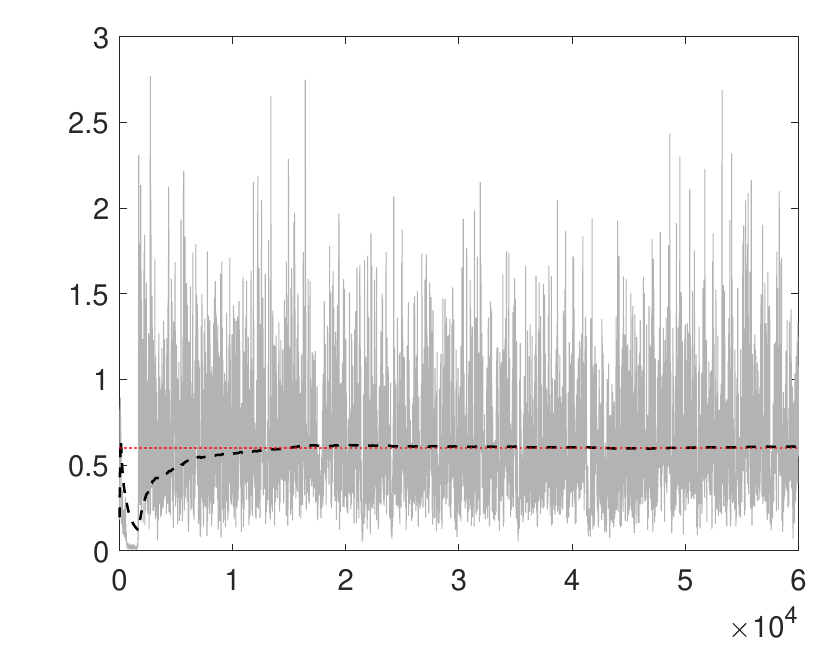}\\ 
			{\vspace{0.8cm}\\$\beta$} & \includegraphics[scale=0.3,trim={3cm 9cm 4cm 9.5cm},clip]{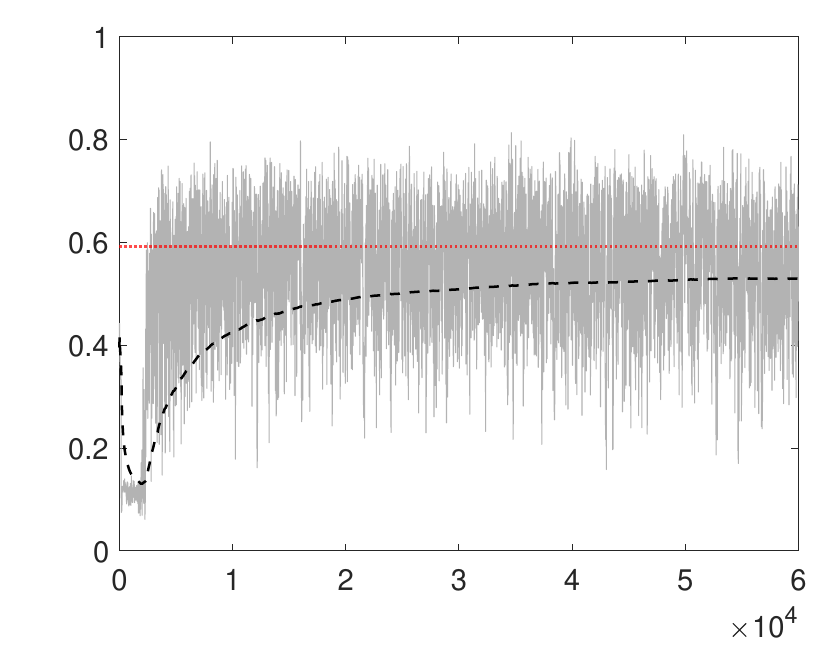} & \includegraphics[scale=0.3,trim={3cm 9cm 4cm 9.5cm},clip]{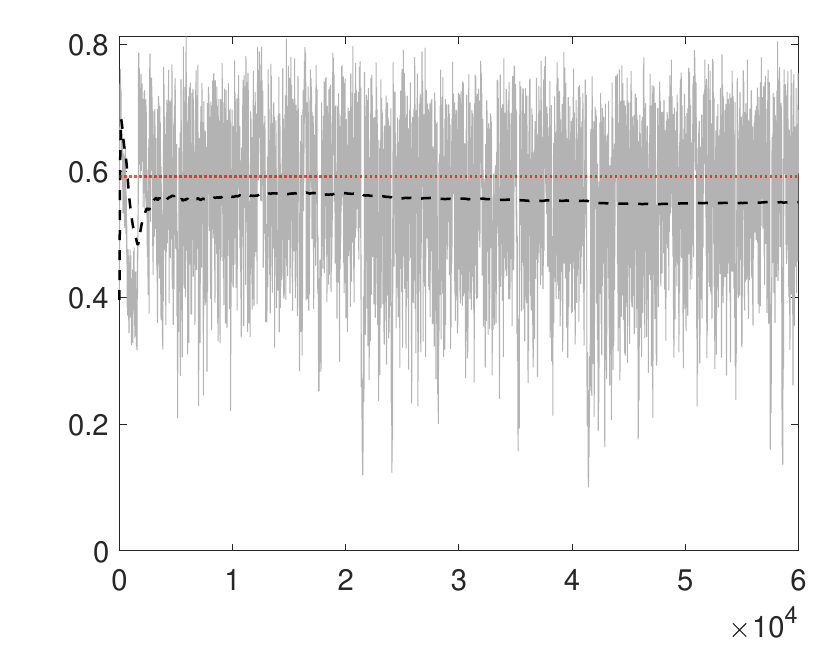}
		\end{tblr}
		
		\caption{\color{black} MCMC output for the parameters of the MS--GLK--INAR(1) with two regimes: High and low persistence and unconditional mean. In all plots, the MCMC draws (gray solid), the progressive MCMC average (dashed black) over the iterations (horizontal axis in thousands), and the true value of the parameter (horizontal red dashed).\label{fig:twoMCMCt}}
	\end{figure}
		
	\begin{figure}[h!]
		\centering
		\begin{tblr}{Q[h,0em]cc}
			& a) Low persistence regime & b) High persistence regime\\
			{\vspace{0.8cm}\\ $\alpha$ } &\includegraphics[scale=0.3,trim={3cm 9cm 4cm 9.5cm},clip]{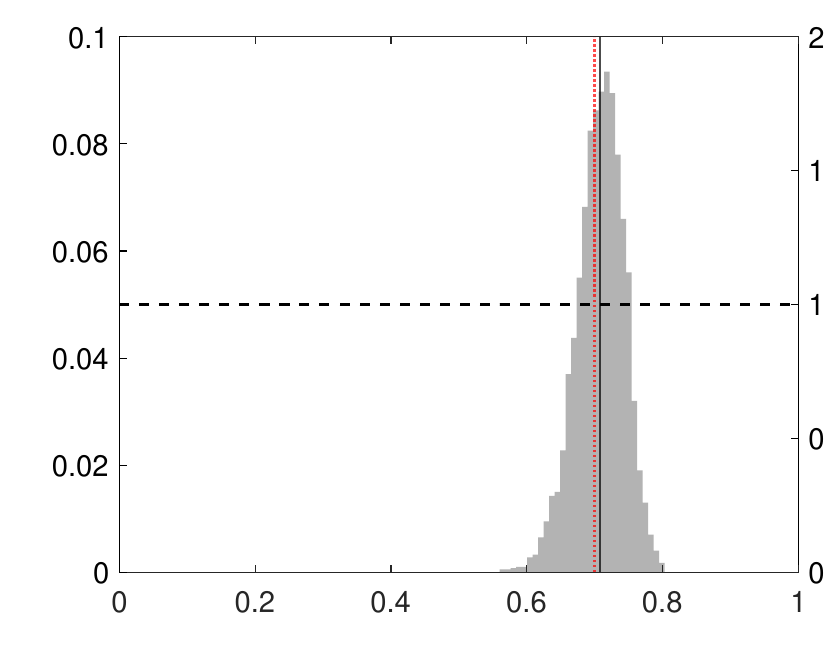} &  \includegraphics[scale=0.3,trim={3cm 9cm 4cm 9.5cm},clip]{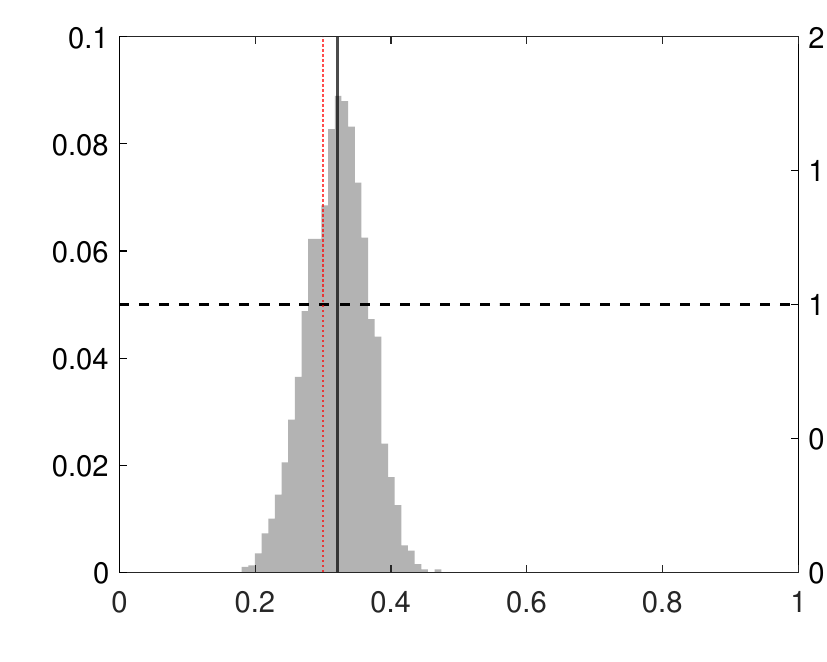}\\ 
			{\vspace{0.8cm}\\$a$} &\includegraphics[scale=0.3,trim={3cm 9cm 4cm 9.5cm},clip]{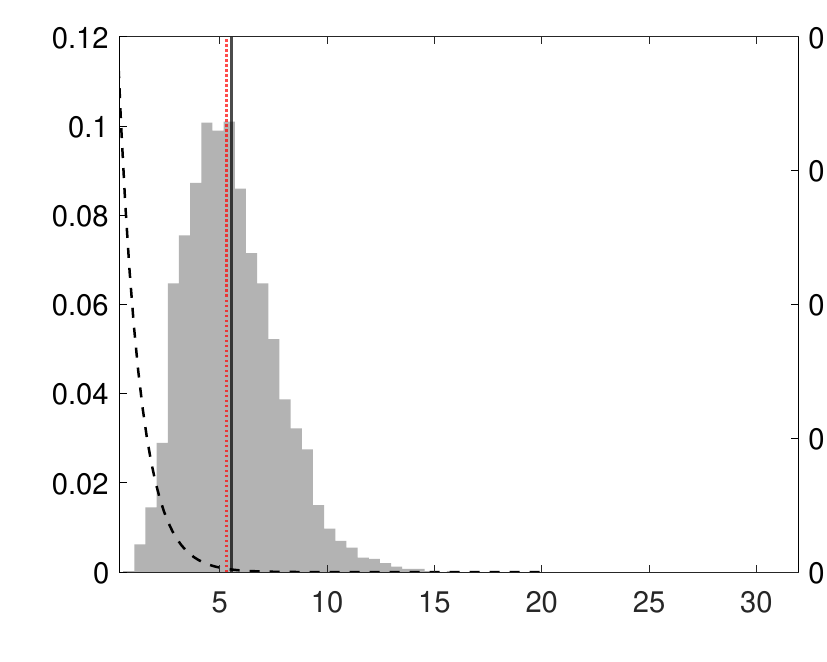} & \includegraphics[scale=0.3,trim={3cm 9cm 4cm 9.5cm},clip]{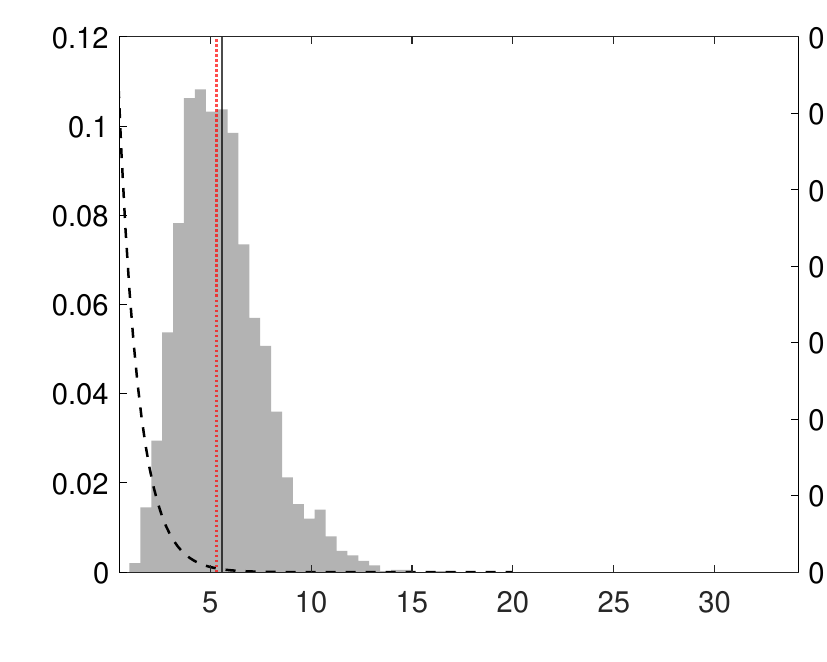}\\ 
			{\vspace{0.8cm}\\$b$}& \includegraphics[scale=0.3,trim={3cm 9cm 4cm 9.5cm},clip]{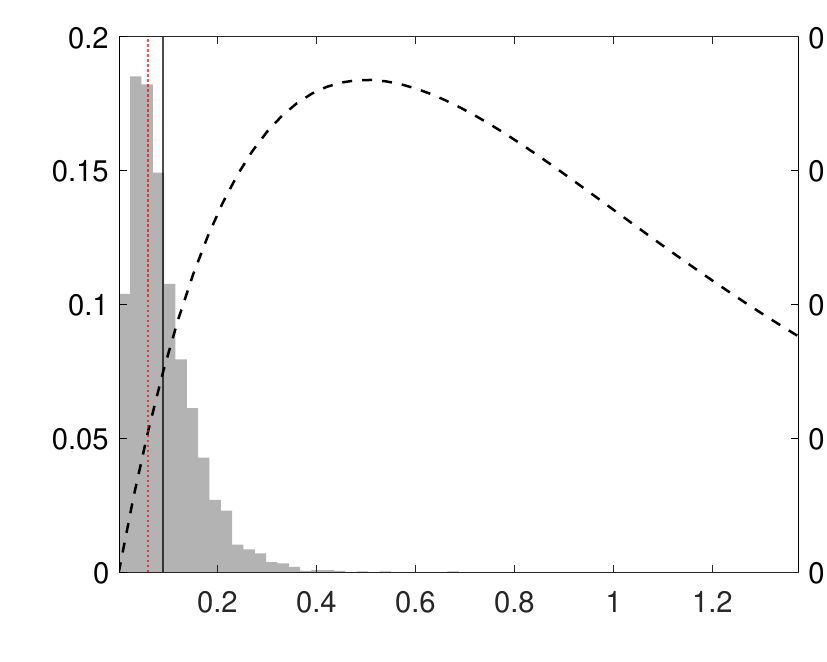} & \includegraphics[scale=0.3,trim={3cm 9cm 4cm 9.5cm},clip]{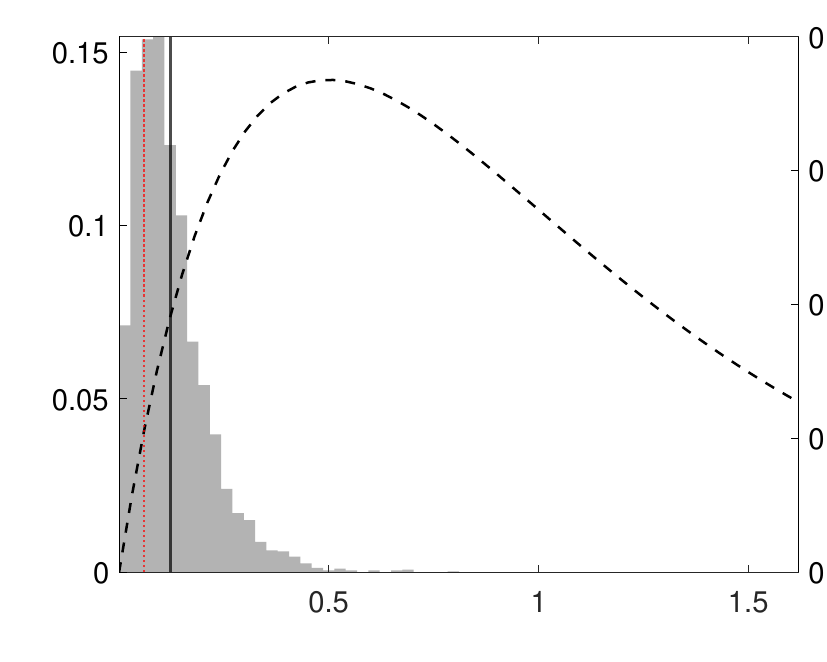} \\ 
			{\vspace{0.8cm}\\$c$} & \includegraphics[scale=0.3,trim={3cm 9cm 4cm 9.5cm},clip]{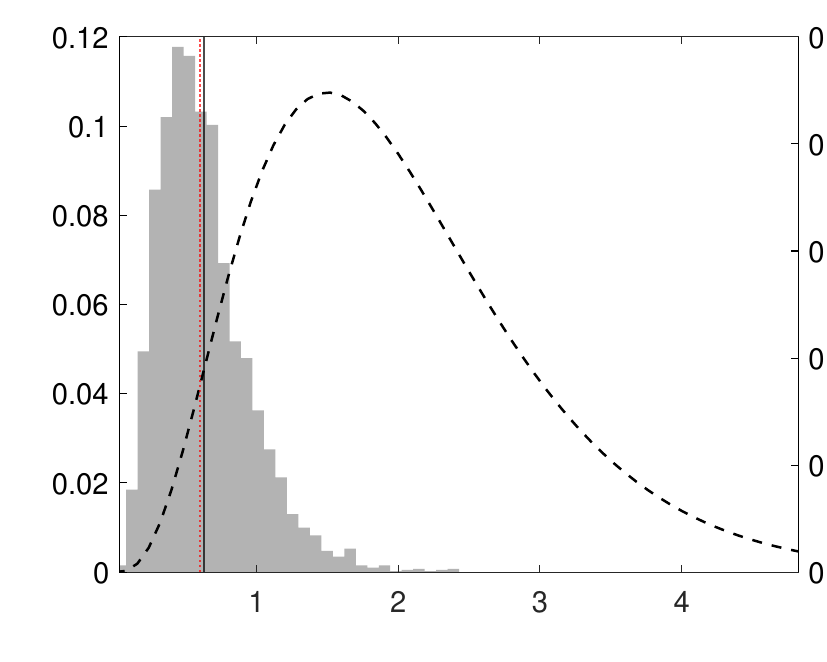} & \includegraphics[scale=0.3,trim={3cm 9cm 4cm 9.5cm},clip]{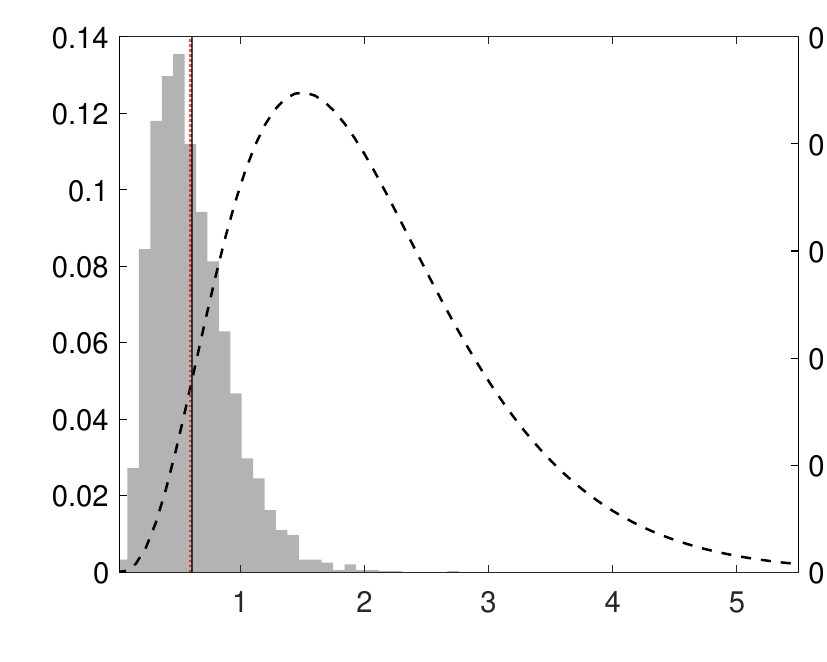}\\ 
			{\vspace{0.8cm}\\$\beta$} & \includegraphics[scale=0.3,trim={3cm 9cm 4cm 9.5cm},clip]{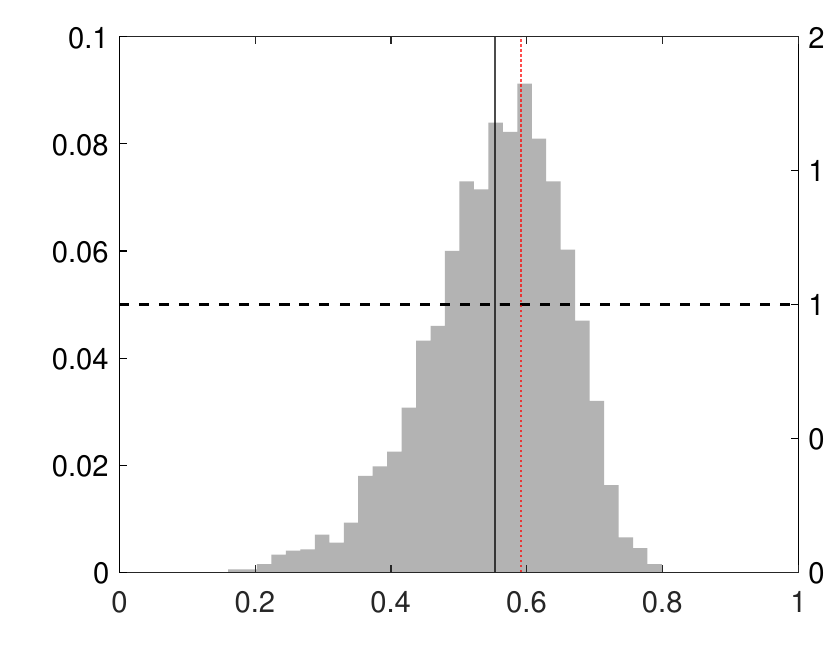} & \includegraphics[scale=0.3,trim={3cm 9cm 4cm 9.5cm},clip]{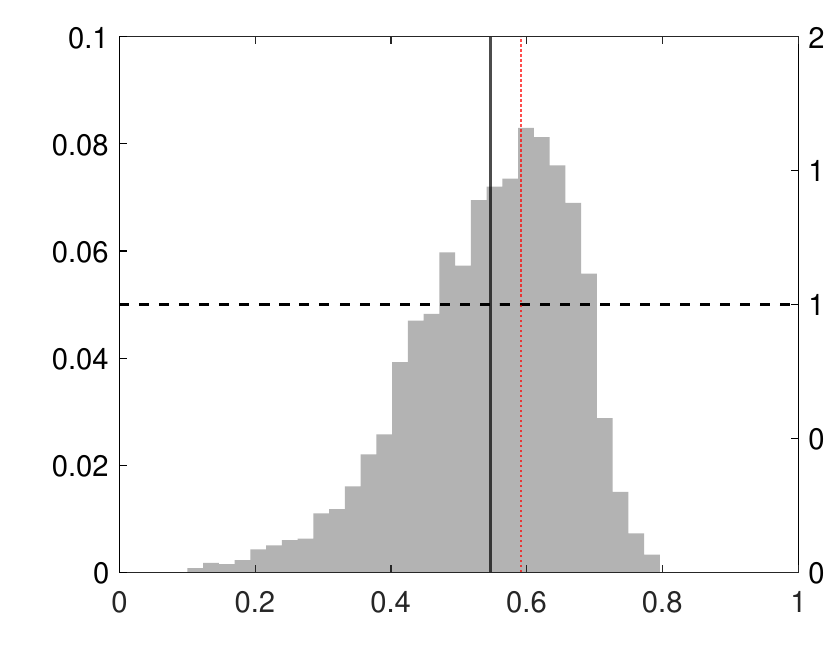}
		\end{tblr}
		
		\caption{ \color{black}MCMC approximation of the posterior distribution (histogram)of the MS--GLK--INAR(1) parameters with two regimes: High and low persistence and unconditional mean. In all plots, the estimated value (vertical black solid), the true value (vertical red dotted) and the prior density (dashed).\label{fig:twoMCMCh}}
	\end{figure}
	
		\begin{figure}[h!]
		\centering
		\begin{tblr}{Q[h,0em]cc}
			{\vspace{0.8cm}\\ $\alpha$ } &  \includegraphics[scale=0.3,trim={3cm 9cm 4cm 9.5cm},clip]{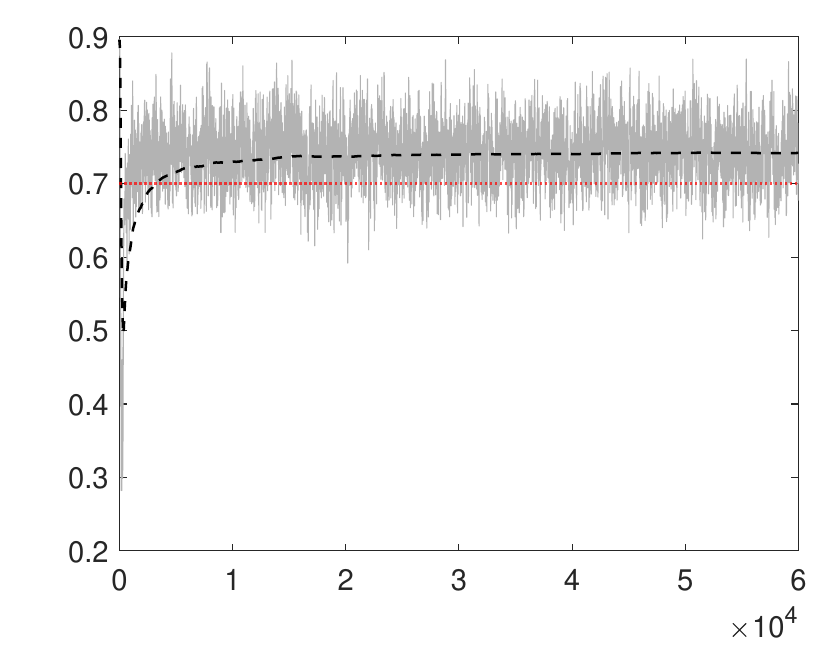} & \includegraphics[scale=0.3,trim={3cm 9cm 4cm 9.5cm},clip]{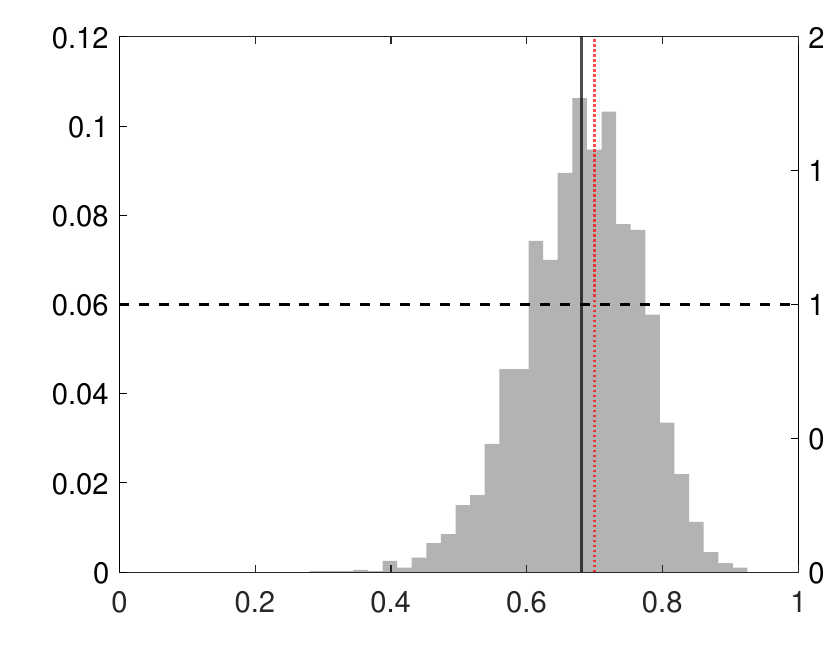}\\ 
			{\vspace{0.8cm}\\$a$} & \includegraphics[scale=0.3,trim={3cm 9cm 4cm 9.5cm},clip]{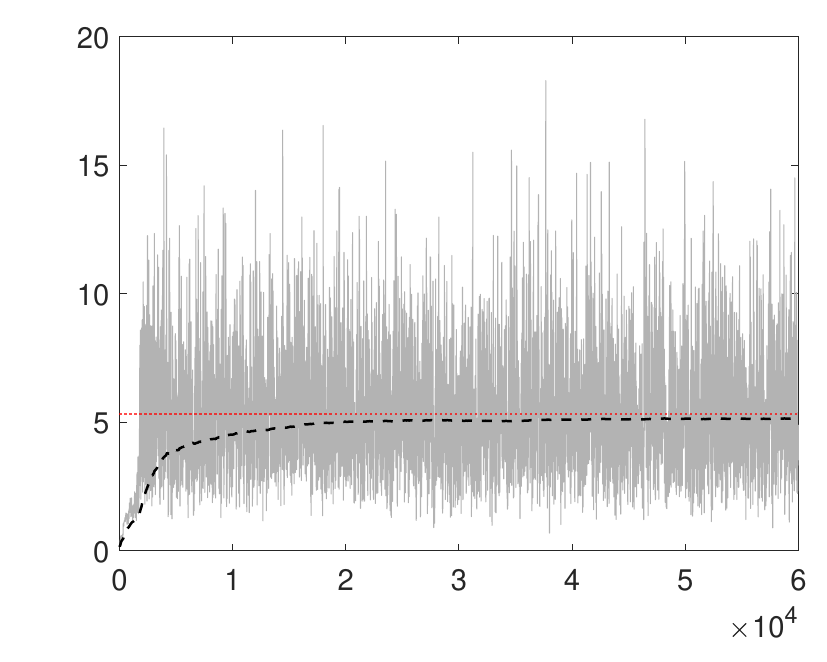} & \includegraphics[scale=0.3,trim={3cm 9cm 4cm 9.5cm},clip]{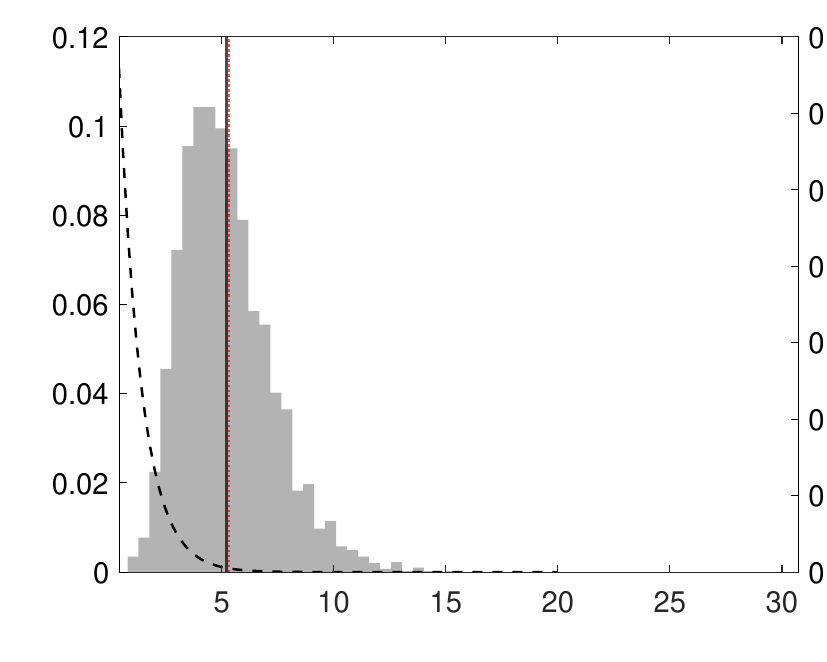}\\ 
			{\vspace{0.8cm}\\$b$}&  \includegraphics[scale=0.3,trim={3cm 9cm 4cm 9.5cm},clip]{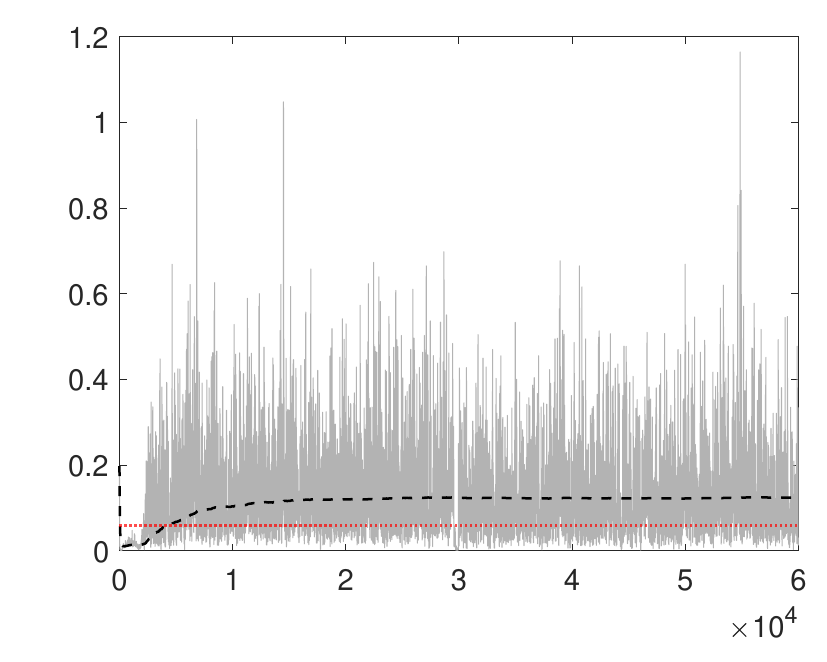} & \includegraphics[scale=0.3,trim={3cm 9cm 4cm 9.5cm},clip]{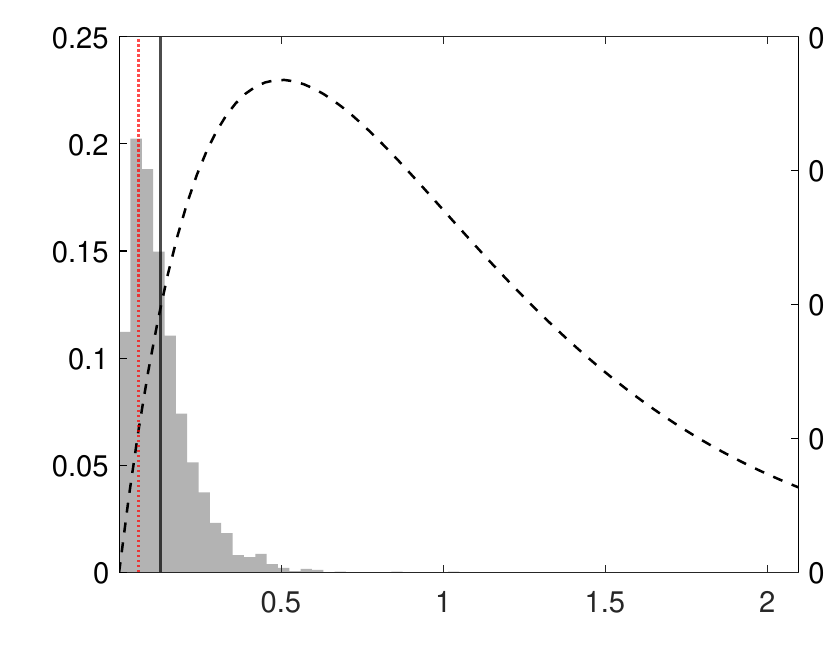} \\ 
			{\vspace{0.8cm}\\$c$} &  \includegraphics[scale=0.3,trim={3cm 9cm 4cm 9.5cm},clip]{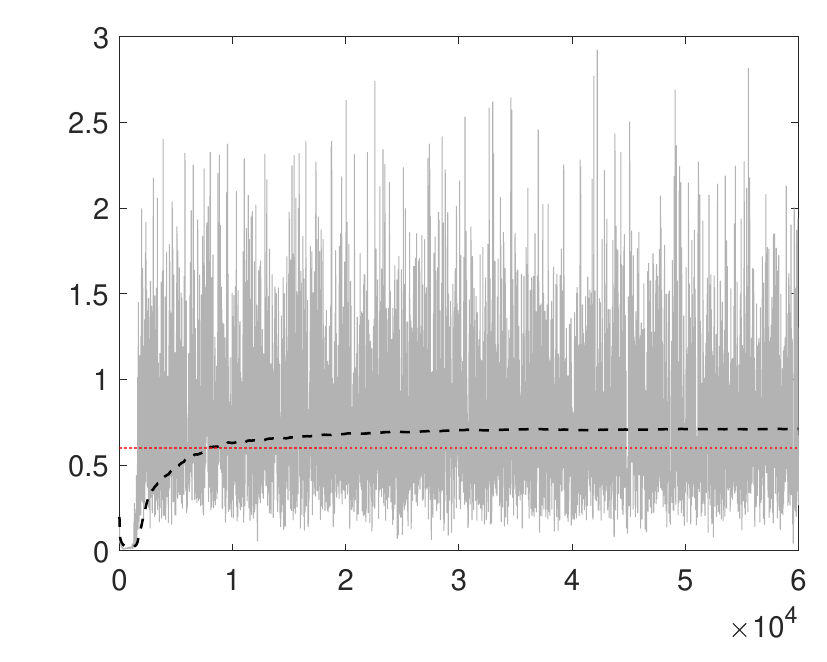}& \includegraphics[scale=0.3,trim={3cm 9cm 4cm 9.5cm},clip]{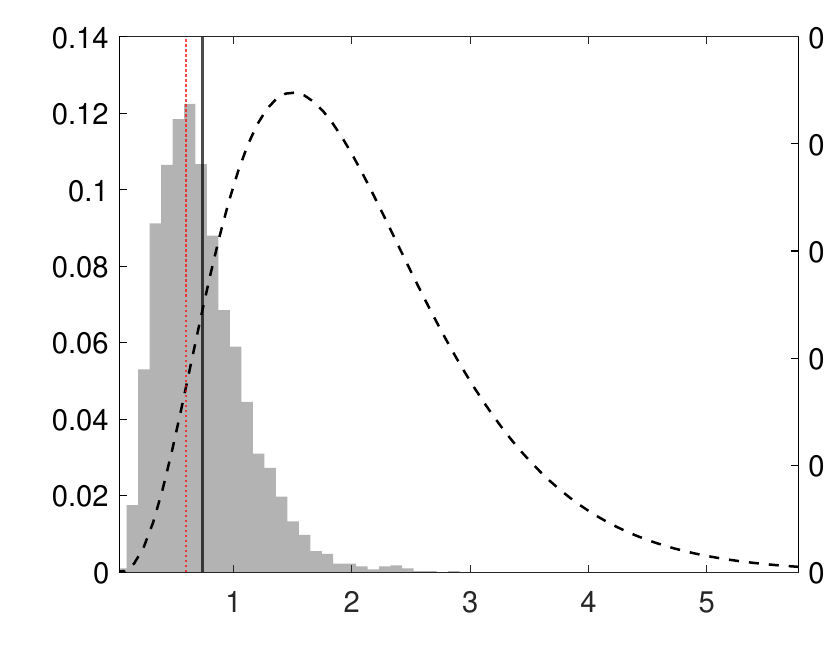}\\ 
			{\vspace{0.8cm}\\$\beta$}   & \includegraphics[scale=0.3,trim={3cm 9cm 4cm 9.5cm},clip]{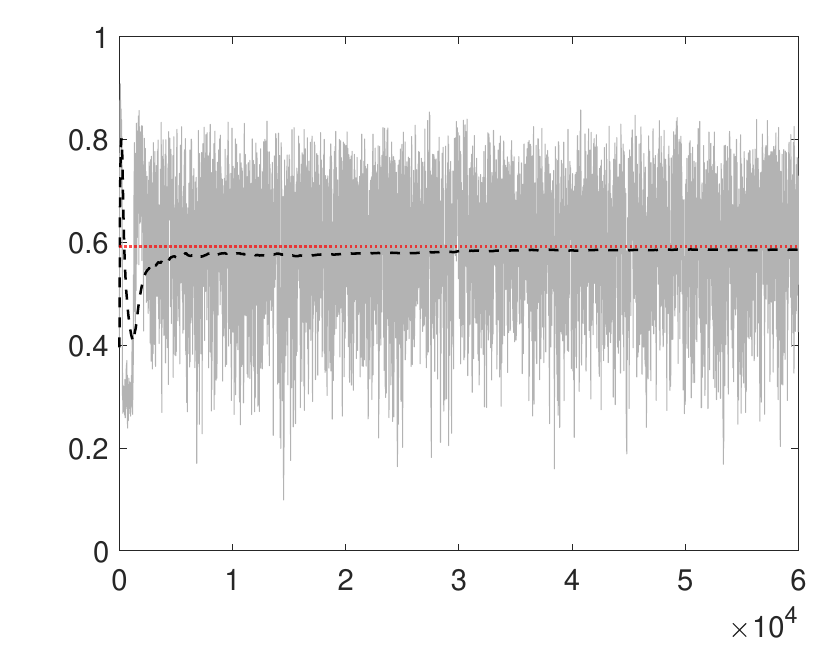} & \includegraphics[scale=0.3,trim={3cm 9cm 4cm 9.5cm},clip]{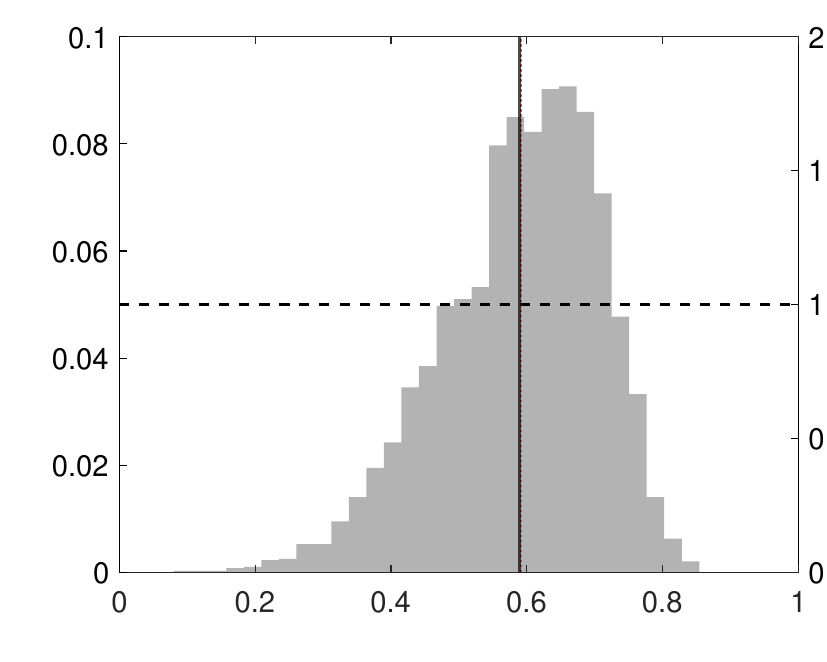}
		\end{tblr}
		
		\caption{ \color{black}MCMC trace plot (left column) and posterior approximation (histograms right column) for the parameters of the MS--GLK--INAR(1) with two regimes: Inflated--zero and High persistence and unconditional mean. In all plots, the MCMC draws (gray solid), the progressive MCMC average (dashed black) over the iterations (horizontal axis in thousands), and the true value of the parameter (red line). \label{fig:zeroMCMC}}
	\end{figure}
}

\pagebreak

\renewcommand\thefigure{D.\arabic{figure}}
\setcounter{figure}{0}
\renewcommand\theequation{D.\arabic{equation}}
\setcounter{equation}{0}
\renewcommand\thetable{D.\arabic{table}}
\setcounter{table}{0}

\section{Further real data results}\label{App:Google}
\vspace{-15pt}
\begin{figure}[h!]
\begin{center}
\renewcommand{\arraystretch}{0.8}
\setlength{\tabcolsep}{0pt}
\begin{tabular}{cc}
\multicolumn{2}{c}{Google search dataset ``Climate Change"}\\
$\alpha$& $a$\vspace{5pt}\\
\includegraphics[scale=0.37]{Figures/INARMCMCthetaHist1Google.eps}&
\includegraphics[scale=0.37]{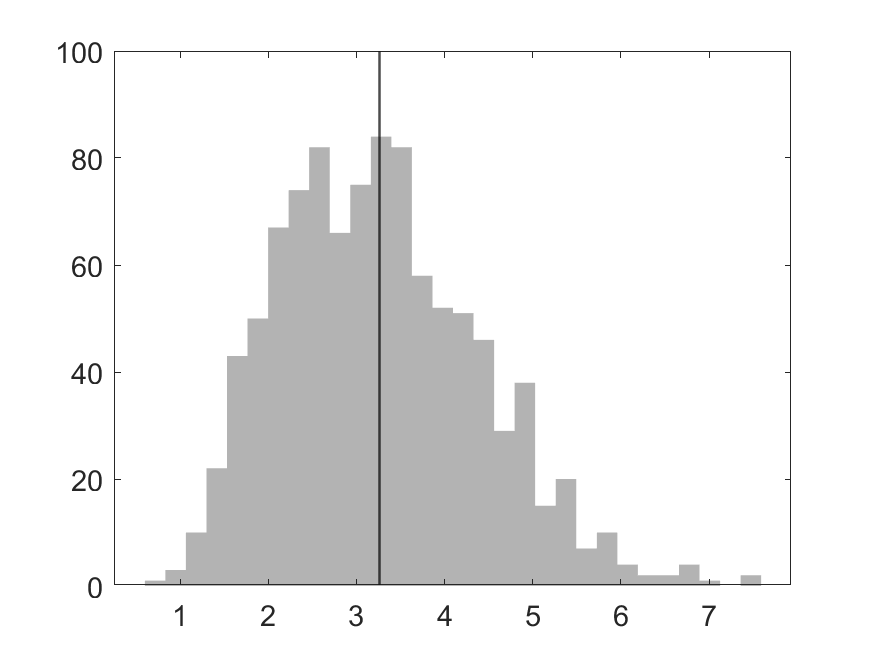}\\
$b$& $c$\vspace{5pt}\\
\includegraphics[scale=0.37]{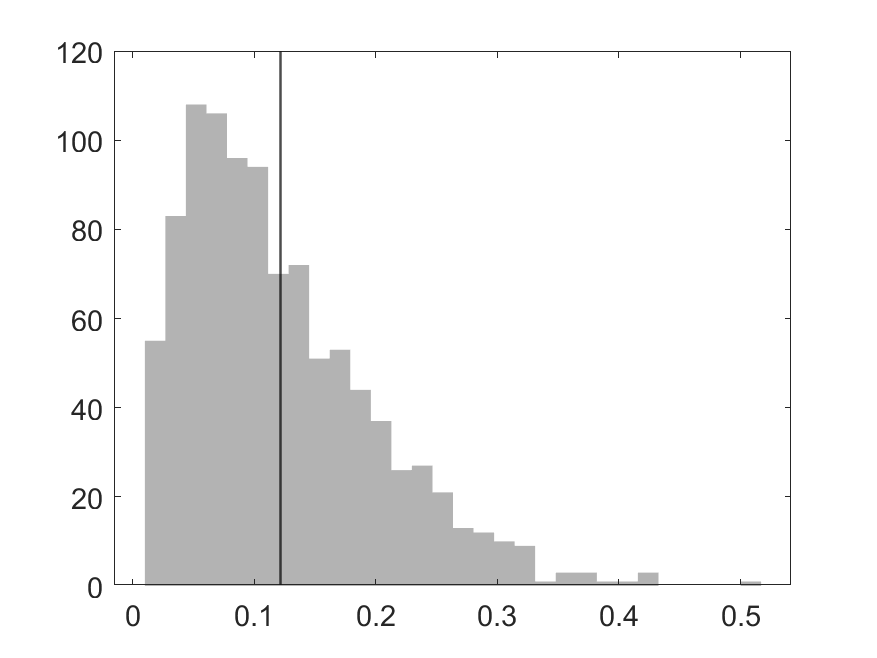}&
\includegraphics[scale=0.37]{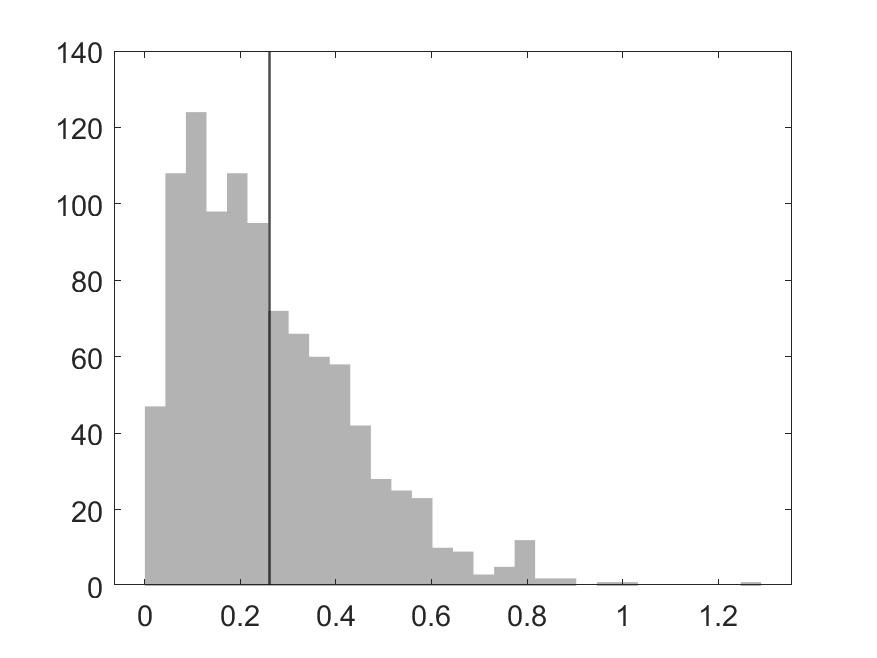}\\
$\beta$& $a\beta/(c-c\beta-\beta b)$\vspace{5pt}\\
\includegraphics[scale=0.37]{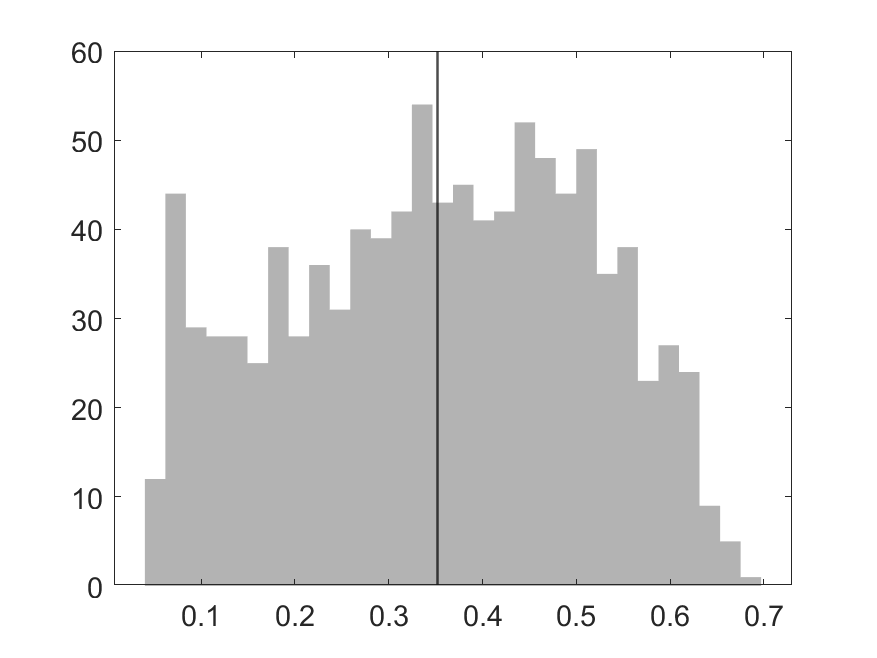}&
\includegraphics[scale=0.37]{Figures/INARMCMCthetaHistMomentGoogle.eps}\\
\end{tabular}
\end{center}
\caption{MCMC approximation of the posterior distribution (histogram) of the parameters. In all plots, the estimated value (vertical black solid).}\label{fig:mcmcHistGoogle}
\end{figure}
\begin{figure}[h!]
\begin{center}
\renewcommand{\arraystretch}{0.8}
\setlength{\tabcolsep}{0pt}
\begin{tabular}{cc}
\multicolumn{2}{c}{Google search dataset ``Global Warming"}\\
$\alpha$& $a$\vspace{5pt}\\
\includegraphics[scale=0.37]{Figures/INARMCMCthetaHist1GoogleWarm.eps}&
\includegraphics[scale=0.37]{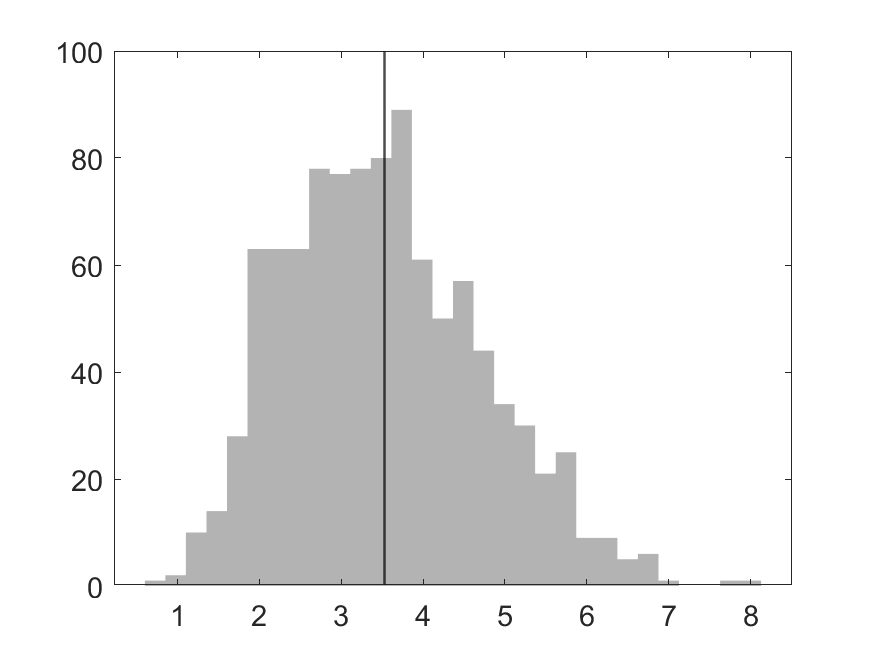}\\
$b$& $c$\vspace{5pt}\\
\includegraphics[scale=0.37]{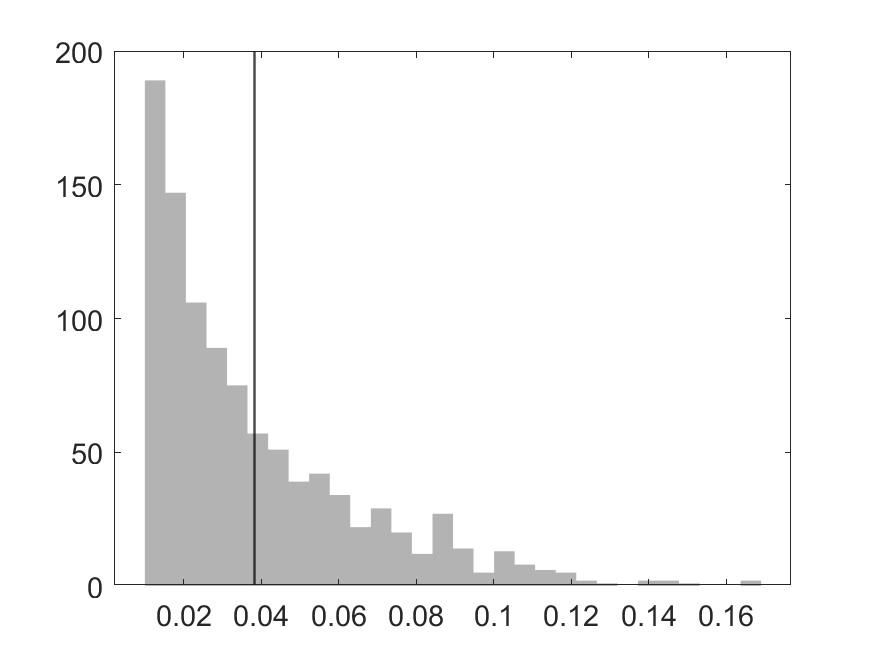}&
\includegraphics[scale=0.37]{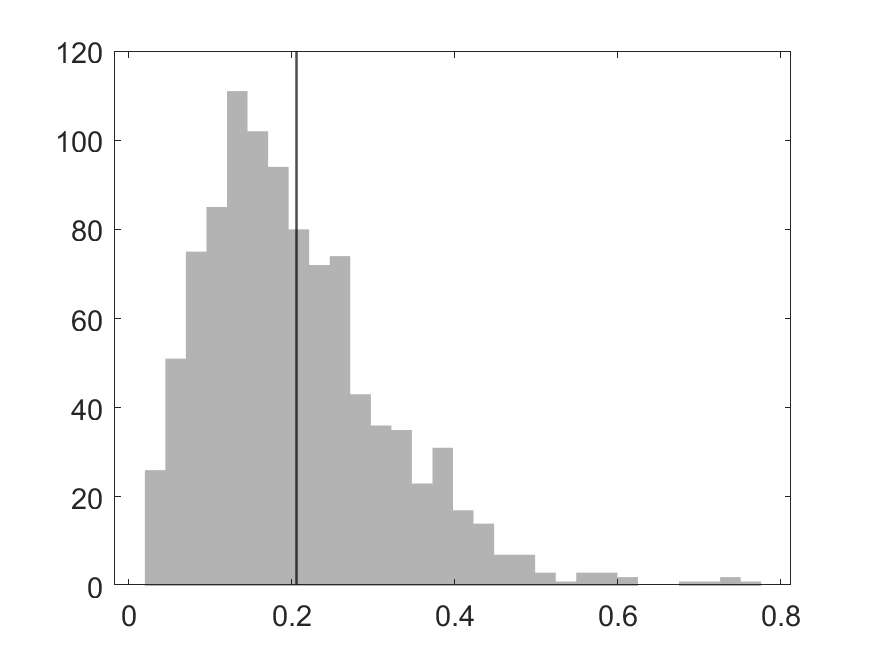}\\
$\beta$& $a\beta/(c-c\beta-\beta b)$\vspace{5pt}\\
\includegraphics[scale=0.37]{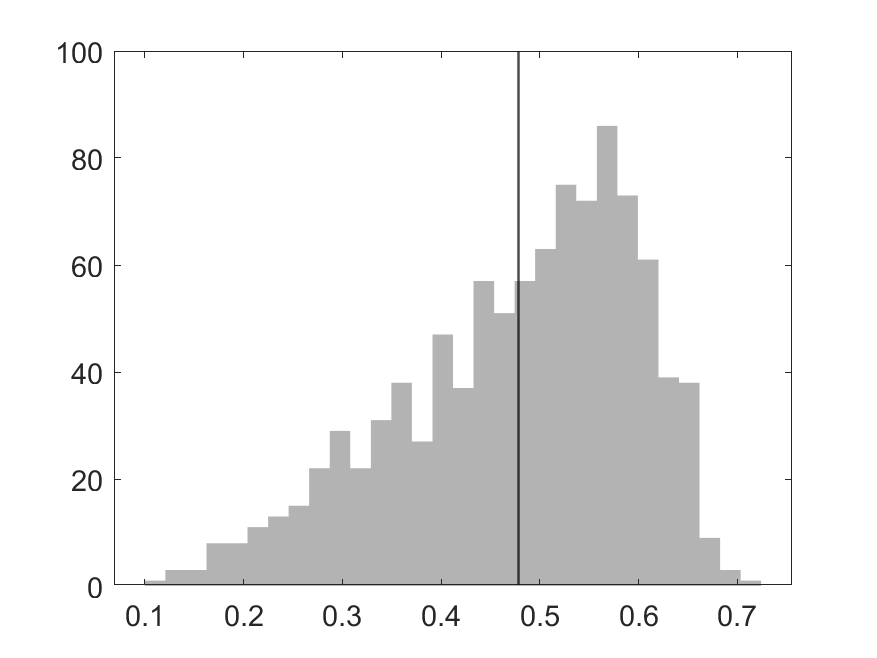}&
\includegraphics[scale=0.37]{Figures/INARMCMCthetaHistMomentGoogleWarm.eps}\\
\end{tabular}
\end{center}
\caption{MCMC approximation of the posterior distribution (histogram) of the parameters. In all plots, the estimated value (vertical black solid).}\label{fig:mcmcHistGoogleWarm}
\end{figure}

\begin{sidewaystable}[p]
\centering
\caption{\label{tab:pers1}Estimated GLK--INAR(1) autoregressive coefficient ($\widehat{\alpha}$) and its 95\% credible interval (CI), and marginal likelihood of the GLK--INAR(1) and the NBINAR(1) models, for the ``Climate Change"  and ``Global Warming" search volumes in different countries. Countries with less than 21\% of zeros in the two series. ``*" indicate the model with the largest marginal likelihood.}
\begin{scriptsize}
\begin{tabular}{c|cccc|cccc}
&\multicolumn{4}{c}{Climate Change dataset}&\multicolumn{4}{c}{Climate Warming dataset}\\
\toprule
Country & $\widehat{\alpha}$ & $CI$ &GLK& NB & $\widehat{\alpha}$ & $CI$ &GLK& NB\\
\hline
Australia&0.547&(0.501,0.589)& -882.43$^{\ast}$& -885.87&0.343&(0.287,0.397)&-1125.99&-1120.90$^{\ast}$\\
Bangladesh&0.018&(0.002,0.053)&-1118.68&-1111.99$^{\ast}$&0.001&(0.001,0.005)&-1166.74&-1162.50$^{\ast}$\\
Brazil&0.011&(0.001,0.029)&-1187.96&-1165.17$^{\ast}$&0.001&(0.001,0.002)&-1027.29&-1024.77$^{\ast}$\\
Canada&0.672&(0.615,0.714)& -711.55$^{\ast}$& -716.70&0.512&(0.468,0.554)&-1003.35&-1000.60$^{\ast}$\\
Emirates&0.010&(0.001,0.029)&-1200.72&-1182.74$^{\ast}$&0.005&(0.001,0.020)&-1158.09&-1149.59$^{\ast}$\\
France&0.184&(0.127,0.248)&-1069.66&-1068.36$^{\ast}$&0.006&(0.001,0.019)&-1140.59&-1130.91$^{\ast}$\\
Germany&0.298&(0.209,0.373)&-1023.87&-1022.87$^{\ast}$&0.021&(0.001,0.060)&-1099.04&-1094.57$^{\ast}$\\
India&0.499&(0.420,0.562)& -922.74$^{\ast}$& -923.57&0.482&(0.417,0.541)&-1019.07&-1018.57$^{\ast}$\\
Indonesia&0.021&(0.003,0.064)&-1123.06&-1108.15$^{\ast}$&0.253&(0.193,0.311)&-1195.34&-1172.30$^{\ast}$\\
Ireland&0.333&(0.279,0.388)& -953.35& -952.64$^{\ast}$&0.001&(0.001,0.002)&-1093.97&-1094.39$^{\ast}$\\
Italy&0.169&(0.090,0.241)& -983.83& -982.26$^{\ast}$&0.001&(0.001,0.004)&-1036.21&-1034.28$^{\ast}$\\
Malaysia&0.002&(0.001,0.008)&-1171.22&-1167.40$^{\ast}$&0.006&(0.001,0.031)&-1135.02&-1129.73$^{\ast}$\\
Mexico&0.009&(0.001,0.031)&-1160.49&-1148.85$^{\ast}$&0.001&(0.001,0.001)&-1098.96&-1097.73$^{\ast}$\\
Netherlands&0.148&(0.075,0.218)&-1064.29&-1061.30$^{\ast}$&0.001&(0.001,0.005)&-1066.26&-1059.54$^{\ast}$\\
NewZealand&0.353&(0.283,0.412)& -894.23$^{\ast}$& -895.29&0.013&(0.001,0.044)&-1151.01&-1136.31$^{\ast}$\\
Nigeria&0.129&(0.068,0.191)&-1043.77&-1041.80$^{\ast}$&0.017&(0.001,0.062)& -974.18& -966.83$^{\ast}$\\
Pakistan&0.212&(0.148,0.272)&-1131.60$^{\ast}$&-1124.74&0.023&(0.001,0.064)&-1135.66&-1128.55$^{\ast}$\\
Philippine&0.409&(0.354,0.460)&-1069.39&-1066.66$^{\ast}$&0.383&(0.327,0.432)&-1150.86&-1138.26$^{\ast}$\\
Singapore&0.159&(0.095,0.222)&-1099.02&-1094.02$^{\ast}$&0.006&(0.001,0.023)&-1132.94&-1122.04$^{\ast}$\\
SouthAfrica&0.413&(0.353,0.467)& -923.33$^{\ast}$& -925.21&0.334&(0.274,0.391)& -865.23$^{\ast}$& -868.48\\
Spain&0.253&(0.193,0.320)& -883.65$^{\ast}$& -886.68&0.001&(0.001,0.002)&-1049.48&-1041.62$^{\ast}$\\
Thailand&0.008&(0.001,0.035)&-1082.74&-1078.49$^{\ast}$&0.004&(0.001,0.012)&-1218.44&-1193.67$^{\ast}$\\
UK&0.535&(0.486,0.587)& -932.80$^{\ast}$& -937.25&0.319&(0.246,0.388)&-1084.84&-1081.81$^{\ast}$\\
US&0.601&(0.549,0.649)& -867.57$^{\ast}$& -871.12&0.606&(0.558,0.649)& -941.39$^{\ast}$& -942.22\\
Vietnam&0.003&(0.001,0.012)&-1189.79&-1178.82$^{\ast}$&0.001&(0.001,0.001)& -987.59& -986.21$^{\ast}$\\
\bottomrule
\end{tabular}
\end{scriptsize}
\end{sidewaystable}

\begin{sidewaystable}[p]
\centering
\caption{\label{tab:pers2}Estimated GLK--INAR(1) autoregressive coefficient ($\widehat{\alpha}$) and its 95\% credible interval (CI), and marginal likelihood of the GLK--INAR(1) and the NBINAR(1) models, for the ``Climate Change"  and ``Global Warming" search volumes in different countries. Countries with more than 21\% of zeros in the two series. ``*" indicate the model with the largest marginal likelihood.}
\begin{scriptsize}
\begin{tabular}{c|cccc|cccc}
&\multicolumn{4}{c}{Climate Change dataset}&\multicolumn{4}{c}{Climate Warming dataset}\\
Country & $\widehat{\alpha}$ & $CI$ &GLK & NB& $\widehat{\alpha}$ & $CI$ &GLK& NB\\
\toprule
Argentina&0.001&(0.001,0.001)&-1022.22$^{\ast}$&-1044.08&0.001&(0.001,0.001)& -803.30$^{\ast}$& -850.10\\
Austria&0.001&(0.001,0.001)&-1085.17&-1082.50$^{\ast}$&0.001&(0.001,0.001)& -717.08$^{\ast}$& -754.18\\
Belgium&0.002&(0.001,0.012)&-1141.59&-1136.92$^{\ast}$&0.001&(0.001,0.001)& -879.09$^{\ast}$& -944.73\\
Colombia&0.001&(0.001,0.002)&-1040.35$^{\ast}$&-1053.91&0.001&(0.001,0.001)& -848.15$^{\ast}$& -882.50\\
Denmark&0.008&(0.001,0.023)&-1123.97&-1101.34$^{\ast}$&0.002&(0.001,0.005)& -973.88& -950.20$^{\ast}$\\
Egypt&0.001&(0.001,0.002)&-1103.36$^{\ast}$&-1114.87&0.001&(0.001,0.002)& -855.84$^{\ast}$& -867.85\\
Ethiopia&0.002&(0.001,0.011)&-1089.74&-1081.68$^{\ast}$&0.001&(0.001,0.001)& -876.11$^{\ast}$& -914.27\\
Finland&0.027&(0.001,0.084)& -987.30& -983.16$^{\ast}$&0.001&(0.001,0.001)& -670.46$^{\ast}$& -679.12\\
Ghana&0.003&(0.001,0.012)& -992.09& -980.8$^{\ast}$3&0.001&(0.001,0.001)& -854.44$^{\ast}$& -922.58\\
Jamaica&0.001&(0.001,0.005)& -995.13& -994.24$^{\ast}$&0.001&(0.001,0.001)& -900.84$^{\ast}$& -947.66\\
Greece&0.001&(0.001,0.001)&-1070.08$^{\ast}$&-1095.69&0.001&(0.001,0.001)& -620.97$^{\ast}$& -687.79\\
HongKong&0.005&(0.001,0.040)&-1116.20&-1104.70$^{\ast}$&0.001&(0.001,0.001)&-1070.33$^{\ast}$&-1076.02\\
Iran&0.001&(0.001,0.002)&-1046.05$^{\ast}$&-1107.53&0.001&(0.001,0.002)& -945.79$^{\ast}$& -992.93\\
Israel&0.001&(0.001,0.001)& -914.20$^{\ast}$& -959.02&0.001&(0.001,0.001)& -726.50$^{\ast}$& -797.23\\
Japan&0.008&(0.001,0.026)&-1209.16&-1193.53$^{\ast}$&0.002&(0.001,0.004)&-1099.15$^{\ast}$&-1109.15\\
Kenya&0.104&(0.049,0.170)&-1100.01&-1092.16$^{\ast}$&0.001&(0.001,0.001)&-1073.55$^{\ast}$&-1096.71\\
Lebanon&0.001&(0.001,0.001)& -761.11$^{\ast}$& -776.44&0.001&(0.001,0.001)& -800.14$^{\ast}$& -850.50\\
Morocco&0.001&(0.001,0.001)& -755.97$^{\ast}$& -839.02&0.001&(0.001,0.002)& -471.85$^{\ast}$& -534.09\\
Mauritius&0.001&(0.001,0.002)& -887.72$^{\ast}$& -926.89&0.001&(0.001,0.001)& -601.12$^{\ast}$& -655.86\\
Myanmar&0.001&(0.001,0.001)& -917.99$^{\ast}$& -979.43&0.001&(0.001,0.002)& -602.14$^{\ast}$& -665.81\\
Nepal&0.001&(0.001,0.001)&-1148.05&-1145.39$^{\ast}$&0.001&(0.001,0.001)&-1027.36$^{\ast}$&-1060.00\\
Norway&0.016&(0.003,0.051)&-1121.71&-1105.17$^{\ast}$&0.001&(0.001,0.001)&-1002.83$^{\ast}$&-1004.64\\
Peru&0.001&(0.001,0.001)& -915.67$^{\ast}$& -950.87&0.001&(0.001,0.001)& -666.93$^{\ast}$& -712.80\\
Polish&0.001&(0.001,0.001)&-1078.86$^{\ast}$&-1090.00&0.001&(0.001,0.001)&-1000.76$^{\ast}$&-1063.99\\
Portugal&0.002&(0.001,0.010)&-1035.62&-1030.57$^{\ast}$&0.001&(0.001,0.001)& -800.35$^{\ast}$& -851.13\\
Qatar&0.001&(0.001,0.001)& -879.41$^{\ast}$& -917.09&0.001&(0.001,0.001)& -674.32$^{\ast}$& -701.04\\
Romania&0.001&(0.001,0.001)& -880.49$^{\ast}$& -901.20&0.001&(0.001,0.001)& -819.31$^{\ast}$& -896.17\\
Russia&0.001&(0.001,0.003)&-1038.70$^{\ast}$&-1050.47&0.001&(0.001,0.001)& -984.09$^{\ast}$&-1023.75\\
StHelena&0.001&(0.001,0.001)& -873.70$^{\ast}$& -914.40&0.001&(0.001,0.002)& -374.81$^{\ast}$& -394.04\\
SouthKorea&0.004&(0.001,0.019)&-1149.02$^{\ast}$&-1142.78&0.001&(0.001,0.003)&-1051.25$^{\ast}$&-1057.01\\
SriLanka&0.001&(0.001,0.001)&-1086.31$^{\ast}$&-1109.74&0.001&(0.001,0.003)& -842.37$^{\ast}$& -863.17\\
Sweden&0.136&(0.067,0.205)&-1031.72&-1026.82$^{\ast}$&0.001&(0.001,0.001)&-1078.30$^{\ast}$&-1089.70\\
Swiss&0.028&(0.005,0.063)&-1055.02&-1046.17$^{\ast}$&0.001&(0.001,0.001)& -835.78$^{\ast}$& -878.17\\
Taiwan&0.001&(0.001,0.001)&-1074.92$^{\ast}$&-1085.60&0.001&(0.001,0.001)& -794.34$^{\ast}$& -815.63\\
TrinidadTobago&0.001&(0.001,0.001)& -920.81$^{\ast}$& -955.62&0.001&(0.001,0.001)& -809.94$^{\ast}$& -858.50\\
Turkey&0.004&(0.001,0.019)&-1095.99&-1091.08$^{\ast}$&0.001&(0.001,0.001)&-1085.46$^{\ast}$&-1091.38\\
Ukraine&0.001&(0.001,0.001)& -902.24$^{\ast}$& -949.78&0.001&(0.001,0.002)& -709.80$^{\ast}$& -745.82\\
Hungary&0.001&(0.001,0.001)& -935.84& -953.58$^{\ast}$&0.001&(0.001,0.001)& -642.03$^{\ast}$& -717.52\\
Zambia&0.001&(0.001,0.003)&-1075.30&-1053.65$^{\ast}$&0.001&(0.001,0.001)& -746.07$^{\ast}$& -789.78\\
Zimbabwe&0.001&(0.001,0.002)&-1130.16&-1128.73$^{\ast}$&0.001&(0.001,0.002)& -658.24$^{\ast}$& -690.68\\
\bottomrule
\end{tabular}
\end{scriptsize}
\end{sidewaystable}

{\color{black}
	\begin{table}[h!]
		\caption{\color{black}In and out--of--sample performance of the Markov--Switching INAR(1) with two and three regimes (K) for the Climate change and Global warming datasets under GLK, Negative Binomial (NB) and Poisson (P) distribution including the Deviance Information Criterion (DIC), Root Mean Squared Error (RMSE), 90\% Credible Interval Coverage (CIcov) and 90\% Credible interval width (CIwidth)  }\label{tab:DICMSEaphor}
		{\color{black}	\begin{tabular}{lllcrrrr}
	\hline 
\multirow{2}{*}{Data} & \multirow{2}{*}{Distr.} & \multirow{2}{*}{K} & In--sample & & \multicolumn{3}{c}{Out--of--sample}\\
\cline{4-4} \cline{6-8}  
&  & & \multicolumn{1}{c}{DIC} & & RMSE & CIcov & CIwidth \\ 
\hline 
\multirow{6}{*}{Climate Change} & \multirow{2}{*}{GLK} & 2 & 1400.81 & & \textbf{2.65} & 1 & 14.00 \\ 
 &  & 3 & \textbf{1365.68} & & 2.68 & 1 & 14.50 \\ 
  & & & & & & & \\
 & \multirow{2}{*}{NB} & 2 & 1399.19 & & 2.68 & 1 & 13.10 \\ 
 &  & 3 & 1379.05 & & 2.68 & 1 & 13.80 \\ 
 & & & & & & & \\
 & \multirow{2}{*}{P} & 2 & 1661.23 & & 4.36 & 1 & 16.80 \\ 
 &  & 3 & 1454.00 & & 2.65 & 1 & 12.70 \\ 
 \hline
\multirow{6}{*}{Global Warming} & \multirow{2}{*}{GLK} & 2 & 1644.95 & & 4.86 & 1 & 22.80 \\ 
 &  & 3 & \textbf{1377.39} & & \textbf{4.67} & 1 & 21.70 \\ 
  & & & & & & & \\
 & \multirow{2}{*}{NB} & 2 & 1647.50 & & 4.80 & 1 & 21.40 \\ 
 &  & 3 & 1590.23 & & 4.69 & 1 & 20.60 \\ 
  & & & & & & & \\
 & \multirow{2}{*}{P} & 2 & 1653.30 & & 5.35 & 1 & 16.80 \\ 
 &  & 3 & 1633.60 & & 4.98 & 1 & 16.90 \\ 
\hline 
\end{tabular}}
\end{table}}

\begin{sidewaystable}
	\centering
	\caption{\color{black}\label{tab:MSallcou1}Estimated MS--GLK--INAR(1) autoregressive coefficient for the non--zero--inflated regime ($\widehat{\alpha}$) and the persistence of the zero--inflated state ($\widehat{\pi}_{11}$), and Deviance Information Criteria of the MS--GLK--INAR(1) and the MS--NBINAR(1) models, for the ``Climate Change"  and ``Global Warming" search volumes in different countries. ``*" indicate the model with the largest marginal likelihood.}
	\begin{tiny}
		{\color{black}
\begingroup\scriptsize
\begin{tabular}{cccccccccccccccc}
  \hline \multirow{4}{3cm}{\centering Country} & \multicolumn{7}{c}{\centering Climate Change dataset} & & \multicolumn{7}{c}{Global Warming dataset} \\ \cline{2-8} \cline{10-16} & \multicolumn{3}{c}{\centering GLK} & & \multicolumn{3}{c}{NB} & & \multicolumn{3}{c}{\centering GLK} & & \multicolumn{3}{c}{NB} \\ \cline{2-4} \cline{6-8} \cline{10-12} \cline{14-16} & $\hat{\alpha}$ & $\hat{\pi}_{11}$ & DIC & & $\hat{\alpha}$ & $\hat{\pi}_{11}$ & DIC & & $\hat{\alpha}$ & $\hat{\pi}_{11}$ & DIC & &$\hat{\alpha}$ & $\hat{\pi}_{11}$ & DIC \\ \hline Argentina & 0.089 & 0.345 & 1619.585$^{\ast}$ &  & 0.099 & 0.345 & 1780.791 &  & 0.039 & 0.607 & 1119.311$^{\ast}$ &  & 0.037 & 0.607 & 1396.970 \\ 
  Australia & 0.598 & 0.509 & 1596.879$^{\ast}$ &  & 0.600 & 0.466 & 1602.322 &  & 0.352 & 0.170 & 2078.651 &  & 0.371 & 0.093 & 2070.579$^{\ast}$ \\ 
  Austria & 0.090 & 0.108 & 1872.957$^{\ast}$ &  & 0.100 & 0.099 & 1933.522 &  & 0.122 & 0.626 & 1074.436$^{\ast}$ &  & 0.127 & 0.627 & 1363.791 \\ 
  Bangladesh & 0.036 & 0.070 & 2045.797$^{\ast}$ &  & 0.043 & 0.071 & 2055.200 &  & 0.074 & 0.097 & 1965.099$^{\ast}$ &  & 0.086 & 0.095 & 2033.827 \\ 
  Belgium & 0.148 & 0.169 & 2007.494$^{\ast}$ &  & 0.170 & 0.160 & 2039.671 &  & 0.070 & 0.550 & 1294.334$^{\ast}$ &  & 0.072 & 0.550 & 1544.908 \\ 
  Brazil & 0.093 & 0.223 & 2053.698$^{\ast}$ &  & 0.105 & 0.220 & 2079.390 &  & 0.074 & 0.193 & 1731.327$^{\ast}$ &  & 0.079 & 0.188 & 1826.657 \\ 
  Canada & 0.672 & 0.508 & 1327.845$^{\ast}$ &  & 0.664 & 0.495 & 1340.503 &  & 0.522 & 0.499 & 1879.075 &  & 0.524 & 0.472 & 1876.222$^{\ast}$ \\ 
  Colombia & 0.090 & 0.398 & 1667.121$^{\ast}$ &  & 0.000 & 0.397 & 1803.969 &  & 0.192 & 0.553 & 1259.727$^{\ast}$ &  & 0.198 & 0.552 & 1498.598 \\ 
  Denmark & 0.107 & 0.255 & 1898.035$^{\ast}$ &  & 0.116 & 0.249 & 1948.963 &  & 0.109 & 0.519 & 1462.477$^{\ast}$ &  & 0.111 & 0.518 & 1659.000 \\ 
  Egypt & 0.066 & 0.244 & 1780.555$^{\ast}$ &  & 0.071 & 0.241 & 1891.883 &  & 0.137 & 0.414 & 1347.583$^{\ast}$ &  & 0.138 & 0.413 & 1563.709 \\ 
  Emirates & 0.160 & 0.071 & 2074.214$^{\ast}$ &  & 0.173 & 0.063 & 2105.633 &  & 0.135 & 0.256 & 1996.096$^{\ast}$ &  & 0.143 & 0.248 & 2039.387 \\ 
  Ethiopia & 0.104 & 0.153 & 1867.472$^{\ast}$ &  & 0.106 & 0.150 & 1924.091 &  & 0.095 & 0.476 & 1337.687$^{\ast}$ &  & 0.092 & 0.475 & 1572.360 \\ 
  Finland & 0.187 & 0.314 & 1724.529$^{\ast}$ &  & 0.204 & 0.291 & 1784.135 &  & 0.057 & 0.645 & 1069.114$^{\ast}$ &  & 0.055 & 0.646 & 1340.069 \\ 
  France & 0.168 & 0.496 & 1935.826 &  & 0.179 & 0.491 & 1933.377$^{\ast}$ &  & 0.114 & 0.244 & 1951.947$^{\ast}$ &  & 0.127 & 0.239 & 2007.952 \\ 
  Germany & 0.293 & 0.514 & 1896.652 &  & 0.294 & 0.498 & 1895.353$^{\ast}$ &  & 0.048 & 0.098 & 2020.428$^{\ast}$ &  & 0.053 & 0.097 & 2025.436 \\ 
  Ghana & 0.151 & 0.240 & 1631.274$^{\ast}$ &  & 0.163 & 0.221 & 1711.621 &  & 0.092 & 0.556 & 1253.993$^{\ast}$ &  & 0.091 & 0.556 & 1512.884 \\ 
  Greece & 0.125 & 0.276 & 1692.857$^{\ast}$ &  & 0.143 & 0.273 & 1837.323 &  & 0.042 & 0.717 & 860.269$^{\ast}$ &  & 0.533 & 0.006 & 1225.069 \\ 
  HongKong & 0.101 & 0.311 & 1901.765$^{\ast}$ &  & 0.114 & 0.307 & 1954.842 &  & 0.163 & 0.315 & 1781.510$^{\ast}$ &  & 0.183 & 0.307 & 1896.437 \\ 
  Hungary & 0.128 & 0.405 & 1476.092$^{\ast}$ &  & 0.132 & 0.406 & 1663.531 &  & 0.033 & 0.663 & 931.925$^{\ast}$ &  & 0.018 & 0.665 & 1256.010 \\ 
  India & 0.534 & 0.491 & 1714.041$^{\ast}$ &  & 0.531 & 0.508 & 1714.829 &  & 0.636 & 0.511 & 1835.222$^{\ast}$ &  & 0.635 & 0.505 & 1839.658 \\ 
  Indonesia & 0.088 & 0.115 & 1977.859$^{\ast}$ &  & 0.099 & 0.108 & 1997.493 &  & 0.361 & 0.162 & 2108.212 &  & 0.359 & 0.424 & 2102.328$^{\ast}$ \\ 
  Iran & 0.105 & 0.362 & 1619.937$^{\ast}$ &  & 0.111 & 0.361 & 1806.291 &  & 0.092 & 0.497 & 1391.738$^{\ast}$ &  & 0.092 & 0.495 & 1625.134 \\ 
  Ireland & 0.395 & 0.114 & 1703.061 &  & 0.431 & 0.047 & 1702.772$^{\ast}$ &  & 0.162 & 0.218 & 1872.783$^{\ast}$ &  & 0.182 & 0.197 & 1947.321 \\ 
  Israel & 0.059 & 0.499 & 1399.355$^{\ast}$ &  & 0.059 & 0.499 & 1623.913 &  & 0.045 & 0.620 & 1070.880$^{\ast}$ &  & 0.045 & 0.619 & 1364.546 \\ 
  Italy & 0.186 & 0.069 & 1779.219$^{\ast}$ &  & 0.211 & 0.050 & 1780.901 &  & 0.149 & 0.252 & 1787.010$^{\ast}$ &  & 0.169 & 0.233 & 1864.037 \\ 
  Jamaica & 0.105 & 0.413 & 1646.154$^{\ast}$ &  & 0.121 & 0.408 & 1765.751 &  & 0.068 & 0.573 & 1313.182$^{\ast}$ &  & 0.067 & 0.573 & 1557.623 \\ 
  Japan & 0.094 & 0.016 & 2097.733$^{\ast}$ &  & 0.105 & 0.017 & 2120.688 &  & 0.071 & 0.264 & 1773.651$^{\ast}$ &  & 0.076 & 0.263 & 1887.743 \\ 
  Kenya & 0.175 & 0.181 & 1929.977$^{\ast}$ &  & 0.183 & 0.178 & 1938.944 &  & 0.103 & 0.361 & 1655.868$^{\ast}$ &  & 0.106 & 0.362 & 1807.085 \\ 
  Lebanon & 0.104 & 0.538 & 1165.567$^{\ast}$ &  & 0.109 & 0.538 & 1420.193 &  & 0.038 & 0.647 & 1187.075$^{\ast}$ &  & 0.036 & 0.649 & 1455.428 \\ 
  Malaysia & 0.182 & 0.158 & 2070.368$^{\ast}$ &  & 0.187 & 0.155 & 2079.151 &  & 0.084 & 0.027 & 2027.732$^{\ast}$ &  & 0.095 & 0.029 & 2042.069 \\ 
  Mauritius & 0.032 & 0.500 & 1326.807$^{\ast}$ &  & 0.029 & 0.500 & 1566.223 &  & 0.131 & 0.704 & 945.556$^{\ast}$ &  & 0.132 & 0.704 & 1272.587 \\ 
  Mexico & 0.099 & 0.097 & 2041.225$^{\ast}$ &  & 0.107 & 0.096 & 2063.796 &  & 0.134 & 0.266 & 1873.485$^{\ast}$ &  & 0.142 & 0.263 & 1942.329 \\ 
  Morocco & 0.036 & 0.671 & 1059.084$^{\ast}$ &  & 0.037 & 0.672 & 1363.009 &  & 0.114 & 0.832 & 625.621$^{\ast}$ &  & 0.127 & 0.832 & 1009.280 \\ 
   \hline
\end{tabular}
\endgroup
}
			\end{tiny}
	\end{sidewaystable}

	\begin{sidewaystable}
			\caption{\color{black}\label{tab:MSallcou2}Estimated MS--GLK--INAR(1) autoregressive coefficient for the non--zero--inflated regime ($\widehat{\alpha}$) and the persistence of the zero--inflated state ($\widehat{\pi}_{11}$), and Deviance Information Criteria of the MS--GLK--INAR(1) and the MS--NBINAR(1) models, for the ``Climate Change"  and ``Global Warming" search volumes in different countries. ``*" indicate the model with the largest marginal likelihood.}
		{\color{black}
		\begin{tiny}
\begingroup\scriptsize
\begin{tabular}{cccccccccccccccc}
  \hline \multirow{4}{3cm}{\centering Country} & \multicolumn{7}{c}{\centering Climate Change dataset} & & \multicolumn{7}{c}{Global Warming dataset} \\ \cline{2-8} \cline{10-16} & \multicolumn{3}{c}{\centering GLK} & & \multicolumn{3}{c}{NB} & & \multicolumn{3}{c}{\centering GLK} & & \multicolumn{3}{c}{NB} \\ \cline{2-4} \cline{6-8} \cline{10-12} \cline{14-16} & $\hat{\alpha}$ & $\hat{\pi}_{11}$ & DIC & & $\hat{\alpha}$ & $\hat{\pi}_{11}$ & DIC & & $\hat{\alpha}$ & $\hat{\pi}_{11}$ & DIC & &$\hat{\alpha}$ & $\hat{\pi}_{11}$ & DIC \\ \hline Myanmar & 0.096 & 0.488 & 1371.038$^{\ast}$ &  & 0.093 & 0.487 & 1609.242 &  & 0.066 & 0.738 & 813.777$^{\ast}$ &  & 0.066 & 0.738 & 1166.376 \\ 
  Nepal & 0.109 & 0.136 & 1965.494$^{\ast}$ &  & 0.116 & 0.132 & 2023.832 &  & 0.104 & 0.432 & 1593.188$^{\ast}$ &  & 0.115 & 0.430 & 1769.996 \\ 
  Netherlands & 0.147 & 0.496 & 1972.367 &  & 0.152 & 0.488 & 1969.887$^{\ast}$ &  & 0.055 & 0.315 & 1792.008$^{\ast}$ &  & 0.000 & 0.314 & 1890.708 \\ 
  NewZealand & 0.362 & 0.492 & 1659.492$^{\ast}$ &  & 0.370 & 0.356 & 1662.115 &  & 0.158 & 0.322 & 1980.746$^{\ast}$ &  & 0.167 & 0.320 & 2033.827 \\ 
  Nigeria & 0.161 & 0.097 & 1932.834$^{\ast}$ &  & 0.170 & 0.084 & 1934.833 &  & 0.106 & 0.122 & 1719.025$^{\ast}$ &  & 0.120 & 0.085 & 1766.515 \\ 
  Norway & 0.137 & 0.259 & 1890.741$^{\ast}$ &  & 0.151 & 0.255 & 1943.169 &  & 0.067 & 0.413 & 1639.325$^{\ast}$ &  & 0.073 & 0.412 & 1794.887 \\ 
  Pakistan & 0.252 & 0.099 & 2043.832 &  & 0.259 & 0.093 & 2043.666$^{\ast}$ &  & 0.114 & 0.068 & 2047.780$^{\ast}$ &  & 0.116 & 0.064 & 2052.311 \\ 
  Peru & 0.123 & 0.465 & 1387.617$^{\ast}$ &  & 0.115 & 0.463 & 1602.407 &  & 0.054 & 0.687 & 937.838$^{\ast}$ &  & 0.052 & 0.688 & 1255.208 \\ 
  Philippine & 0.497 & 0.493 & 1921.648 &  & 0.498 & 0.494 & 1917.078$^{\ast}$ &  & 0.409 & 0.065 & 2064.834 &  & 0.411 & 0.060 & 2063.957$^{\ast}$ \\ 
  Polish & 0.069 & 0.229 & 1767.794$^{\ast}$ &  & 0.077 & 0.229 & 1882.671 &  & 0.090 & 0.401 & 1560.757$^{\ast}$ &  & 0.096 & 0.400 & 1752.885 \\ 
  Portugal & 0.041 & 0.277 & 1727.197$^{\ast}$ &  & 0.041 & 0.273 & 1827.148 &  & 0.031 & 0.587 & 1215.954$^{\ast}$ &  & 0.030 & 0.586 & 1478.256 \\ 
  Qatar & 0.058 & 0.520 & 1291.855$^{\ast}$ &  & 0.064 & 0.521 & 1539.752 &  & 0.042 & 0.554 & 1006.056$^{\ast}$ &  & 0.041 & 0.551 & 1307.701 \\ 
  Romania & 0.097 & 0.458 & 1351.055$^{\ast}$ &  & 0.098 & 0.457 & 1558.692 &  & 0.053 & 0.626 & 1211.659$^{\ast}$ &  & 0.052 & 0.625 & 1477.310 \\ 
  Russia & 0.075 & 0.368 & 1553.053$^{\ast}$ &  & 0.075 & 0.370 & 1720.946 &  & 0.082 & 0.449 & 1524.936$^{\ast}$ &  & 0.081 & 0.448 & 1715.720 \\ 
  Singapore & 0.199 & 0.165 & 2010.912 &  & 0.207 & 0.158 & 2009.881$^{\ast}$ &  & 0.072 & 0.073 & 2002.081$^{\ast}$ &  & 0.087 & 0.066 & 2035.269 \\ 
  SouthAfrica & 0.522 & 0.388 & 1630.689$^{\ast}$ &  & 0.530 & 0.435 & 1632.432 &  & 0.418 & 0.120 & 1570.317$^{\ast}$ &  & 0.441 & 0.084 & 1571.766 \\ 
  SouthKorea & 0.054 & 0.020 & 2025.651$^{\ast}$ &  & 0.061 & 0.021 & 2046.260 &  & 0.067 & 0.415 & 1599.901$^{\ast}$ &  & 0.071 & 0.415 & 1755.566 \\ 
  Spain & 0.246 & 0.518 & 1651.155$^{\ast}$ &  & 0.253 & 0.414 & 1655.710 &  & 0.126 & 0.312 & 1759.649$^{\ast}$ &  & 0.133 & 0.306 & 1850.803 \\ 
  SriLanka & 0.050 & 0.345 & 1670.538$^{\ast}$ &  & 0.050 & 0.341 & 1818.467 &  & 0.044 & 0.517 & 1268.525$^{\ast}$ &  & 0.042 & 0.516 & 1515.381 \\ 
  StHelena & 0.132 & 0.484 & 1338.414$^{\ast}$ &  & 0.133 & 0.482 & 1572.449 &  & 0.080 & 0.817 & 594.577$^{\ast}$ &  & 0.583 & 0.007 & 860.226 \\ 
  Sweden & 0.160 & 0.262 & 1883.905$^{\ast}$ &  & 0.169 & 0.257 & 1899.062 &  & 0.060 & 0.347 & 1829.726$^{\ast}$ &  & 0.066 & 0.345 & 1935.821 \\ 
  Swiss & 0.083 & 0.087 & 1888.159$^{\ast}$ &  & 0.090 & 0.075 & 1913.949 &  & 0.098 & 0.551 & 1283.276$^{\ast}$ &  & 0.098 & 0.550 & 1529.698 \\ 
  Taiwan & 0.116 & 0.324 & 1714.102$^{\ast}$ &  & 0.119 & 0.322 & 1839.746 &  & 0.126 & 0.460 & 1245.494$^{\ast}$ &  & 0.119 & 0.460 & 1483.356 \\ 
  Thailand & 0.132 & 0.022 & 1928.947$^{\ast}$ &  & 0.143 & 0.023 & 1948.096 &  & 0.027 & 0.154 & 2014.468$^{\ast}$ &  & 0.029 & 0.152 & 2072.000 \\ 
  TrinidadTobago & 0.077 & 0.447 & 1359.122$^{\ast}$ &  & 0.076 & 0.446 & 1587.225 &  & 0.114 & 0.585 & 1217.544$^{\ast}$ &  & 0.109 & 0.586 & 1477.778 \\ 
  Turkey & 0.096 & 0.373 & 1854.282$^{\ast}$ &  & 0.106 & 0.370 & 1927.291 &  & 0.135 & 0.480 & 1793.179$^{\ast}$ &  & 0.148 & 0.478 & 1905.162 \\ 
  UK & 0.571 & 0.510 & 1696.845$^{\ast}$ &  & 0.568 & 0.454 & 1701.288 &  & 0.322 & 0.507 & 2016.025 &  & 0.324 & 0.498 & 2012.407$^{\ast}$ \\ 
  Ukraine & 0.111 & 0.457 & 1349.152$^{\ast}$ &  & 0.113 & 0.458 & 1578.258 &  & 0.049 & 0.597 & 1002.874$^{\ast}$ &  & 0.048 & 0.595 & 1311.657 \\ 
  US & 0.632 & 0.520 & 1623.722$^{\ast}$ &  & 0.624 & 0.506 & 1634.120 &  & 0.636 & 0.509 & 1761.454$^{\ast}$ &  & 0.632 & 0.511 & 1765.975 \\ 
  Vietnam & 0.100 & 0.149 & 2053.407$^{\ast}$ &  & 0.109 & 0.147 & 2091.799 &  & 0.212 & 0.180 & 1777.152$^{\ast}$ &  & 0.336 & 0.035 & 1804.133 \\ 
  Zambia & 0.102 & 0.262 & 1736.355$^{\ast}$ &  & 0.111 & 0.254 & 1849.733 &  & 0.131 & 0.589 & 1111.358$^{\ast}$ &  & 0.123 & 0.588 & 1396.448 \\ 
  Zimbabwe & 0.170 & 0.308 & 1808.413$^{\ast}$ &  & 0.192 & 0.303 & 1913.692 &  & 0.050 & 0.659 & 862.586$^{\ast}$ &  & 0.049 & 0.661 & 1197.766 \\ 
   \hline
\end{tabular}
\endgroup

		\end{tiny}}
	\end{sidewaystable}

\end{document}